\documentclass[twocolumn,aps]{revtex4}

\usepackage{graphicx}
\usepackage{epstopdf}
\usepackage{physics}
\begin{document}

\title{\bf Recent experimental results in sub- and near-barrier  heavy ion fusion reactions}

\author{ Giovanna Montagnoli$^{1}$, Alberto M. Stefanini$^2$}
\affiliation{\medskip
$^1$ Dipartimento di Fisica e Astronomia, Universit\`a di Padova, and INFN, Sez. di Padova, I-35131 Padova, Italy\\
$^2$ INFN, Laboratori Nazionali di Legnaro, I-35020 Legnaro (Padova), Italy
}
\date{\today}

\begin{abstract}
Recent advances obtained  in the field of 
near and sub-barrier heavy-ion fusion reactions are reviewed. Emphasis is given to the results obtained in the last decade,
and focus will be mainly on the experimental work performed concerning
the  influence of transfer channels on fusion cross sections and the hindrance 
phenomenon far below the barrier. 
Indeed, early data of sub-barrier fusion taught us that cross sections may
strongly depend on the low-energy collective modes of the colliding nuclei, and, possibly,  on 
couplings to transfer channels. The coupled-channels (CC) model has been quite successful in the interpretation of the experimental evidences. Fusion barrier distributions often yield the fingerprint of the
relevant coupled channels. 
Recent results obtained by using radioactive beams are reported.
At deep sub-barrier energies, the slope of the excitation function in a semi-logarithmic plot keeps increasing in many
cases and standard CC calculations overpredict the cross sections. 
This was named a hindrance phenomenon, and its physical origin is still a matter of debate. Recent theoretical developments suggest that this effect, at least partially, may be a consequence of the Pauli exclusion principle.
The hindrance may have far-reaching consequences in astrophysics where fusion of light systems determines stellar evolution during the carbon and oxygen burning stages, and yields important information for exotic reactions that take place in the inner crust of accreting neutron stars.

\end{abstract}
\par
\maketitle

\tableofcontents

\section{Introduction}

This review is a selection of the most significant results obtained in the study of the dynamics of heavy-ion fusion $\it{near}$ the Coulomb barrier, with a particular emphasis on recent achievements and results, and on the experimental set-ups and methods employed for fusion cross section measurements. This review will be essentially limited to the field of fusion dynamics in medium-mass systems. We will not deal with neighbouring fields such as fusion of light and weakly bound nuclei and the production of superheavy elements. However, a few hints will be given to consequences of the fusion of light systems far below the barrier for astrophysics.

Heavy-ion fusion is a quite complex phenomenon whose study has been involving several experimental
and theoretical efforts, after large Tandem electrostatic accelerators have been put into operation, and have allowed to produce medium-mass heavy-ion beams with sufficient energy to overcome the Coulomb barrier in collisions with targets of nearly all elements.
A specific interest in the study of the fusion dynamics evolved in the seventies, following the awareness that
fusion reactions between heavy stable nuclei 1) can produce exotic nuclei away from stability on the proton-rich side of the mass
valley and 2) they are crucial for the synthesis of very heavy elements. 

There was actually optimism that superheavy elements could be produced in such a way. Soon, it was recognised that this was not so simple as naively anticipated (or better hoped), nevertheless plenty of success was reached in this field in the following decades. 

Measured fusion excitation functions of light heavy-ion systems essentially follow the predictions of the well-known Wong formula based on the quantal penetration of the barrier~\cite{Wong}, but
experimental and theoretical studies on near- and sub-barrier heavy-ion fusion received a strong push in the late 70's because two basic kinds of experimental evidences  were discovered: on one side, experimenters found the first hints of generalised very large enhancements of cross sections with respect to the simple predictions of the Wong formula~\cite{Wong}. On the other side, shortly after, measurements gave evidence of strong isotopic effects, that is, fusion excitation functions of near-by systems may differ substantially in magnitude and shape.
\par
This indicated that a close connection exists between the sub-barrier fusion dynamics and the low-lying collective structure of the two colliding nuclei, and the coupled-channels (CC) model was developed in order to reproduce the experimental evidences. In the following decade several experiments were performed aiming at clarifying this link in various experimental situations.

Subsequently, Neil Rowley suggested~\cite{rowBD} that the fusion barrier distributions (BD) originated by channel couplings could be obtained from the second  derivative of the energy-weighted excitation functions with respect to the energy, and a second sequence of measurements started in the early 90's, aiming at extracting the shape of the BD for several different systems, as a fingerprint of channel couplings in the various cases. Measurements of this kind are very delicate and are still being performed nowadays.

Around ten years later, an experiment performed at Argonne Nat. Lab.~\cite{Jiang0} indicated that fusion cross sections of $^{60}$Ni + $^{89}$Y have an unexpected behaviour far below the barrier, i.e., they drop much faster than predicted by standard CC calculations. This opened a new area of research, and this phenomenon (named ``hindrance") was soon recognised as a general effect, even if with different aspects whose origin is still a matter of debate and research in the community.  
Such studies on the ``competition" between the opposite trends originated from near-barrier enhancements and hindrance at lower energies, will certainly receive great impulse from the upcoming availability of heavy radioactive beams with high intensity and good quality. 

Several review articles were published along the years on the various features  of heavy-ion fusion and on the new results and developments, starting from the early 80's~\cite{Vaz,Beck85,Beck88,Stead,Vanden,Reis,AnnuRev,Balan,Liang}. More recent reviews can be found in Refs.~\cite{BBB,PTP}, where the first one  covers  all aspects of heavy-ion fusion, and the second one contains a detailed theoretical treatment of many-body quantum tunnelling. A further interesting review can be found in 
Ref.~\cite{Mis_rev} that is specifically dedicated to the topic of hindrance in HI fusion. 

A series of Conferences has been dedicated to this field, the first one~\cite{MITConf} being the ``International Conference on Fusion reactions below the Coulomb barrier", held at MIT, Cambridge (Massachusetts, USA) in June 1984, while the most recent one~\cite{Fusion17} (the `` International Conference on Heavy-Ion Collisions at Near-Barrier Energies") has taken place in Hobart, (Tasmania, Australia) in February 2017.

This paper is organised as follows:
Sect.~\ref{Tunnenh} illustrates the phenomena of tunnelling effects and cross section enhancements in medium-heavy systems, and a brief description will be given of  the coupled-channels model. Sect.~\ref{FusBD} introduces the concept of fusion barrier distributions, and points out the ways the barrier distribution (BD) can be extracted from the data, and a few  experimental results will be shown. The specific influence of couplings to transfer channels is pointed out in Sect.~\ref{InfTransf} and recent results in this sub-field are reported. Early experiments and  new  evidences  concerning the phenomenon of fusion hindrance far below the barrier are illustrated in Sect.~\ref{Hindrance}, as well as the various theoretical models developed for this effect, and its consequences for stellar evolution. Sect.~\ref{oscilla} illustrates the recent evidence of above-barrier oscillations in the excitation function of medium-mass systems, in relation to the analogous effects already observed in light-ion cases. The following Sect.~\ref{radioactive} briefly outlines the extra-features of heavy-ion fusion that can be investigated by using exotic beams, with a special emphasis on the prospects offered by heavy radioactive beams and by new or recently installed experimental set-ups.   Concluding remarks are given in Sect.~\ref{concluding} with a final summary of the article, and an outlook for the future.

\section{Tunnelling and Enhancement}
\label{Tunnenh}

   In  Sect.~\ref{expresu} we present a selection of experimental results obtained  in several laboratories,  on sub-barrier fusion enhancements and corresponding isotopic effects. The development of the CC model was triggered by early measurements,  that could not be explained by one-dimensional barrier penetration. The CC model is briefly illustrated in Sect.~\ref{CCmodel}.
 
 \subsection{Experimental evidences}
  \label{expresu}
  
A significant example of the strong  isotopic differences observed in the fusion cross sections of near-by systems, is given by the series of measurements concerning $^{16}$O + $^{148,150,152,154}$Sm~\cite{Stok78}, which showed a large variation in the energy dependence of the fusion cross sections when going from the spherical $^{148}$Sm target to heavier and statically deformed Sm isotopes. 

The experiments were performed by implanting the fusion-evaporation residues (ER) into a catcher foil, and, subsequently, by off-line detection (using a small Ge spectrometer) of the characteristic X-rays of the residual nuclei following the EC, $\beta^+$ decay of ER.

\begin{figure}[h]
\centering
\resizebox{0.38\textwidth}{!}{\includegraphics{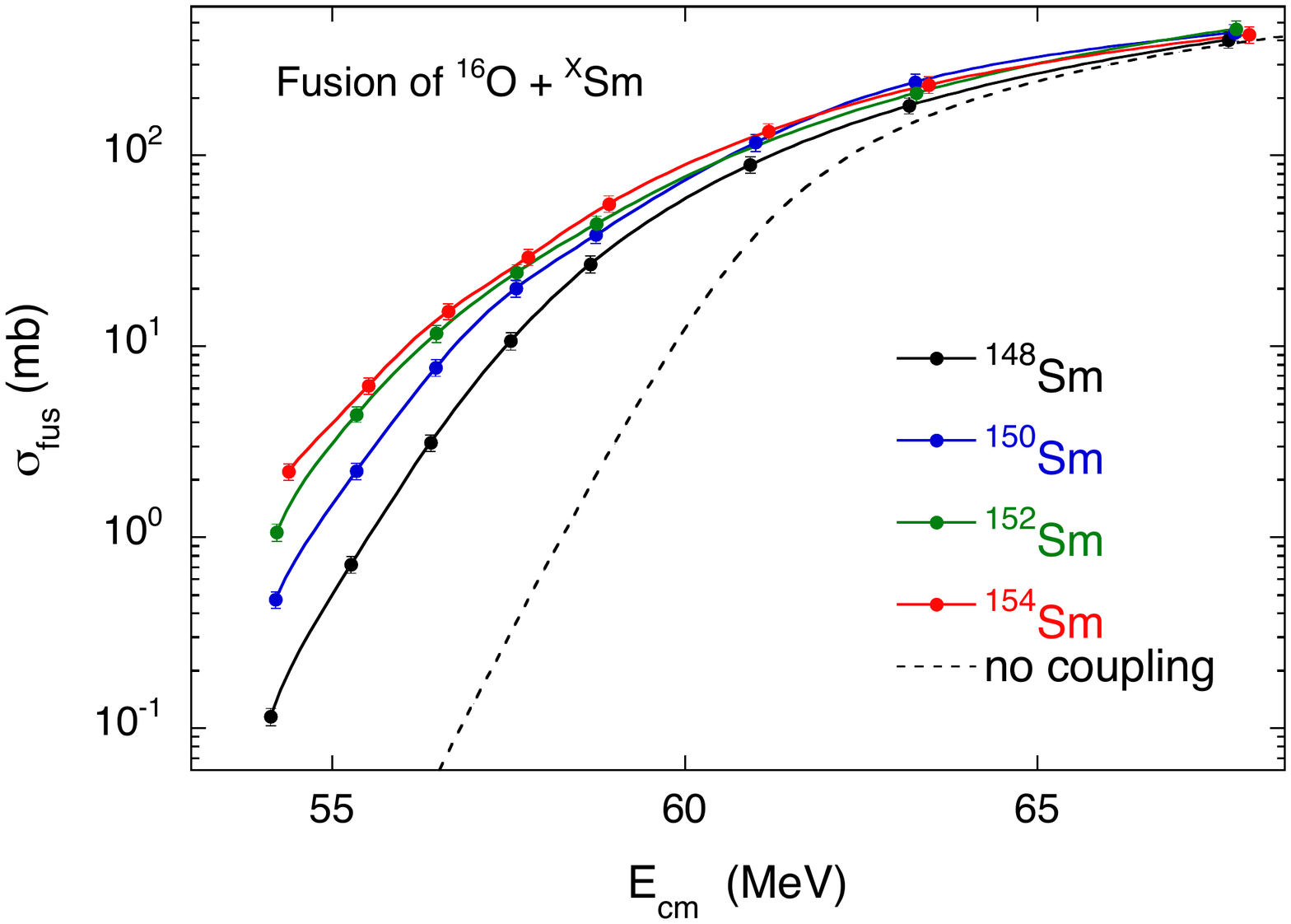}}
\caption{Fusion cross sections vs. bombarding energy for $^{16}$O + $^{148,150,152,154}$Sm~\cite{Stok78}. All excitation functions are strongly enhanced with respect to the ``no coupling" calculation, and  
the more deformed Sm isotopes have larger cross sections below the barrier.}
\label{Stok}  
\end{figure}

The results of those measurements are summarized in Fig.~\ref{Stok} where, additionally, the prediction based on the penetration  of a one-dimensional potential barrier is shown as a dashed line for the case of $^{16}$O + $^{154}$Sm. The corresponding  ``no coupling" calculations for the other targets are very close to this case, because the variation of the Coulomb barrier is very small (0.5 MeV when going from $^{154}$Sm to $^{148}$Sm).
\par
\begin{figure}[h]
\centering
\resizebox{0.45\textwidth}{!}{\includegraphics{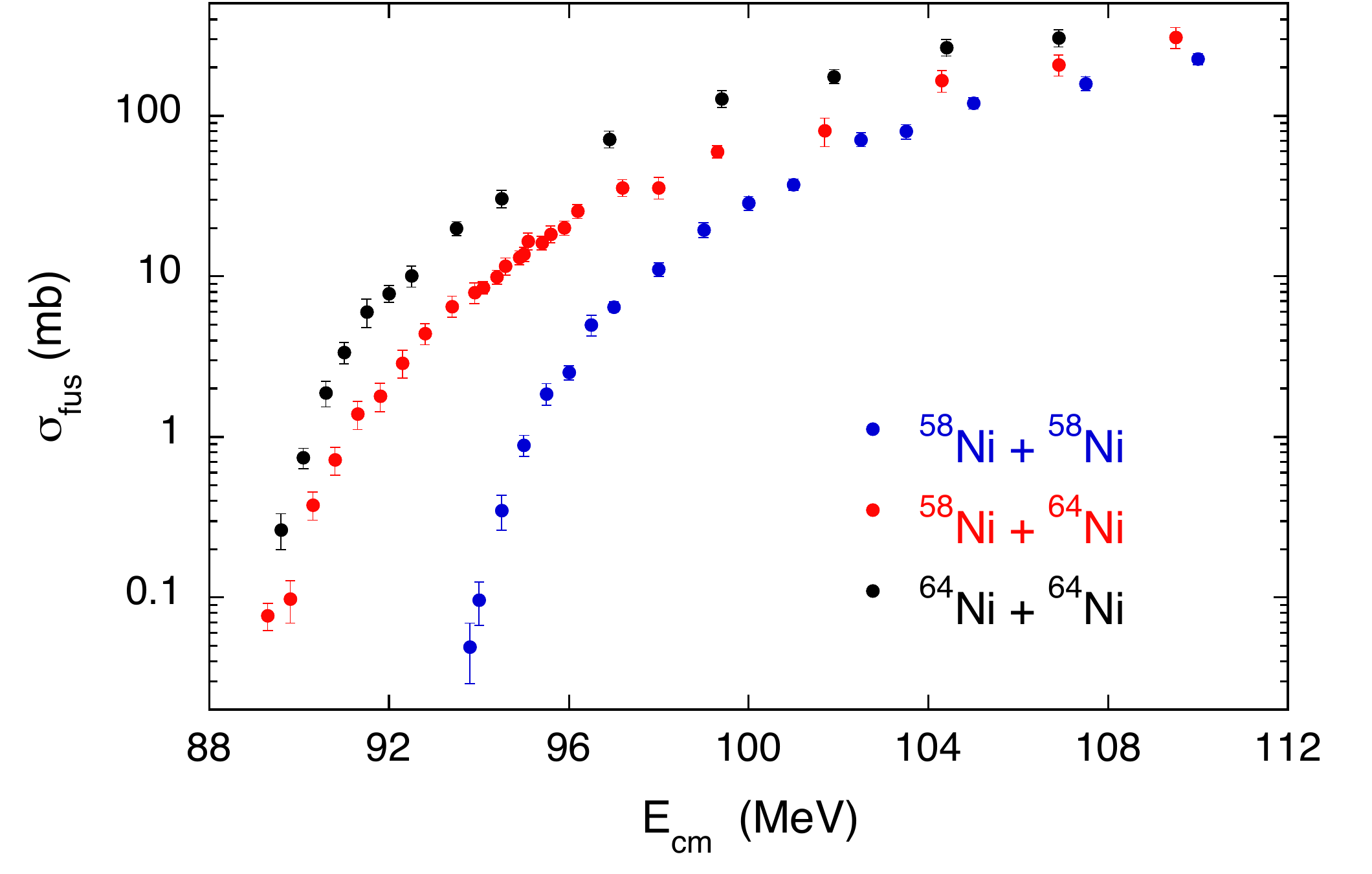}}
\caption{Fusion excitation functions of various Ni + Ni systems as measured by Beckerman {\it et al.}; figure redrawn from Ref.~\cite{BeckNiNi}.}
\label{NINI}     
\end{figure}

The experiments performed at MIT by the group led by M. Beckerman  on the fusion of various combinations of nickel isotopes~\cite{BeckNiNi}  gave clear evidence of isotopic effects in cases where the structure of the interacting nuclei does not  change much when going from one system to another. The results that by now are well-known  indicated for the first time the possible influence of transfer reactions on near- and sub-barrier cross sections, and opened up a new field of research. Indeed, the role of transfer couplings has never been unambiguously identified, after those early indications of their influence, with reliable and systematic calculations. We shall come back to this point later in this review.

We report in Fig.~\ref{NINI} the measured excitation functions of the three systems  $^{58}$Ni + $^{58}$Ni, $^{58}$Ni + $^{64}$Ni and $^{64}$Ni + $^{64}$Ni. Besides the trivial differences due to the varying Coulomb barriers, the remarkable feature is the contrasting slope of the asymmetric system  $^{58}$Ni + $^{64}$Ni, when compared to the other two symmetric cases. 
\begin{figure}[h]
\centering
\resizebox{0.40\textwidth}{!}{\includegraphics{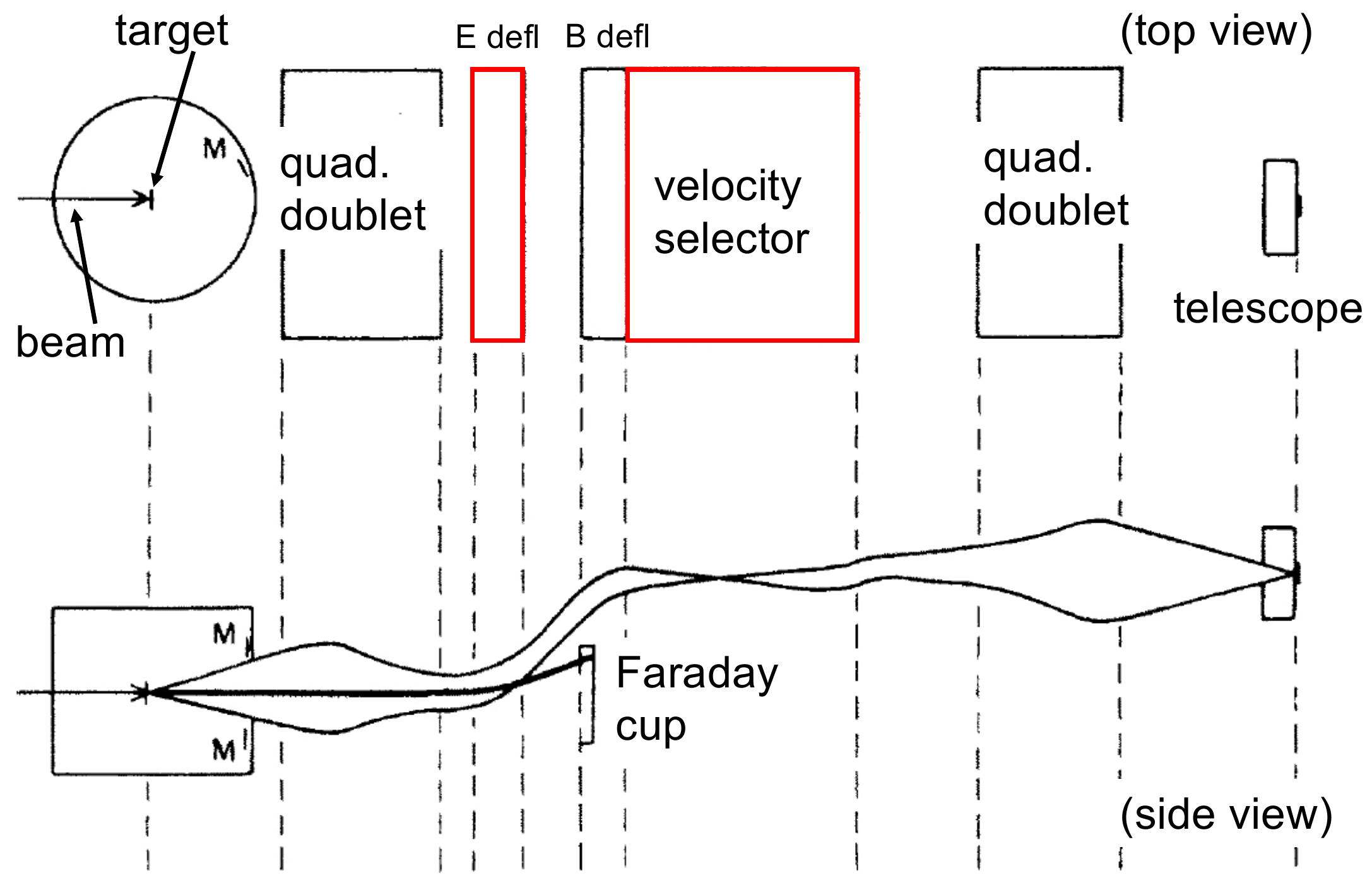}}
\caption{Layout of the MIT-BNL Recoil Mass Selector consisting of a first magnetic quadrupole doublet, a small electrostatic deflector (separating the ER away from the beam), a Wien filter and a final quadrupole doublet focusing the ER onto a detector telescope. Figure modified from Ref.~\cite{MITFilter}. }
\label{MIT}     
\end{figure}

Indeed, the cross sections of $^{58}$Ni + $^{64}$Ni decrease much slower with decreasing energy. Shortly after, this was associated~\cite{Bro83} with the availability, only in this system, of neutron transfer channels with positive $Q$-values.

The set-up used in those experiments is shown schematically in Fig.~\ref{MIT}. The fusion cross sections were measured by direct  detection of the ER at 0$^o$ and at small angles with respect to the beam, after separating out the beam and beam-like particles by using an electrostatic deflector and a  E$\times$B crossed-field velocity selector. ER were detected and identified in a $\Delta$E-E telescope consisting of a proportional chamber filled with isobutane and a 450 mm$^2$ silicon
surface barrier detector mounted at the rear of the gas chamber.
\par
\begin{figure}[h]
\centering
\resizebox{0.40\textwidth}{!}{\includegraphics{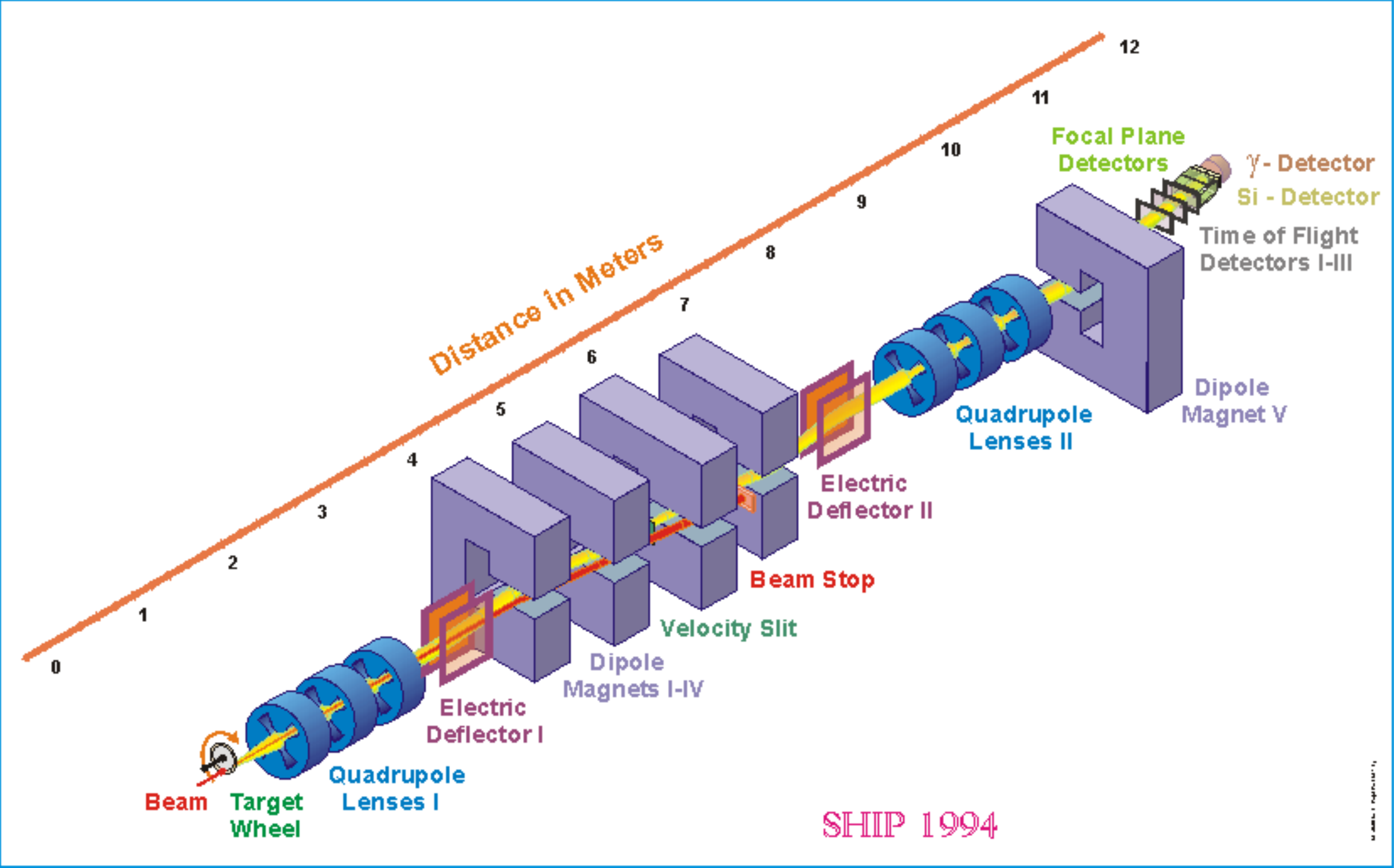}}
\caption{Schematic view of the velocity filter SHIP, presently installed at GSI~\cite{Ship,Ship2}. The set up consists of a sequence of magnetic and electric fields 
having  a very high capability of beam rejection, and transporting ER with high efficiency down to the focal plane detectors.}
\label{SHIP}     
\end{figure}
\begin{figure}[h]
\centering
\resizebox{0.45\textwidth}{!}{\includegraphics{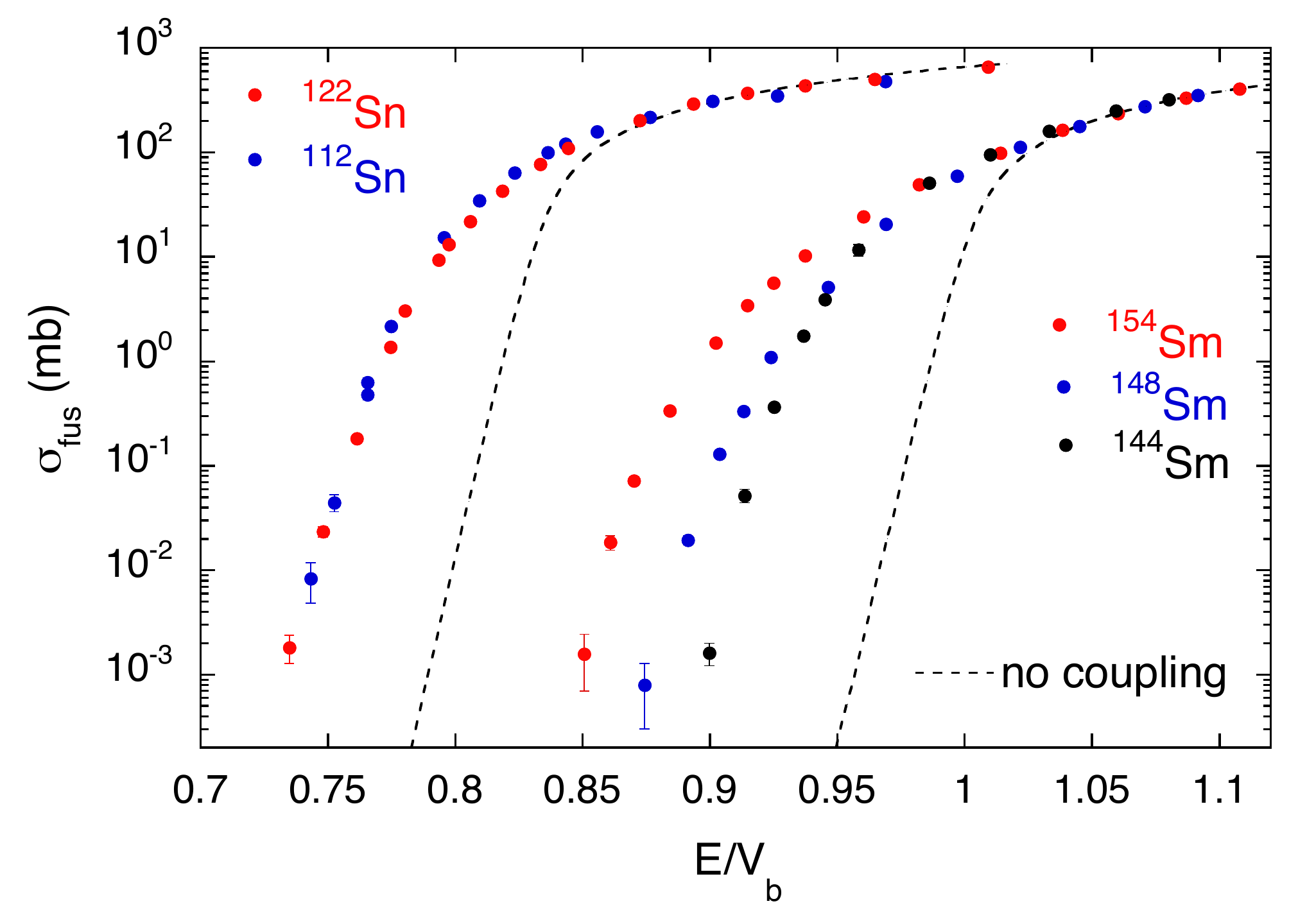}}
\caption{Fusion excitation functions of $^{40}$Ar + $^{112,122}$Sn, $^{144,148,154}$Sm~\cite{Reis83}. The abscissa is the energy relative to the Aky\"uz-Winther  Coulomb barrier~\cite{Akyuz} for all systems.}
\label{Reis}     
\end{figure}

We now illustrate a few experiments performed at the velocity filter SHIP at GSI (see Fig.~\ref{SHIP}), that was installed with the main purpose of synthetizing superheavy elements. Great success was achieved in this field~\cite{Hofmun}, and very significant results were also obtained concerning the dynamics of heavy-ion fusion in the 80's. We show an example of these measurements in Fig.~\ref{Reis}. We observe that, in reduced energy scales, the excitation functions of $^{40}$Ar + $^{112,122}$Sn essentially coincide, while those of $^{40}$Ar + $^{144,148,154}$Sm~\cite{Reis83} are largely different below the barrier, $^{154}$Sm having the largest enhancement. This is due, on one side, to the very similar low-energy nuclear structure of the tin isotopes, and, on the other hand, to the well-known shape change when going from $^{144}$Sm (spherical) to $^{154}$Sm having a stable prolate deformation.

\begin{figure}[h]
\centering
\resizebox{0.40\textwidth}{!}{\includegraphics{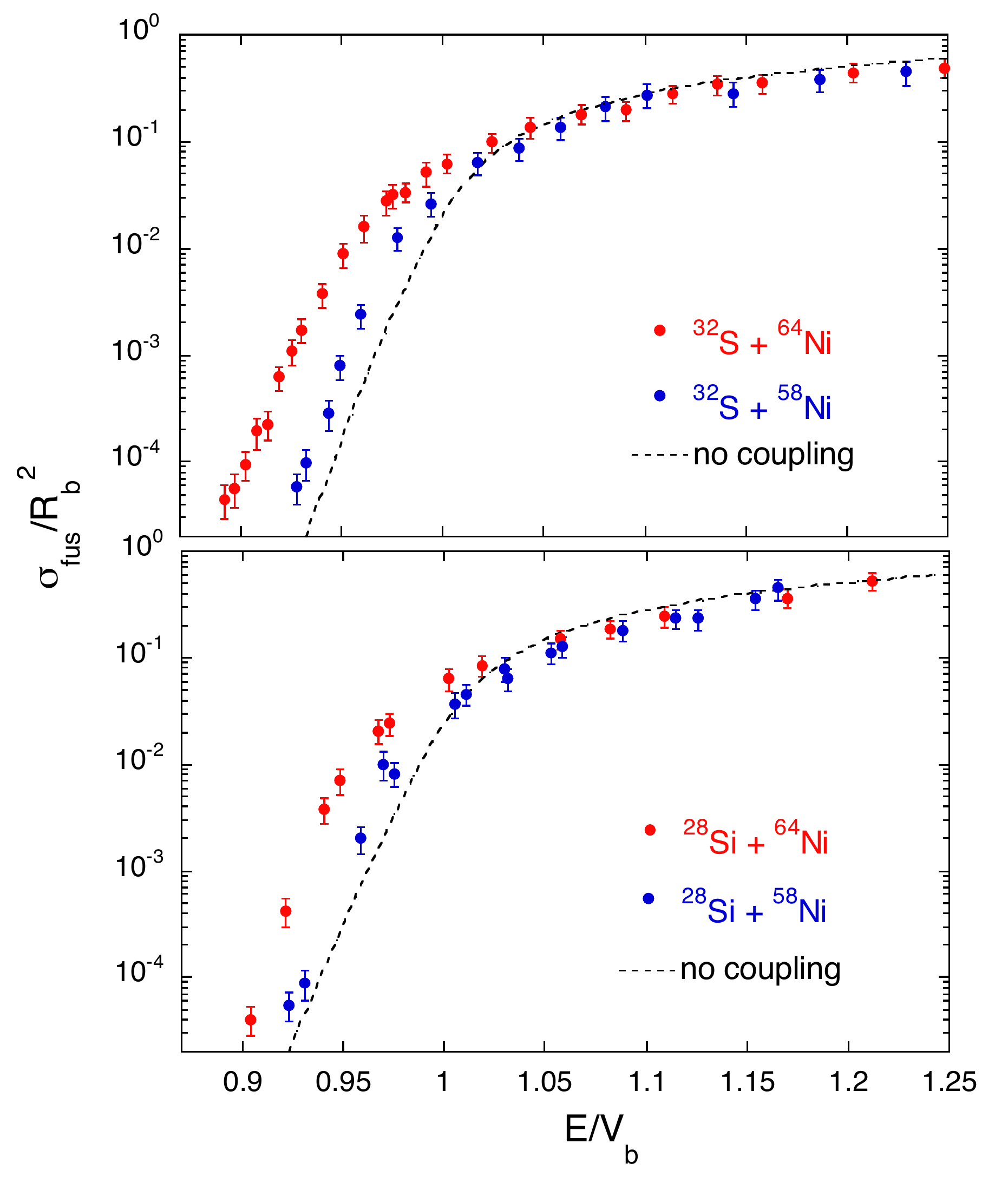}}
\caption{Reduced excitation functions for fusion of several Si and S + Ni systems~\cite{SNi86}. The energy scale is relative to the AW Coulomb barrier and the cross sections are divided  by the square of the barrier radius R$_b$.}
\label{SNI}     
\end{figure}

The study of excitation functions of several Si+Ni and S+Ni systems was performed at Laboratori Nazionali di Legnaro (LNL) of INFN. Some of the results are reported in Fig.~\ref{SNI}. In particular, the figure shows the measured excitation functions for 
$^{28}$Si + $^{58,64}$Ni (lower panel) and for $^{32}$S + $^{58,64}$Ni (upper panel). For each case, the expected cross sections in the one-dimensional barrier penetration limit (no coupling limit), calculated~\cite{CCFULL} using the Aky\"uz-Winther (AW) potential~\cite{Akyuz}, are also reported with dashed lines.
One notices that measured cross sections are systematically and largely enhanced with respect to the no-coupling limit. Moreover, the two systems $^{28}$Si,$^{32}$S + $^{64}$Ni display larger enhancements and smoother slope with decreasing energy, when compared to the two near-by cases $^{28}$Si,$^{32}$S + $^{58}$Ni. This was attributed to the effect of couplings to the neutron pick-up channels having positive ground state Q-values only for the first two systems.

\begin{figure}[h]
\centering
\resizebox{0.40\textwidth}{!}{\includegraphics{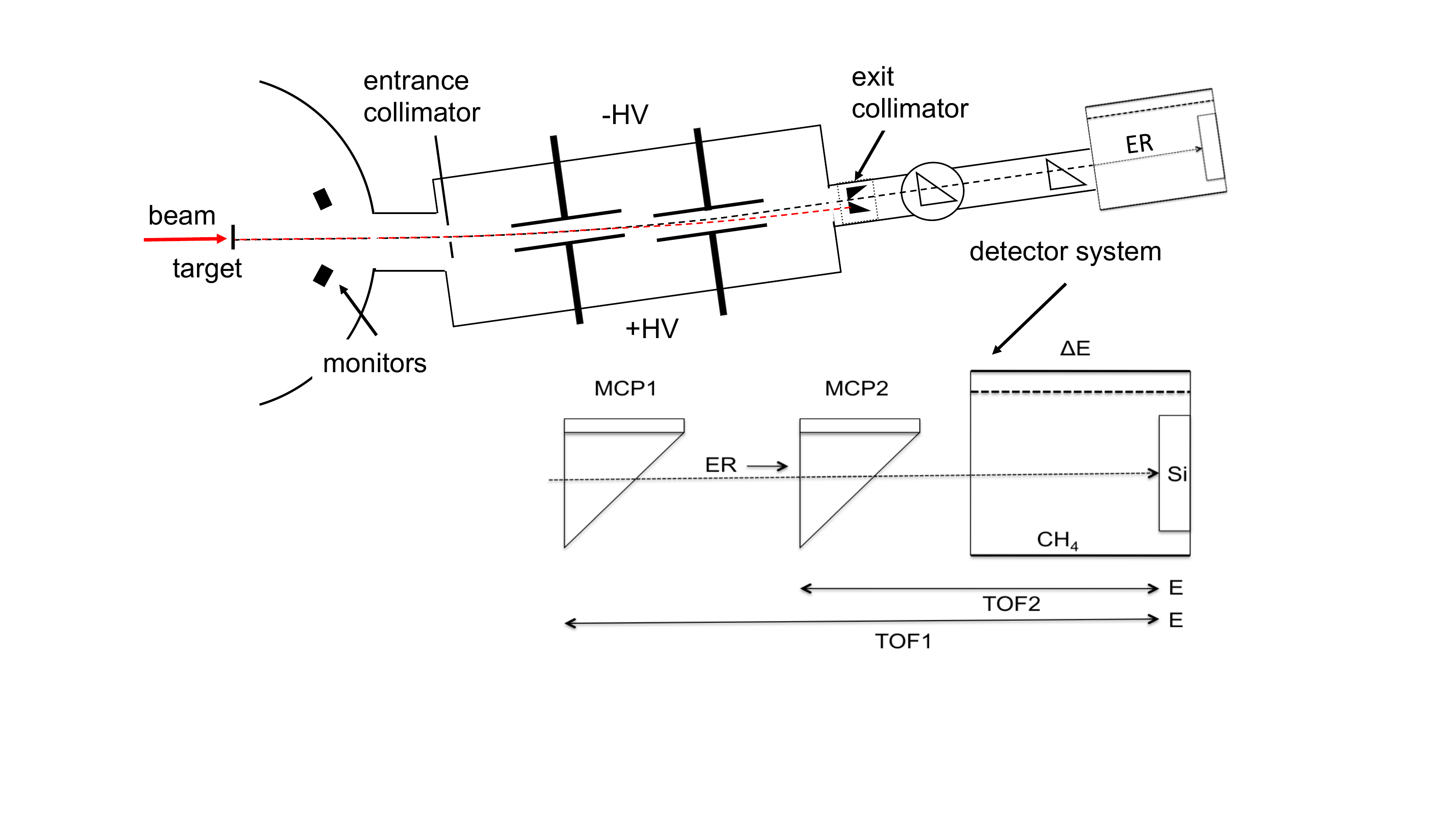}}
\caption{Present layout of the set-up used for fusion measurements at LNL~\cite{NiFedeep}. A two-stages electrostatic deflector follows an entrance collimator, most of beam and beam-like particles are stopped on one side of the exit collimator which allows ER to enter the Energy-ToF-$\Delta$E detector telescope. Partially redrawn from Ref.~\cite{NiFedeep}. }
\label{MAIA}     
\end{figure}

Those experiments were performed using the first version of the set-up based on the beam electrostatic separator installed at LNL~\cite{Beghini85}, whose present layout is schematically shown in Fig.~\ref{MAIA}. Indeed, the original set-up underwent various upgrades in recent years~\cite{NiFedeep}, aimed at improving its sensitivity to very small cross sections (down to 0.5-1 $\mu$b). That set-up is still in operation at LNL.

\begin{figure}[h]
\centering
\resizebox{0.40\textwidth}{!}{\includegraphics{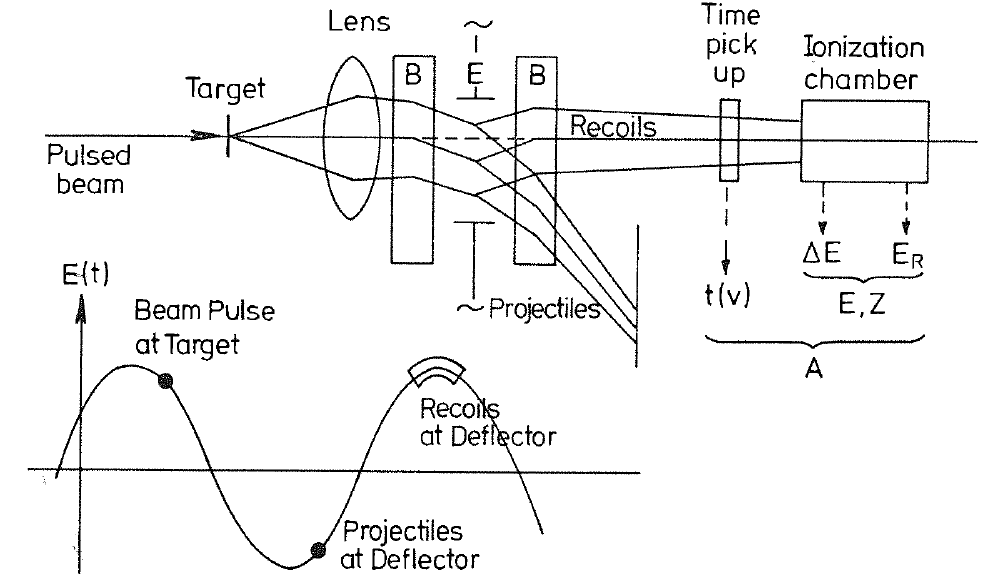}}
\caption{Schematic set-up of the Munich recoil spectrometer~\cite{Rud}. The electric field oscillates in phase with the
beam pulsing. In lower part of the picture one sees arrival times of the projectiles and ER as qualitative examples,  depending 
on the particular experiment. Partially redrawn from Ref.~\cite{Rud}, courtesy of R. Pengo.}
\label{RF_Mu}     
\end{figure}

Several systems were  investigated using the radio frequency recoil spectrometer (see Fig.~\ref{RF_Mu})~\cite{Rud}  built at the MP Tandem accelerator in Munich. 
This  spectrometer was a unique arrangement of a quadrupole lens, a
Wien-type velocity filter with a radio-frequency electric field, a time-of-flight
detector and a $\Delta$E - E gas telescope.

\begin{figure}[h]
\centering
\resizebox{0.42\textwidth}{!}{\includegraphics{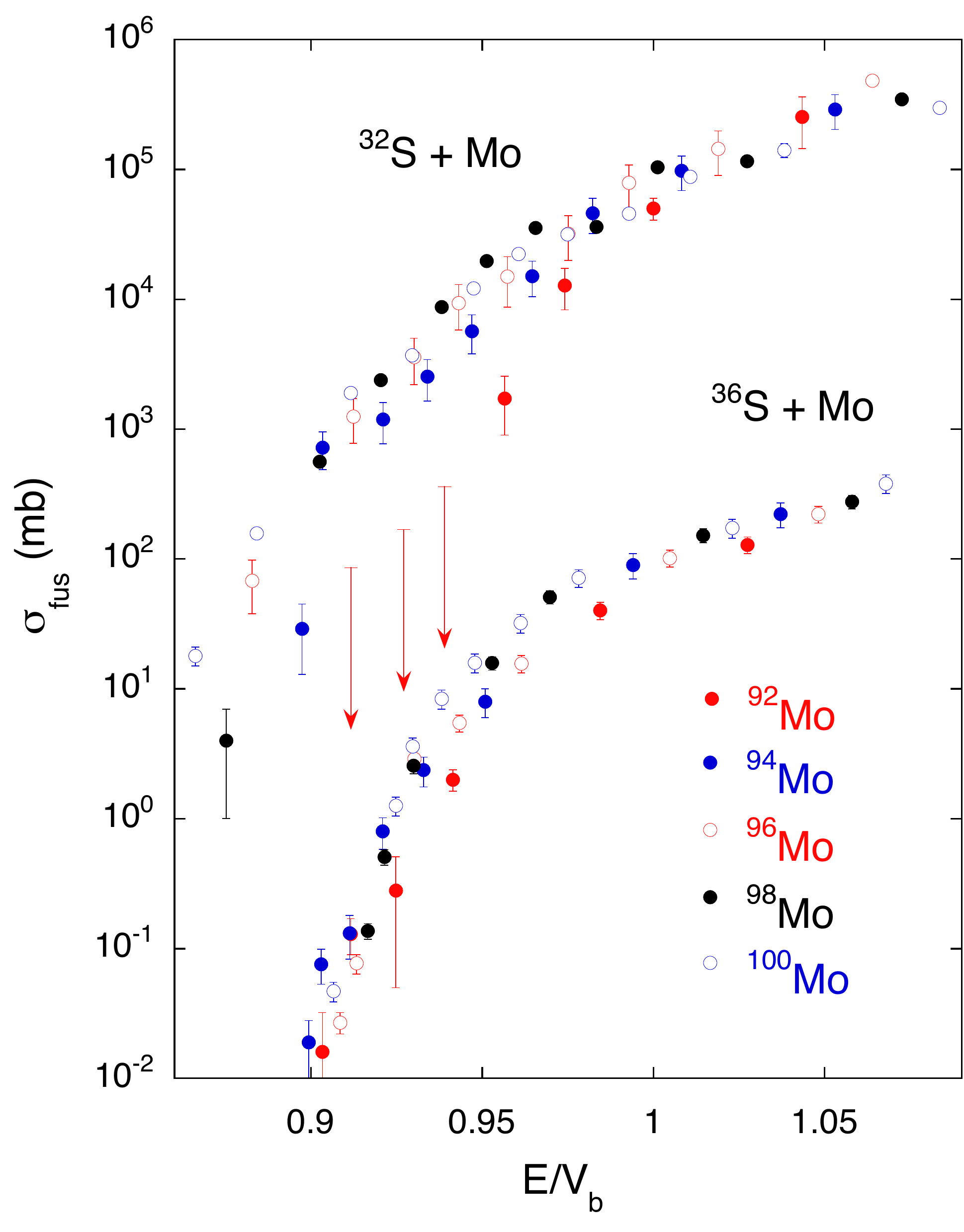}}
\caption{Fusion excitation functions of $^{32,36}$S + Mo isotopes where the energy scale is normalised to the AW Coulomb barrier~\cite{Akyuz}. 
The cross sections of  $^{32}$S + Mo have been multiplied by 10$^3$ in the figure for the sake of clarity. 
The red arrows indicate the upper limits determined for $^{32}$S + $^{92}$Mo at the lowest energies.
}
\label{Dati_Pengo}     
\end{figure}

Using  $^{32,36}$S beams on series  of isotopes of Mo, Ru, Rh, and Pd  different sub-barrier fusion behaviours were evidenced~\cite{Pengo}. The enhancements and isotopic effects 
were discussed in terms of nuclear structure and of  the influence of neutron transfer channels with positive Q-values. See Fig.~\ref{Dati_Pengo} for the results concerning  $^{32,36}$S + Mo. One can notice that all systems with $^{32}$S as beam have sub-barrier cross sections decreasing with the energy more slowly than the corresponding cases where $^{36}$S is involved. This was qualitatively attributed to the coupling to transfer channels with Q$>$0 which are only available for the $^{32}$S + Mo systems. The exception to this general trend is the case of $^{32}$S + $^{92}$Mo where, indeed,  all neutron pick-up channels have negative Q-values.

\begin{figure}[h]
\centering
\resizebox{0.45\textwidth}{!}{\includegraphics{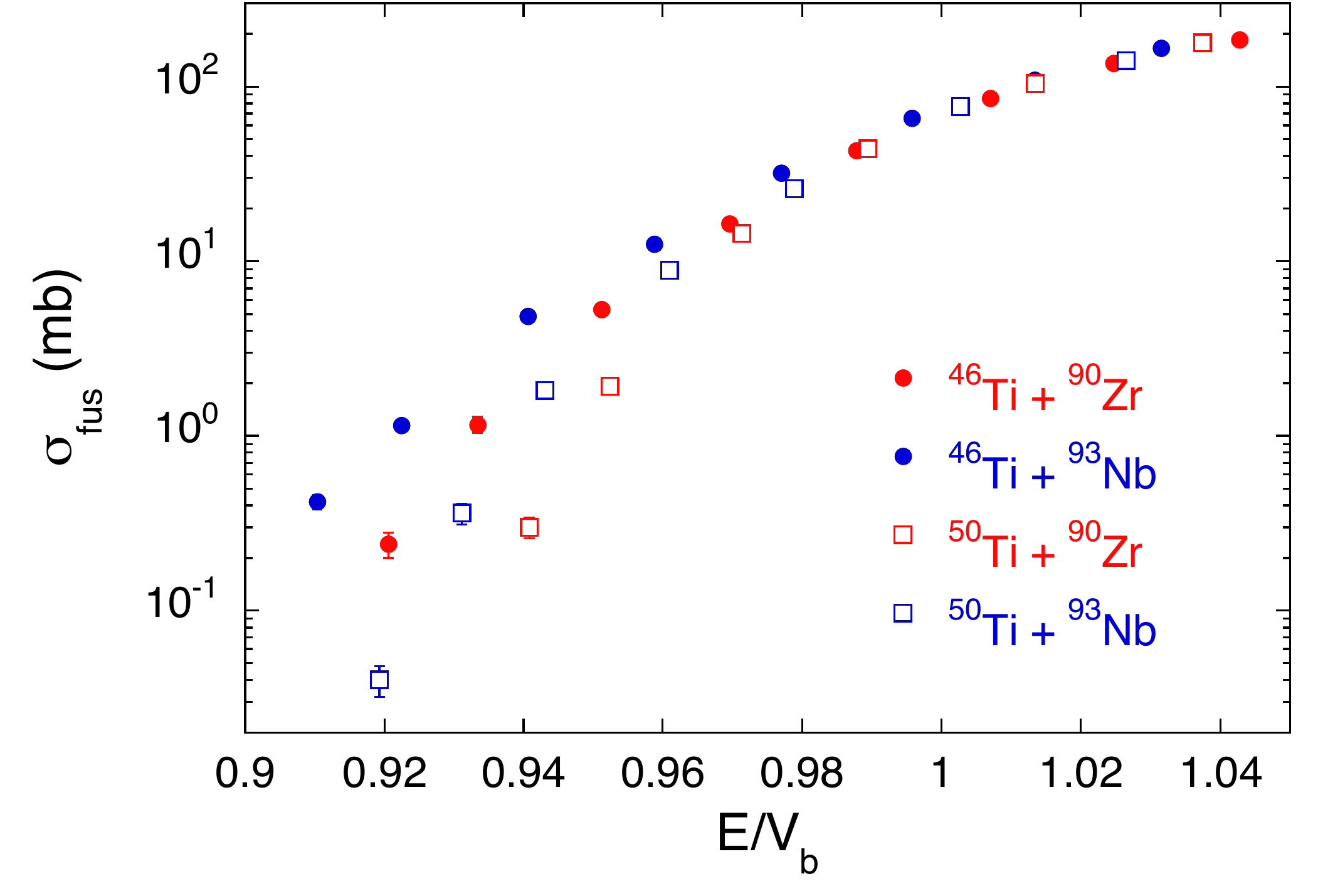}}
\caption{Fusion excitation functions of $^{46,50}$Ti + $^{90}$Zr,$^{93}$Nb~\cite{Stelson}. The energy scale is relative to the AW Coulomb barrier.}
\label{Dati_ORNL}     
\end{figure}

\begin{figure}[h]
\centering
\resizebox{0.45\textwidth}{!}{\includegraphics{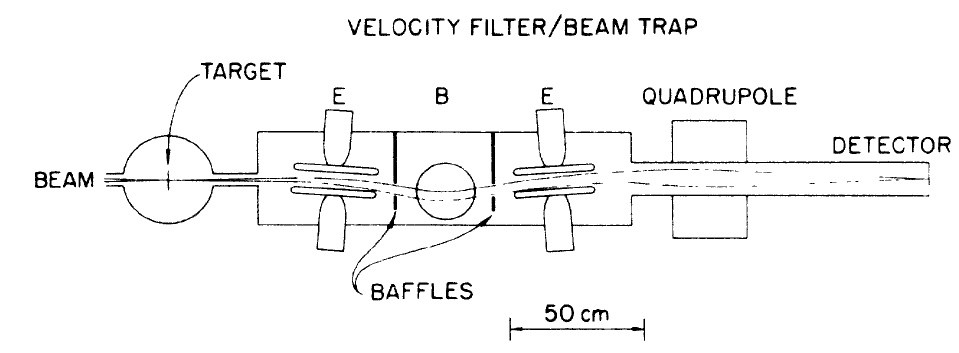}}
\caption{Layout of the velocity filter that was used at ORNL for the experiments reported in the text~\cite{Stelson}. The velocity filter consisted of two electrostatic deflectors separated by a dipole magnet. A quadrupole doublet followed the deflectors to focus the ER onto the focal plane detector (a $\Delta$E-E ionisation chamber followed by a silicon detector).  This set-up could rotate around the target position to measure the ER angular distribution. Its  solid angle was $\approx$1msr. Figure from Ref.~\cite{Stelson},  \copyright~American Physical Society (APS). }
\label{VF_ORNL}     
\end{figure}

Large variations of the fusion cross sections were measured at Oak Ridge National Laboratory (ORNL) for the systems $^{46,50}$Ti + $^{90}$Zr,$^{93}$Nb~\cite{Stelson},   as reported in Fig.~\ref{Dati_ORNL}, using the set-up shown in Fig.~\ref{VF_ORNL}. These strong isotopic effects in the sub-barrier region were attributed to very different degrees of collectivity in the colliding nuclei, and the authors suggested that neck formation is playing an important role in the fusion dynamics.
Moreover, they put  these observations in relation with the concept of the  barrier distribution which was fully developed only a few months later in Ref.~\cite{rowBD}.

When comparing the sub-barrier  trend of different systems, as shown in the previous figures, it is customary to normalize
the energy scale in order to take into account  their different 
Coulomb barriers.
To this end,  the Aky\"uz-Winther (AW) potential~\cite{Akyuz} is often used 
(the Bass potential~\cite{Bass} usually gives very similar  barriers).
\begin{figure}[h]
\centering
\resizebox{0.40\textwidth}{!}{\includegraphics{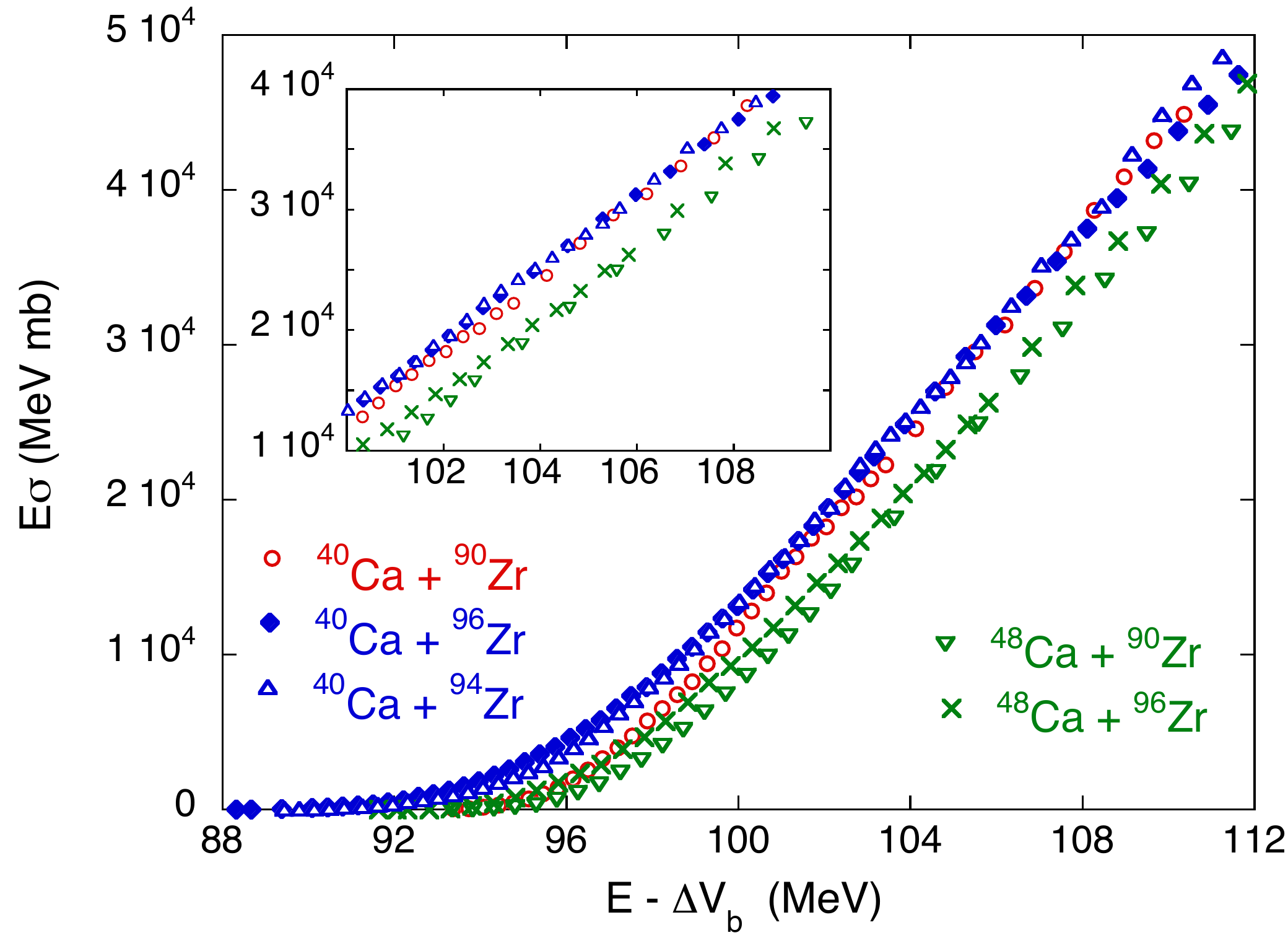}}
\caption{Energy weighted  excitation functions of several Ca + Zr systems. The abscissa is $E-\Delta$V$_b$
where, for each system, $\Delta$V$_b$ is 
the difference of the AW Coulomb barrier with respect to the case
of $^{40}$Ca + $^{90}$Zr. The insert shows an expanded region above the Coulomb barrier.}
\label{CaZr_AW}     
\end{figure}

Fig.~\ref{CaZr_AW} reports the energy weighted excitation functions of several Ca + Zr systems  where the (linear) energy scales are  adjusted following  the case of $^{40}$Ca + $^{90}$Zr  taken as reference~\cite{4094}.
\par
We observe that using this abscissa the systems are clearly divided into two groups, one 
 including the  three cases with $^{40}$Ca as projectile and the second one, shifted in energy, with  $^{48}$Ca (see also the expanded view in the insert).
 The two Ca isotopes have actually high-energy octupole modes with very different strengths; this implies that a larger adiabatic renormalisation of the potential~\cite{adiabatic} is produced by the stronger vibration in $^{40}$Ca, and, as a result, one obtains a lower Coulomb barrier.

We have to be conscious that the AW (and others) parametrisations  were introduced to reproduce the average behavior of the nuclear potential for several systems, when mass number and Z  extend over a large range of values.
Therefore, local variations of the potential due to different nuclear excitations (as  is the case of the octupole modes in the two calcium isotopes) cannot be obviously reproduced. We also notice that
the five excitation functions shown in Fig.~\ref{CaZr_AW} have very similar slopes. This means that  the variation of barrier radius  $R_b$ has little importance in this context. 

In view of all this, and of the example of the Ca+Zr systems in particular, a good degree of carefulness should be used when applying the concept of ``reduced excitation functions", and little quantitative significance, in general, has to be given to considerations and conclusions based on them.

\subsection{The coupled-channels model} 
\label{CCmodel}

The clear evidences of strong fusion cross section enhancements with conspicuous isotopic effects were successfully reproduced in various cases by the coupled-channels (CC) model which associates the nuclear structure of the two interacting nuclei to their relative motion. The model was originally developed in Refs.~\cite{Dasso,Dasso2,Dasso3,Broglia}. 
\par
A special role was shown to be played by strong couplings to the low-lying collective modes of the two nuclei and, possibly, to quasi-elastic transfer channels having positive Q-values.
The application of an ingoing-wave boundary condition allowed such coupling effects in the barrier region to be studied, 
and it was shown that the effect of couplings is actually to enhance the near- and sub-barrier fusion cross sections.

Here below we briefly illustrate the basic principles and the framework of the CC model, in the limit where the incident energy is large compared to the excitation energies of the relevant coupled states and to the coupling interaction. This is usually called the ``sudden limit"~\cite{Dasso2}.
We consider two colliding nuclei whose internal structures are described by the variable $\xi$, so that the total Hamiltonian of the system can be written 

\begin{equation}
H = H_k +V_l(r) +H_o(\xi) + V_{int}(\vec{r},\xi)
\end{equation}

where H$_k$= -($\hbar^2$/2$\mu$) $\nabla^2$ is the kinetic energy, $\mu$ is the reduced mass and V$_l$(r) is the ion-ion potential for the $l^{th}$ wave, while H$_o$($\xi$) describes the internal structure of projectile and target nuclei, and V$_{int}\vec{r},\xi$) is the potential that couples their relative motion to the reaction channels.

By introducing the eigenstates $\ket{n}$ of H$_o$  and  by expanding the total wave function $\Psi$ in terms of those eigenstates, that is 
\begin{equation}
\nonumber
H_o(\xi)\ket{n}= \epsilon_n\ket{n}
\end{equation}
\begin{equation}  
\nonumber
\Psi= \sum_n\chi_n(r)\ket{n}
\end{equation}

the stationary Schr\"odinger equation is equivalent to the set of coupled equations for the wave functions $\chi_n$  of the relative motion

\begin{equation}
\Big[-\frac{\hbar^2}{2\mu}\nabla^2 +V_l(r)-E \Big]\chi_n(r)=
\end{equation}
\begin{equation}
\nonumber
= -\sum\limits_{m}\Big[\epsilon_n\delta_{nm}+\bra{n} V_{int}(\vec{r},\xi) \ket{m}\Big]\chi_m(r)
\end{equation}

We  impose the usual boundary conditions for a scattering problem 
and we consider the case where  the coupling interaction can be factorised as follows 
\begin{equation}
\bra{n} V_{int}(\vec{r},\xi) \ket{m}=F(r)\bra{n}G(\xi)\ket{m}=F(r)G_{nm}
\end{equation}
where $F(r)$ is the form factor associated to the relative motion and $G_{nm}$ only depends on the intrinsic variables describing nuclear structure. 
So we can write 
\begin{equation}
\Big[-\frac{\hbar^2}{2\mu}\nabla^2 +V_l(r)-E \Big]\chi_n(r)= -\sum\limits_{m}M_{nm}\chi_m(r) 
\end{equation}
where
\begin{equation}
M_{nm}=\epsilon_n\delta_{nm}+F(r)G_{nm}
\end{equation}
By treating the form factor $F(r)$ as a constant $F_o$  and then $M_{nm}$=$\epsilon_{n}\delta_{nm}+F_oG_{nm}$ we can decouple the equations by using the unitary transformation $U$ that diagonalises the matrix $M$, i.e., 
\begin{equation}
\sum\limits_{ik}U_{ni}M_{ik}U_{km}^{-1}=\lambda_m\delta_{nm}
\end{equation}
We obtain the uncoupled equations
\begin{equation}
\Big[-\frac{\hbar^2}{2\mu}\nabla^2 +V_l(r)+\lambda_m-E \Big]Y_m(r)= 0 
\end{equation}
where
$Y_m(r)=\sum\limits_{n}U_{mn}\chi_n(r)$.
It follows that (see again Ref.~\cite{Dasso2}) the total  transmission coefficient $T_l$ for a given angular momentum $l$ 
is  a sum of contributions coming from all coupled channels, that is
\begin{equation}
T_l (E)=\sum\limits_{m}\mid U_{mo}\mid^2T_l[E,V_l(r)+\lambda_m]
\label{Telle}
\end{equation}
and 
\begin{equation}
\sigma_f(E)=\sum\limits_{l,m}(2l+1)\mid U_{mo}\mid^2T_l[E,V_l(r)+\lambda_m]
\end{equation}

In other words the effect of couplings for each $l$ is:
 1) to replace the original barrier $V(r)$ with a set of effective barriers $[V(r)+\lambda_m]$ seen by the entrance flux where
 2) the transmission for each barrier is weighted  by the overlaps $\mid U_{mo}\mid^2=\mid\bra{m}\ket{0}\mid^2$ of the entrance  state with the eigenvectors of the $M$ matrix corresponding to the eigenvalues  $\lambda_m$.
Therefore the cross section is enhanced if at least one of the $\lambda_m$ is negative which always is the case.
Alternatively one can write
\begin{equation}
\sigma_f(E) = \sum\limits_{m}w_m \sigma_f^{m}(E)
\label{sigmaCC}
\end{equation}
  
that is, the fusion cross section is
given as a weighted sum~\cite{Naga1,Naga2}
 of the cross sections for uncoupled eigenchannels, and  the weights are $w_m=\mid U_{mo}\mid^2$.

We wish to point out that, although the previous simplified treatment of the coupled channels model contains various approximations and assumptions,
 it correctly highlights the way nuclear structure influences fusion dynamics near the barrier.
More detailed and rigorous analyses can be found elsewhere~\cite{BBB,PTP} and are anyway outside the scope of the present review.
 
\section{Fusion Barrier Distributions}
\label{FusBD}

The concept of a barrier distribution is closely associated with the coupled-channels (CC) model of fusion as already illustrated  by Dasso et al.~\cite{Dasso}. They considered the effect of strong channel couplings in the vicinity of the Coulomb barrier and showed that, within some approximation,  everything happens as if the nominal barrier (unperturbed barrier) would split into two or more, where the presence of barriers lower than the original one enhances the transmission at low energies.

In principle, all  information on the fusion dynamics is already contained in the excitation function, however
extracting the fusion barrier distributions from careful and detailed measurements  proved to be particularly useful in the identification of the nature of couplings responsible for cross section enhancements. 
Different shapes of barrier distribution are predicted (and often observed) for the coupling to low-energy
surface excitations  and to the transfer of one or more nucleons.

The fusion excitation functions reported in Fig.~\ref{Reis} were actually analysed in Refs.~\cite{Reis83,Reis85} within the CC model exploiting theoretical barrier distributions. This confirmed the important role of nuclear structure in determining the sub-barrier fusion cross sections.
Those calculated barrier distributions allowed to obtain good fits of the experimental data. 

The GSI group actually analysed also the fusion excitation functions of symmetric $^{90}$Zr-induced reactions by using the concept of barrier distribution that they obtained by an unfolding procedure of fusion probability data~\cite{Keller}. The resulting barrier distribution for $^{90}$Zr + $^{90}$Zr was found to be  much narrower than that for other projectile-target combinations. This is the consequence of the semi-magic nature of $^{90}$Zr. However this approach to get the experimental BD, was not followed afterwards.

\subsection{The basic concept of barrier distributions}

A decisive step forward was taken when
Rowley et al.~\cite{rowBD} proposed  that the barrier distribution can  be extracted from the measured excitation function by taking 
its  second derivative multiplied by the energy (the so-called ``energy-weighted excitation function'')

\begin{equation}
B(E) = \frac{d^2(E \sigma_f)}{dE^2} 
\label{cent}
\end{equation}

\noindent where $E$ is the center-of-mass energy.

We would like to illustrate briefly this method, starting  from the classical fusion cross section 

\begin{equation}
\sigma_f(E) = \pi R_b^2\Big(1-\frac{V_b}{E}\Big) \hspace{5mm} for   \hspace{2mm}E>V_b
\end{equation}
\begin{equation}
\sigma_f(E)=0 \hspace{24mm}  for  \hspace{2mm}E<V_b
\end{equation}

\begin{figure}[h]
\centering
\resizebox{0.45\textwidth}{!}{\includegraphics{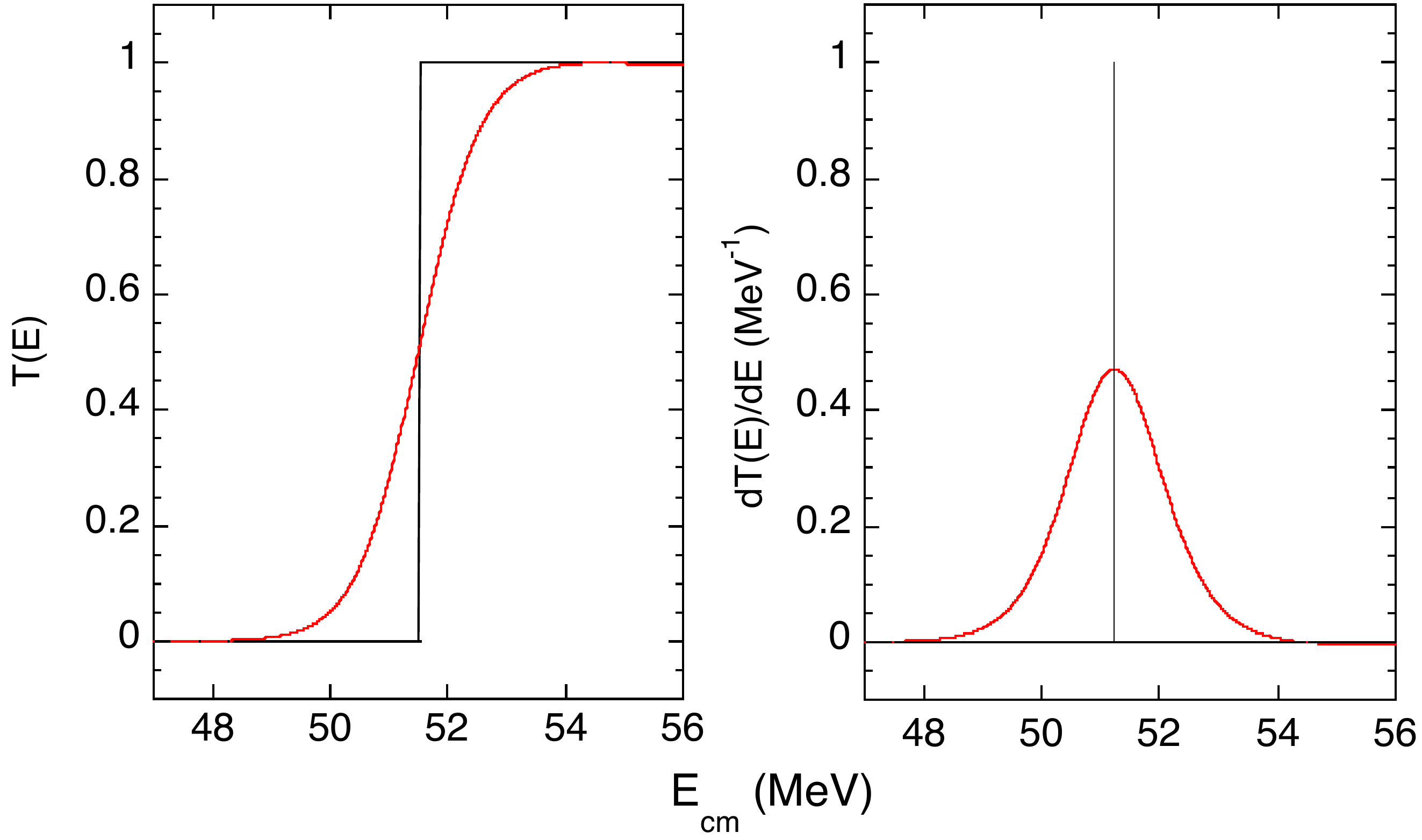}}
\caption{Transmission function and its energy derivative calculated in the classical limit (black lines) and quantum mechanically (red lines), for the system $^{48}$Ca + $^{48}$Ca.
The code CCFULL~\cite{CCFULL} and the AW potential~\cite{Akyuz} have been used.}
\label{T(E)}     
\end{figure}

It follows that the first derivative of $E\sigma_f$ is proportional to the classical penetrability T(E) for a one -- dimensional barrier of height V$_b$, that is T(E)=0 for E$<$V$_b$ and T(E)=1 for E$>$V$_b$ (see left panel of Fig.~\ref{T(E)}, black line)

\begin{equation}
\frac{d}{dE}[E \sigma_f(E)] = \pi R_b^2\hspace{1mm}T(E)
\end{equation}

and that the second derivative is proportional to a delta function

\begin{equation}
 \frac{d^2}{dE^2}[E \sigma_f(E)] = \pi R_b^2\delta(E-V_b)
\end{equation}

Quantum tunneling was taken into account by Wong~\cite{Wong} who used the Hill-Wheeler transmission coefficients~\cite{HW} valid when the barrier is approximated by an inverted parabola, to deduce the following expression for the fusion cross section

\begin{equation}
\sigma_f(E) = \Big(\frac{R_b^2 \hbar\omega}{2E}\Big)ln\Big\{1+exp\Big[\frac{2\pi(E-V_b)}{\hbar\omega}\Big]\Big\}
\label{Wong}
\end{equation}
 
assuming that the position of the Coulomb barrier R$_b$ and its curvature $\hbar\omega$ are independent of the angular momentum $l$.
Let us define $x$=$2\pi (E-V_b)/\hbar\omega$.
It follows that the first and second derivatives of $E\sigma_f$ can be written as 
\begin{equation}
 \frac{d}{dE}[E \sigma_f(E)] = \pi R_b^2\frac{e^x}{1+e^x} = \pi R_b^2\hspace{1mm}T(E)
\label{Wong_1}
\end{equation}

\begin{equation}
 \frac{d^2}{dE^2}[E \sigma_f(E)] = \pi R_b^2\frac{2\pi}{\hbar\omega}\frac{e^x}{(1+e^x)^2} = \pi R_b^2\hspace{1mm} \frac{dT(E)}{dE}
\label{Wong_2}
\end{equation}

In analogy with $\delta(E-V_b)$  and as  shown in Fig.~\ref{T(E)}, $dT(E)/dE$ is centered at E=V$_b$, is symmetric around V$_b$, it has unit area and FWHM$\sim$0.56$\hbar\omega$. 

When couplings to various channels are effective, 
we know from  Sect.~\ref{CCmodel} that in the sudden limit of the CC model,  the fusion cross section is
 a weighted sum of uncoupled  cross sections (see Eq.~(\ref{sigmaCC})).



Consequently $B(E)$ (see Eq.~(\ref{cent})) that is  usually referred to as the fusion barrier distribution,  is a weighted sum of 
individual barriers with the same weight factors  appearing in Eq.~(\ref{sigmaCC}), i.e.

\begin{equation}
B(E)= \frac{d^2}{dE^2}[E \sigma_f(E)] =  \sum\limits_{m}w_m \frac{d^2}{dE^2}[E \sigma^{m}_f(E)] 
\label{BD_final}
\end{equation}

It should be clear that, only in the sudden approximation limit,  this barrier distribution representation,
 has a direct physical meaning. Nevertheless, this method~\cite{rowBD} has been used also for systems where 
the excitation energies are relatively large.  Indeed it has led to a considerable improvement in our understanding of the mechanisms underlying the sub-barrier fusion in several cases, and in clarifying the intimate links between nuclear structure and reaction dynamics~\cite{AnnuRev}.

\subsubsection{Spin distributions}

Measurements of angular momentum  dependence of fusion cross sections  (spin distributions) provide an independent observable to get detailed information on the dynamic of the process.  The spin distribution (or the average angular momentum) can, for example, help for discriminating  between different theoretical models of the phenomenon of fusion hindrance (see Sect.~\ref{models}).
 
In the CC model the transmission coefficient for the different partial waves (see Eq.~\ref{Telle}) contains explicitly the effect of the relevant coupled channels, so that this detailed information is directly transferred to the $\sigma_l(E)$ that are written as
\begin{equation}
\sigma_l(E)= \frac{\pi}{k^2}(2l+1)T_l(E)
 \label{sigmal0}
\end{equation}
where $k$ is the wave number. 

A detailed review was dedicated  by Vandenbosch~\cite{Vanden} to this topics, covering both theoretical and experimental aspects  of spin distributions. 
Recently spin distribution measurements were performed for the reaction $^{64}$Ni + $^{100}$Mo at three energies around and above the barrier. The results indicate the increasing fission competition with particle evaporation at higher beam energies~\cite{Singh}.

We point out that, as already noted by Sahm et al. and Ackermann et al.~\cite{Sahm,Ackermann}, under certain approximations the energy dependence of fusion cross sections determines the spin distribution uniquely. By assuming that:
1) the cross sections correspond to the flux transmitted by one or more barriers  which are energy-independent, 2) the angular momentum only increases the barrier by an amount corresponding to the centrifugal term
\begin{equation}
V_l(E)= \frac{l(l+1)\hbar^2}{2\mu R_b^2}
\label{Vl}
\end{equation}
where $\mu R_b^2$ is $l$- and $E$-independent,
3) $\mu$ and $R_b$ are taken as the asymptotic reduced mass and unperturbed barrier radius, 
one obtains
 
\begin{equation}
\sigma_l(E)= \frac{(2l+1)}{k^2R_b^2}\Big[E'\frac{d\sigma}{dE}\Big|_{E=E'}+\sigma(E')\Big].
 \label{sigmal}
\end{equation}
and 
\begin{equation}
E'=E- \frac{l(l+1)\hbar^2}{2\mu R_b^2}.
\label{Eprimo}
\end{equation}
This approach was applied, for example, to the analysis of fusion data for the systems $^{16}$O + $^{112}$Cd, $^{28}$Si + $^{94,100}$Mo and $^{58,64}$Ni + $^{64}$Ni~\cite{Ackermann,Ackermann2}.

\subsection{Extracting barrier distributions from fusion data}
\label{extract}

The point-difference formula is very useful for approximating the second derivative $B(E)$ (see Eq.~(\ref{BD_final})) and, consequently, to extract the fusion barrier distribution from sets of data (excitation functions). It has been routinely used especially in cases where the energy spacing between consecutive energy points is constant (or just about constant). Let's consider such a set of data $\sigma_f(E)$ with a fixed energy spacing $\Delta$E. Then the point-difference formulae for the first and second energy derivative of $E\sigma_f(E)$ read respectively

\begin{equation}
 \frac{d}{dE}[E \sigma_f(E)] = \frac{(E \sigma_f)_2 -(E \sigma_f)_1}{\Delta E}                                     
 \label{2points}
\end{equation}
\begin{equation}
\frac{d^2}{dE^2}[E \sigma_f(E)] =  \Big[\frac{(E \sigma_f)_3-2(E \sigma_f)_2+(E \sigma_f)_1}{\Delta E^2}\Big]
\label{3points}
\end{equation}
Next we consider the uncertainties in the derivatives obtained in this way. Let us suppose that the relative (statistical) error affecting the n-th measured cross sections is ($\Delta\sigma$)$_n$ = $f_n\sigma_n$. Then it is easy to show~\cite{AnnuRev,Row92} that the error in the first derivative is
\begin{equation}
 \delta\Big[\frac{d}{dE}(E \sigma_f(E))] = \Big[(f\sigma E)^2_{n+1} + (f\sigma E)^2_{n}\Big]^{1/2}/\Delta E                                 
 \label{err2points}
\end{equation}

which is approximated, if all the $f$ are the same, by 

\begin{equation}
 \delta\Big[\frac{d}{dE}(E \sigma_f(E))]  \approx\sqrt 2fE\sigma_f(E)/\Delta E                                    
 \label{errapprox2points}
\end{equation}

and, for the second derivative

\begin{equation}
 \delta\Big[\frac{d^2}{dE^2}(E \sigma_f(E))\Big]  \approx\sqrt 6fE\sigma_f(E)/(\Delta E)^2                                    
 \label{errapprox3points}
\end{equation}

This expression tells us that the error on the barrier distribution extracted from a measurement of the fusion excitation function increases with 1/$\Delta E^2$. Therefore, it is convenient to use relatively large energy steps provided that
 $\Delta E\leq$0.56$\hbar\omega$ MeV, i.e. $\simeq$2-3 MeV, even if this may cause some degree of smoothing in the shape of the distribution introduced by quantum tunnelling. This does not mean that one should not measure excitation functions with smaller energy steps. Indeed, it is usually important to perform experiments with significantly smaller steps, because this provides with two (or even more) distinct data sets yielding 
independent measurements of the barrier distribution. Experimental groups routinely use this procedure.

\begin{figure}[h]
\centering
\resizebox{0.40\textwidth}{!}{\includegraphics{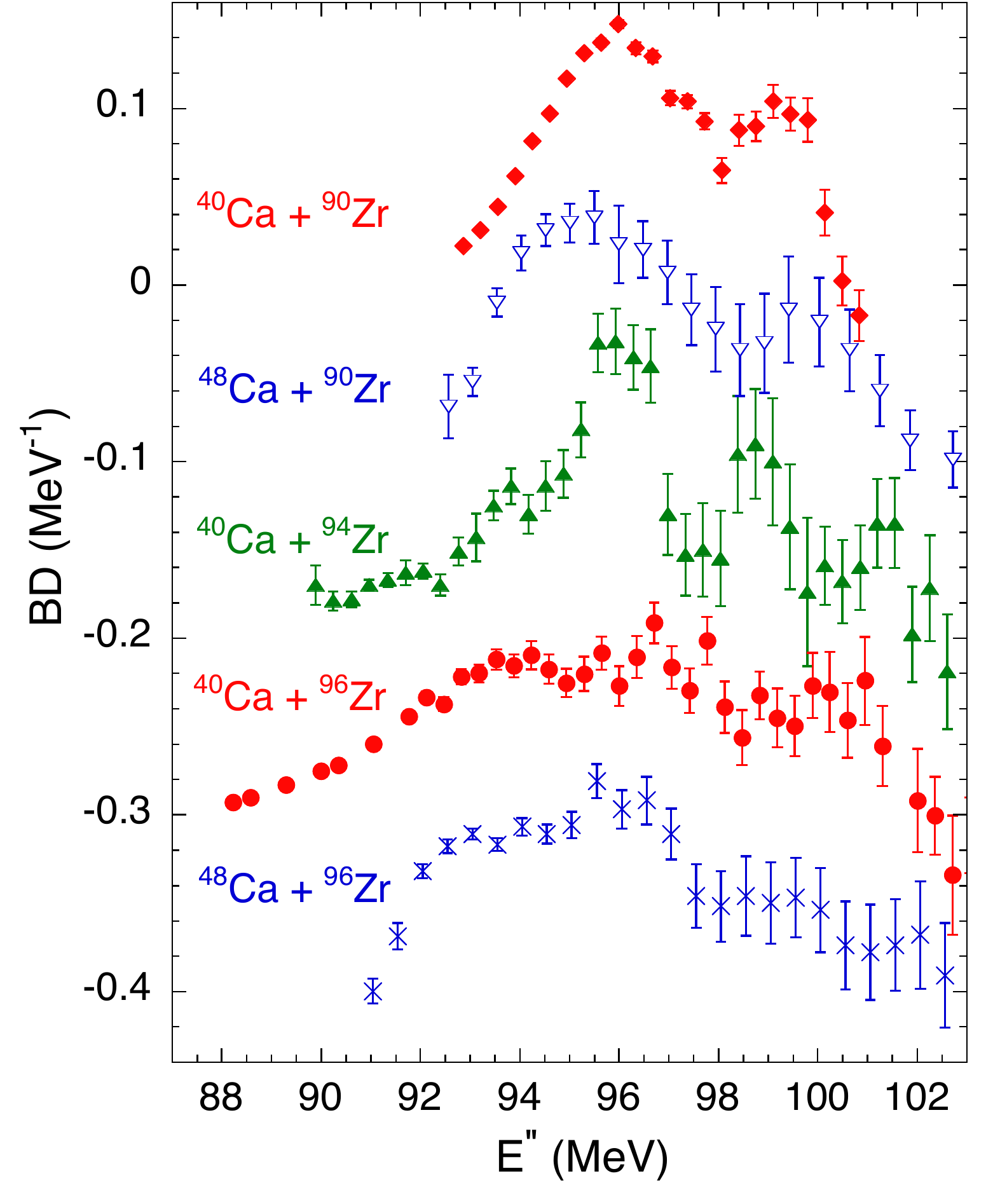}}
\caption{Examples of fusion barrier distributions for several Ca+Zr systems, obtained with the three-point difference formula~\cite{4094}. The ordinates for the various cases are diminished by successive 0.1 MeV$^{-1}$ steps with respect to $^{40}$Ca + $^{90}$Zr. The energy scale E" is modified so that the excitation functions of all systems approximately coincide above the barrier. Figure redrawn from Ref.~\cite{4094}.}
\label{BD_all}     
\end{figure}

From  the formula~(\ref{errapprox3points}) it is also evident that the error in the derivative increases linearly with the 
cross section $\sigma_f(E)$ and with the energy E. However in a typical experiment the energy is varied by only  $\simeq$  10-15$\%$ 
around the barrier, so that the main effect is introduced by the strong variation of the cross section in that limited energy range. 
With respect to the sub-barrier region, the cross section increases considerably above the barrier while the second derivative of the excitation function tends to vanish  in the classical limit. 
It follows that the error (see Eq.~(\ref{errapprox3points})) may become much larger than the derivative itself.

For the same reason, the  statistical uncertainties in  the measured cross sections $\sigma_f$ should be kept as small as possible.  One should keep in mind anyway, that having a 1$\%$ statistical error requires
accumulating ten thousand fusion events, which may not always be so easy, depending on the experimental conditions.

Fig.~\ref{BD_all} shows the  barrier distributions  obtained by  applying  the point-difference formula to the fusion excitation functions of several Ca + Zr systems~\cite{4094}. 
It appears that the various  BD have different shapes reflecting different coupling effects. In particular the long low-energy tails observed for $^{40}$Ca + $^{94,96}$Zr originate from strong couplings to positive Q-value transfer channels available in those two systems only. 
This behaviour will be discussed in more detail in Sect.~\ref{InfTransf}.
We notice that the errors are systematically larger at higher energies as indicated by Eq.~(\ref{errapprox3points}).

The following  Figs.~\ref{Leigh},~\ref {5860} also show  that the uncertainties on the extracted barrier distributions become very large at high energies so that this method gradually loses its sensitivity to the underlying fusion dynamics, and its usefulness drops above the barrier.

\par
\subsection{Fusion barrier distributions for several systems}
\label{expBD}

\begin{figure}[h]
\centering
\resizebox{0.40\textwidth}{!}{\includegraphics{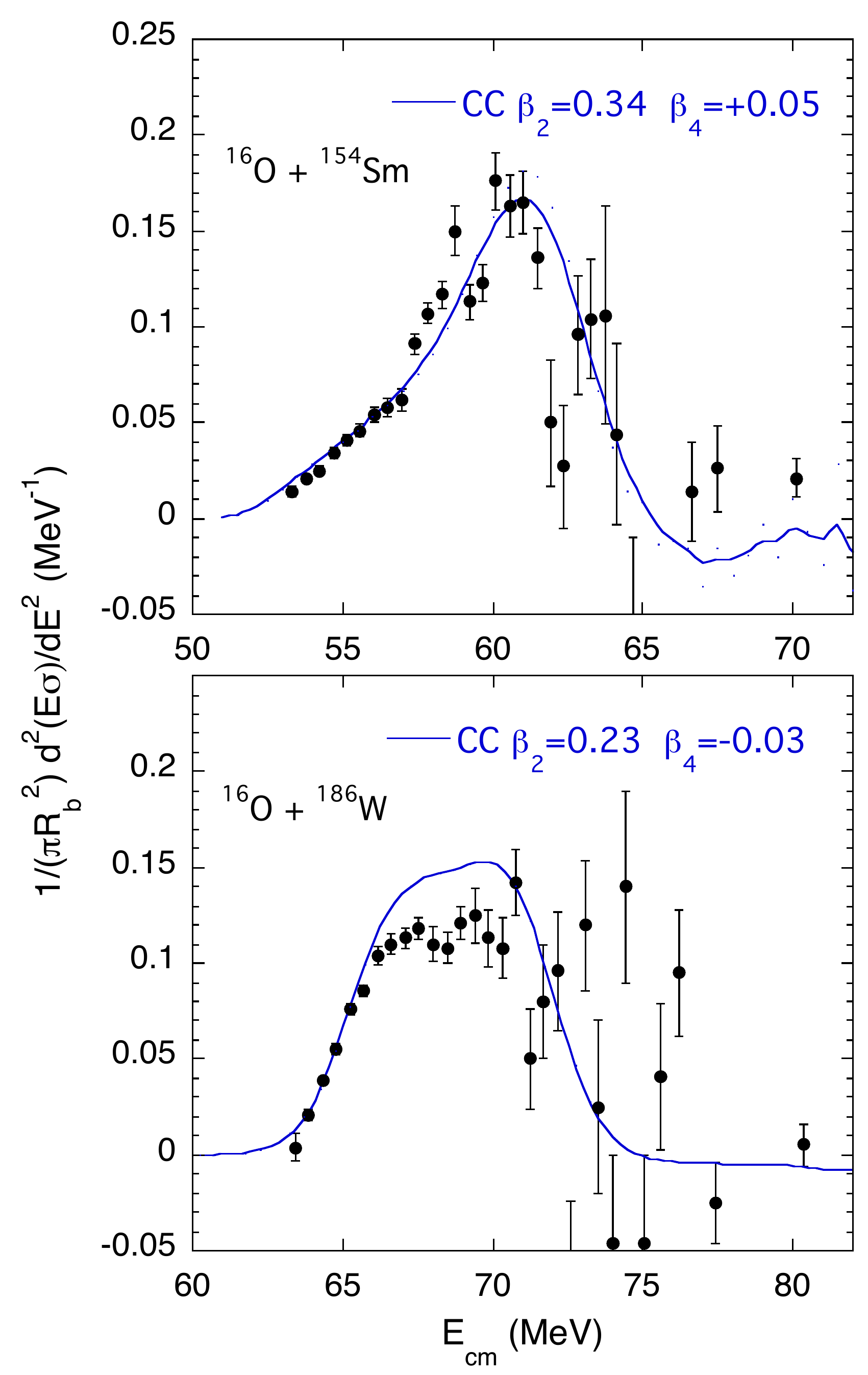}}
\caption{Fusion barrier distributions of $^{16}$O + $^{154}$Sm (upper panel), and for $^{16}$O +$^{186}$W (lower panel)~\cite{Leigh95}. The lines are CC calculations including the ground-state rotational bands of the two heavy nuclei, up to the 8$^+$ level. One can clearly notice the sensitivity of the  BD shape of the two systems to the different deformation parameters optimising the CC calculations.}
\label{Leigh}     
\end{figure}

The study of fusion reactions between medium-mass nuclei is particularly attractive because channel couplings are often strong enough to produce wide fusion barrier distributions and, consequently, large sub-barrier cross section enhancements. Peaks and structures can be identified in the barrier distributions, and are signatures of the couplings involved. At the same time, measurements of fusion cross sections between medium-heavy nuclei are relatively simple, since fusion-evaporation is, in most cases, the only relevant reaction channel following capture inside the Coulomb barrier.

\begin{figure}[h]
\centering
\resizebox{0.45\textwidth}{!}{\includegraphics{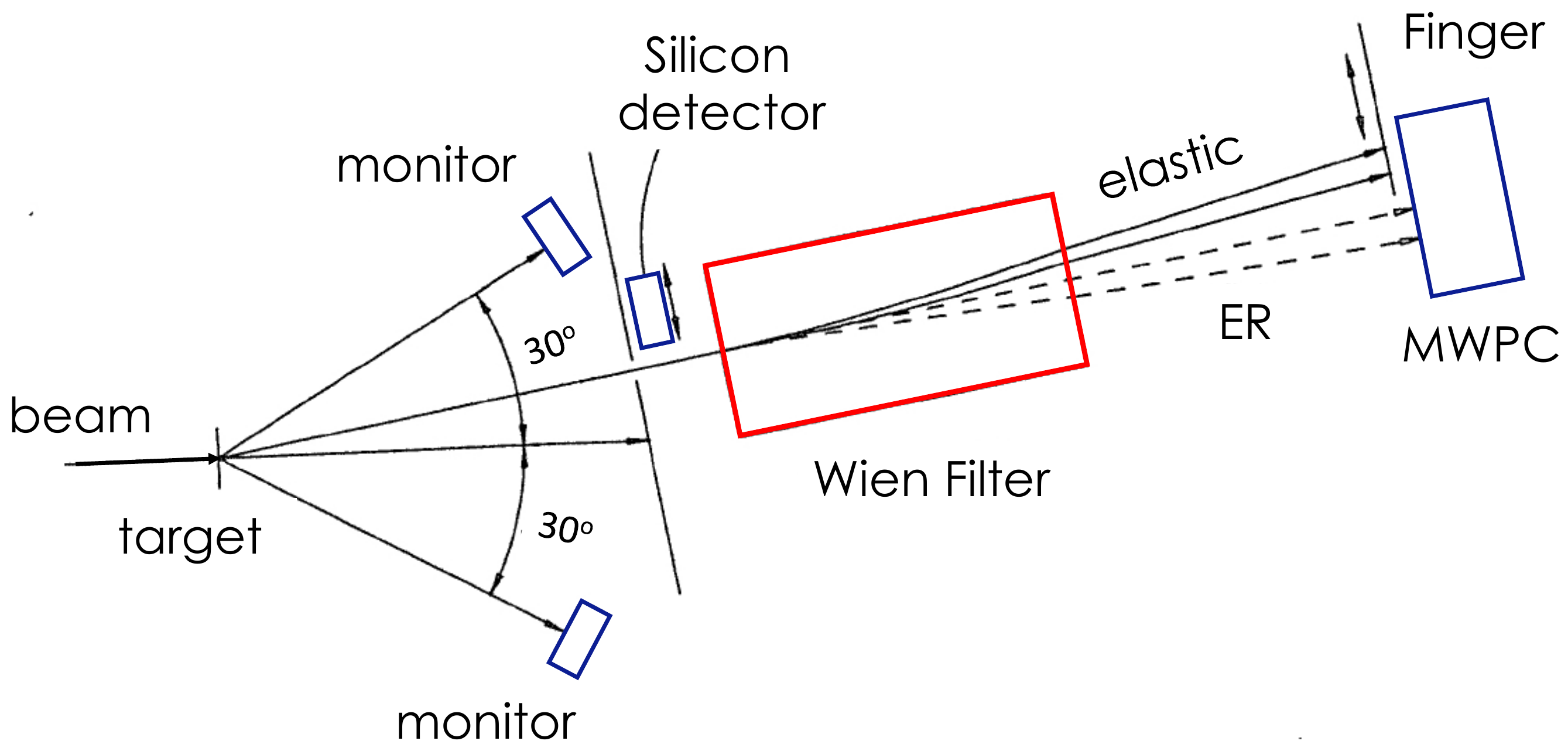}}
\caption{The set-up based on a velocity filter at the Australian National University. Figure modified from Ref.~\cite{WeiFilter}.}
\label{ANU}     
\end{figure}

The sensitivity of the shape of barrier distributions to the static nuclear deformation was evidenced in early experiments performed by the Canberra group. Indeed, Fig.~\ref{Leigh} shows the results of those detailed measurements of excitation functions for the two systems $^{16}$O + $^{154}$Sm, $^{186}$W~\cite{Leigh95}, where $^{16}$O can be considered inert because of its very stiff  structure (the lowest 3$^-$ and 2$^+$ excitations are at 6.13 and 6.92 MeV, respectively). 
The experiments were performed at ANU using the simple set-up shown in Fig.~\ref{ANU}, based on a velocity filter to separate the ER from the overwhelming flux of beam-like and elastic scattering events at very small angles.

The remarkable feature emerging from those results is the sensitivity of the barrier distribution not only to the static quadrupole deformation $\beta_2$, but also to the hexadecapole deformation $\beta_4$. As a matter of fact, the CC calculations reported 
in Fig.~\ref{Leigh} reproduce quite nicely the difference between the two cases, mainly related to the opposite sign of $\beta_4$. In the figure, and in the following ones, the BD is normalised so to have unit area, by dividing the second derivative of $E\sigma$ by $1/\pi R_b^2$ (see Eq.~(\ref{Wong_2})).
\par

Multi-phonon excitations
have been shown to become dominant for medium-heavy nuclei~\cite{Esben05} and produce
complex fusion barrier distributions, in some cases with discrete structures. In this sense, 
the experimental study of $^{58}$Ni + $^{60}$Ni~\cite{PRL95} revealed for the first time
the existence of a barrier distribution with several well-defined peaks that could only be explained by multiphonon couplings. 

\begin{figure}[h]
\centering
\resizebox{0.40\textwidth}{!}{\includegraphics{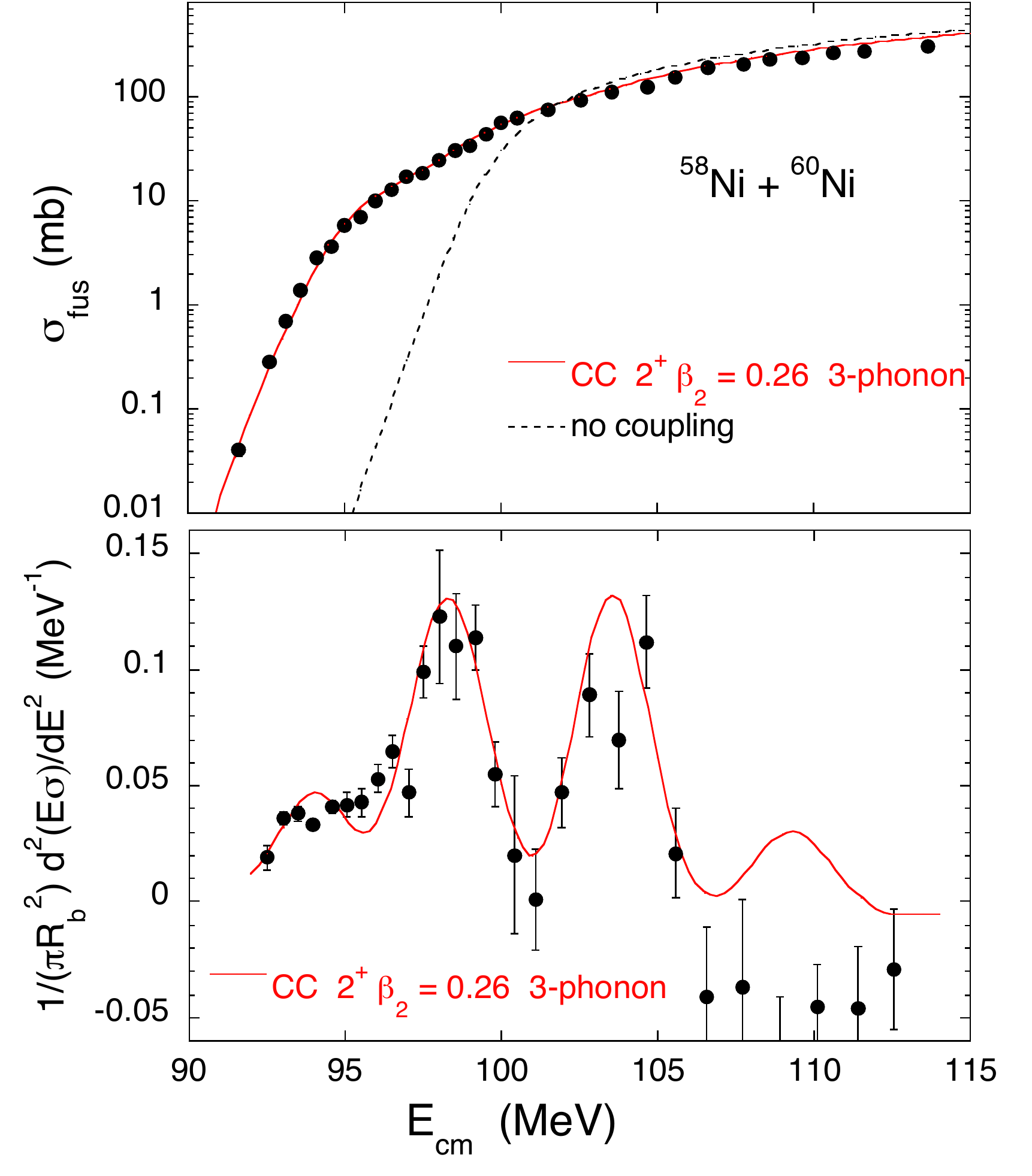}}
\caption{Fusion excitation function (top panel) and experimental barrier distribution (bottom panel) of $^{58}$Ni + $^{60}$Ni~\cite{PRL95}, compared to CC calculations including three quadrupole phonons (see text). The dashed line shows the calculated cross sections in the no-coupling limit. The calculations used  a Woods-Saxon potential with parameters  V$_o$= 180.5 MeV, r$_o$= 1.0 fm and a = 0.90 fm.}
\label{5860}     
\end{figure}

The essential outcome of those measurements is summarized in Fig.\ref{5860}. The fusion cross sections near and below the barrier show a huge enhancement with respect to the calculation in the one-dimensional barrier limit (``no coupling" line in the top panel), otherwise the excitation function seems to be quite structureless and it is not easy to identify the channel coupling(s) producing such enhancement.
This is best revealed by the barrier distribution (bottom panel), that displays a clear  structure with  three well-defined peaks.

The CC calculations in that figure have been performed using the code CCFULL~\cite{CCFULL} with an ion-ion potential very similar to that employed in the original article~\cite{PRL95}. Up to three quadrupole  phonons of the form  $\ket{21}$
 and  $\ket{12}$ have been included,  built by mutual excitation of the single- and double-phonon states in either nucleus.
It is worth while pointing out that the original calculations  were performed using a CC code precursor  of CCFULL, where  fitting  the data required  two phonons in both nuclei (four quadrupole phonons overall). Using CCFULL  we obtain a comparably good agreement with only three phonons.

As an alternative approach, Hagino and Yao~\cite{HagYao} applied the multi-reference density functional theory, in combination with CC calculations, to study the sub-barrier fusion of $^{58}$Ni + $^{60}$Ni. Anharmonicity effects were also taken into account, which tend to smear the fusion barrier distribution. 


The interest in the study of $^{58}$Ni + $^{60}$Ni was originally triggered by the search for a specific influence of the two neutron elastic transfer channel on the near-barrier fusion behaviour. 
Indeed, strong coupling
to a single channel with zero Q-value is predicted to produce a characteristic fusion barrier distribution with 
two peaks, one on each side of the original uncoupled
Coulomb barrier. 
The experimental results, however, and the evidences coming from the comparison with theoretical models indicate that multi-phonon excitations determine the sub-barrier fusion yields and the shape of the barrier distribution to a large extent, so that the possible effect of the elastic transfer (if any) is overcome and not observable in this system.
\par


In a further study, the possibility to observe a double-peaked barrier distribution was investigated in the case of $^{42}$Ca + $^{40}$Ca.
$^{40}$Ca is a magic nucleus with double shell closure, having a strong octupole vibration at high energy (3.737 MeV). Its effect on sub-barrier fusion is expected to be mainly a potential renormalisation ``rigidly" shifting the barrier distribution to lower energies, but without essentially affecting its shape. The other nucleus
$^{42}$Ca, with two neutrons in the 1f$_{7/2}$ shell, is rather stiff with a weak 2$^+$ quadrupole excitation at 1.524 MeV.
This situation makes it ``a priori" plausible that the influence of the elastic 2-neutron pair transfer can be recognised in the barrier distribution.
\par
\begin{figure}[h]
\centering
\resizebox{0.38\textwidth}{!}{\includegraphics{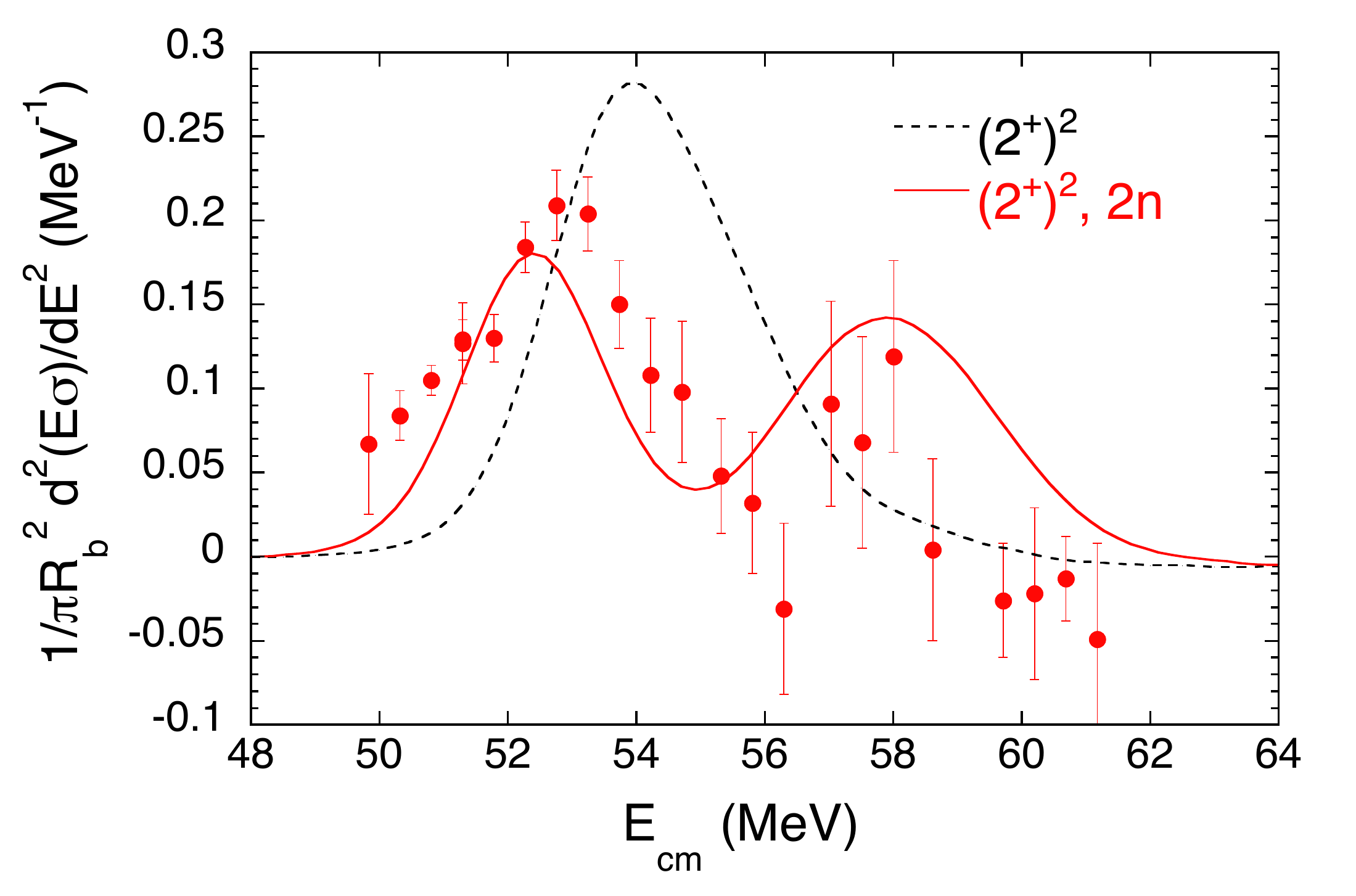}}
\caption{Extracted barrier distribution of  $^{42}$Ca + $^{40}$Ca~\cite{4240}, compared to the results of CC calculations (see text). 
The black dashed line is the result of the calculation when only quadrupole excitations are considered. Including additionally the coupling of the 2n elastic transfer channel  one obtains the red line showing  two quasi-symmetric peaks closely resembling the experimental distribution.}
\label{4240}     
\end{figure}

The BD extracted from the data is shown in Fig.~\ref{4240}, where two nice peaks can actually be observed~\cite{4240}, and this has been compared to the predictions of CC calculations~\cite{CCFULL}. In this code
the 2n pair transfer is schematically described (or simulated) by the form factor~\cite{EsbenFric} 
$$ V_t = - \sigma_t dU(r)/dr$$ In those calculations $\sigma_t$ has been given the value 0.39 fm best fitting the two-neutron transfer coupling in $^{40}$Ca + $^{48}$Ca~\cite{4040}.
When quadrupole surface excitations and the additional 2n-transfer mode are considered,  an encouraging two-peak structure is predicted for $^{42}$Ca + $^{40}$Ca, closely resembling the experimental BD.
However, by including also the octupole modes additional oscillations are calculated (not shown in Fig.~\ref{4240}) and the good agreement is lost. Hence, while we can anticipate that the effect of transfer is surely large, the evidence for an {\it elastic} transfer coupling is not clear even in the case of this system $^{42}$Ca + $^{40}$Ca. One should not forget that the expression of the 2n form factor is very rough, and its strength is arbitrary in the calculations to a large extent, 
so that the requirement is felt to place theoretical predictions for this and other systems where transfer is expected, on a more solid basis. Probably the best way to accomplish this would be to measure the transfer cross sections and calibrate the models to reproduce them.

\begin{figure}[h]
\centering
\resizebox{0.40\textwidth}{!}{\includegraphics{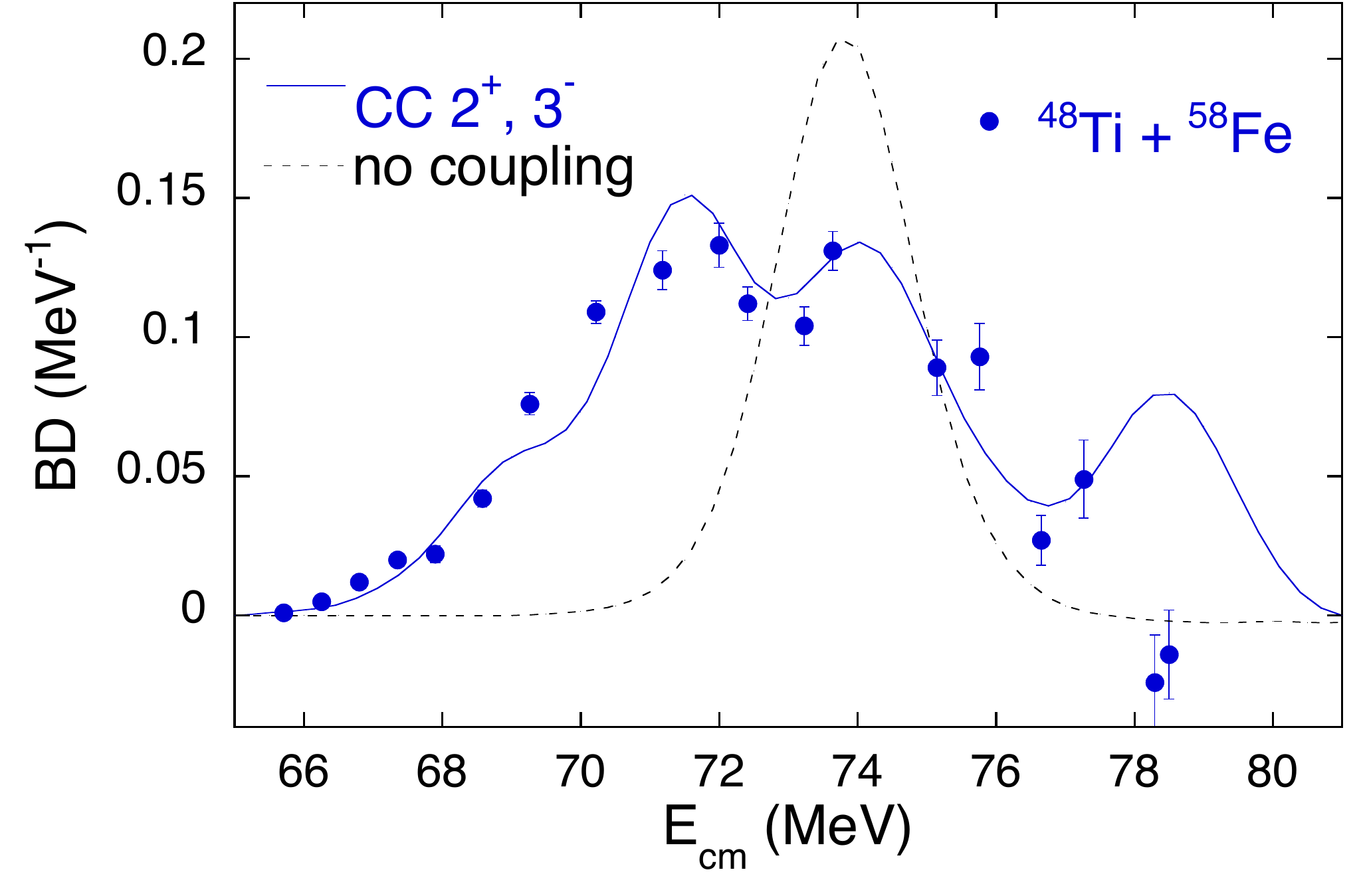}}
\caption{Fusion barrier distribution for $^{48}$Ti + $^{58}$Fe compared with CC calculations (see text)~\cite{48+58}. The single peak calculated in the  no coupling limit (its ordinate is divided by two) is shown  to emphasise the  strong effect of low-energy quadrupole (and octupole) vibrations.}
\label{BD_4858}     
\end{figure}

When the influence of several weaky coupled inelastic states or of several transfer channels is relevant, the shape of the barrier distribution may undergo some smoothing (see the following Sects.~\ref{BDqe} and ~\ref{InfTransf} for specific examples). 
Otherwise, couplings to low-energy collective modes of the two nuclei often produce a clear fingerprint in the structure of the BD.
 One of the most recent examples of such a situation is 
 the  barrier distribution for the system $^{48}$Ti + $^{58}$Fe, shown  in Fig.~\ref{BD_4858}~\cite{48+58}. It  has a complex structure with various partially resolved peaks (resembling the case of $^{58}$Ni + $^{60}$Ni discussed above). This is mainly due to the strong quadrupole vibrations existing in both soft nuclei, while the octupole modes are rather weak and at excitation energies above 3.3 MeV, anyway. This is confirmed by the standard CC calculations using a WS potential, which reproduce the shape of the BD quite well. 
 It will be  further pointed out in Sect.~\ref{subhindrance}
where this system is compared to the behaviour of $^{58}$Ni + $^{54}$Fe in relation to the hindrance effect.

An analysis of the moments of fusion-barrier distributions has recently been presented~\cite{Rehm}. Rather than extracting the second derivative of the excitation function (see Eq.~(\ref{cent})),  it is proposed in that work that one simply obtains the successive moments of the distribution by fitting the energy-weighted cross sections. It is shown that the fusion radius and the height of the Coulomb barrier are given by the zeroth and first moments, the second moment is the width of the distribution, and the third moment is its skewness. It is pointed out that a small fusion radius and a large width are systematically observed for systems where transfer couplings are strong, and prolate (oblate) target deformations are correlated with a negative (positive) skewness of the distribution. 

This kind of analysis may be useful for comparing the barrier distributions measured for different systems, and, more importantly, for obtaining overall information on the barrier distribution for cases where the excitation function has not been measured with small energy steps, possibly orienting future more detailed studies on those systems.

Umar and Oberacker originally introduced the
density - constrained Time Dependent Hartree Fock method (TDHF) to
calculate the energy-dependent ion-ion interaction potential~\cite{Umar,Umar2}, and successful investigations on
 sub-barrier fusion have been performed for various systems~\cite{Obera,Obera2,Keser,Sime0}.
 \par
\begin{figure}[h]
\centering
\resizebox{0.45\textwidth}{!}{\includegraphics{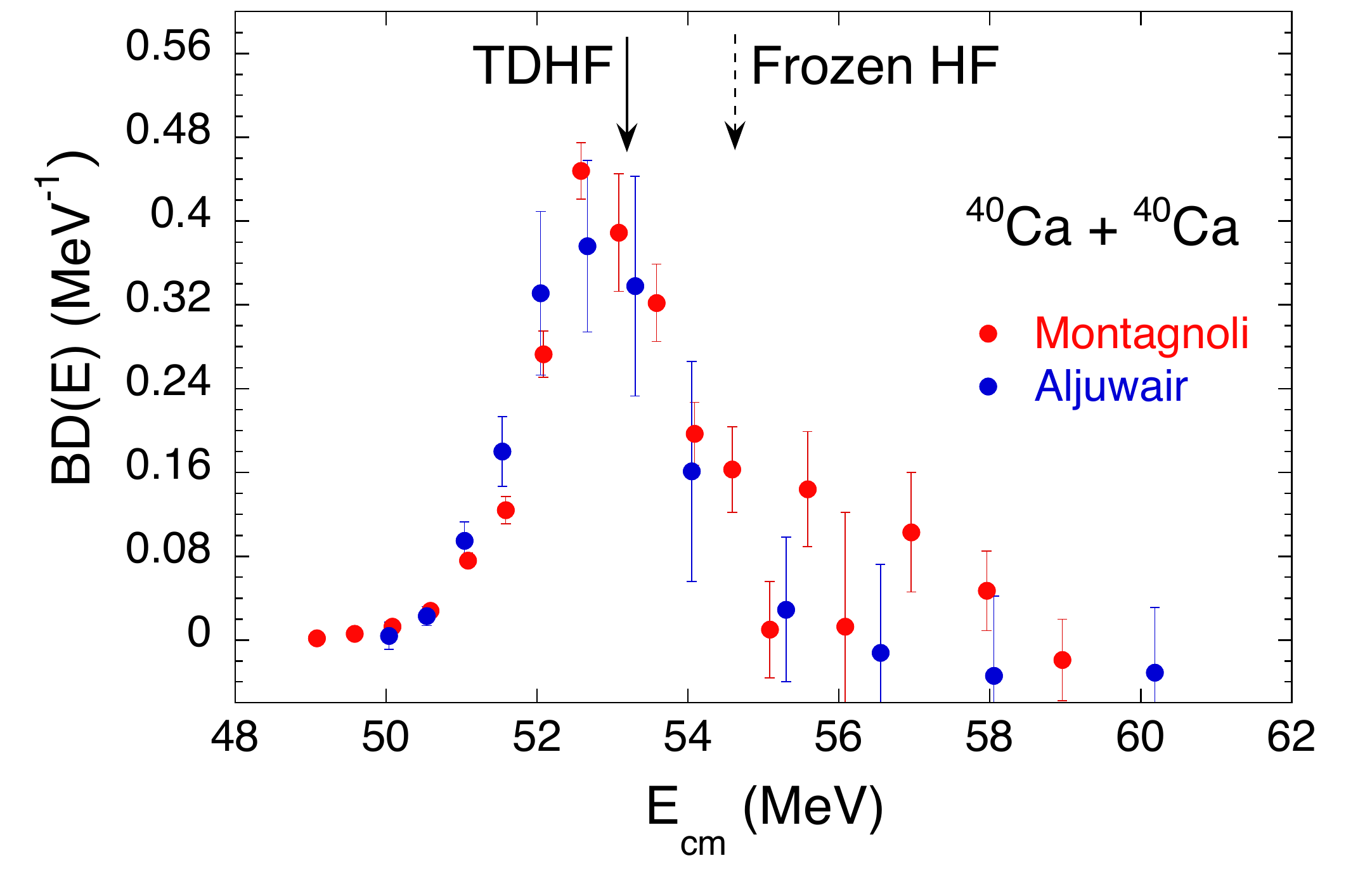}}
\caption{Fusion barrier distribution for
$^{40}$Ca + $^{40}$Ca from the data of Montagnoli et al.~\cite{4040} and of Aljuwair et al.~\cite{Alju}. 
The dashed arrow marks the barrier position according to the frozen Hartree-Fock method, while the solid arrow indicates
the TDHF fusion threshold, that is, the expected position of the centroid
of the barrier distribution. Couplings to low-lying collective excitations are responsible for the difference between the 
TDHF and frozen HF barriers~\cite{Sime}.}
\label{TDHF_Simenel}     
\end{figure}

 The applications and the usefulness of the method are growing, in the field of low-energy heavy-ion reactions, thanks also to the increasing computational capabilities.
 We want to mention, as an example, the results recently obtained with the TDHF method for $^{40}$Ca + $^{40}$Ca and $^{56}$Ni + $^{56}$Ni~\cite{Sime}. The barrier position obtained in that work for $^{40}$Ca + $^{40}$Ca is compared in Fig.~\ref{TDHF_Simenel} to the outcomes of two independent experiments, and the result is quite good. 
One may say that most excitation and transfer processes are correctly described by the TDHF mean-field dynamics, even if this is achieved in an average way, at variance to what happens in CC calculations~\cite{Sime0}.

\subsection{Barrier distributions from back scattering }
\label{BDqe}

It was proposed to extract barrier distributions produced by the channel couplings also  from pure elastic and quasi elastic scattering cross sections  at backward angles~\cite{RowleyE,TimmersQE}.
A further detailed study by Hagino and Rowley~\cite{HagRow} was dedicated to comparing the quasi elastic barrier distribution $D_{qe}(E)$ with the corresponding distribution for fusion. They obtained a justification of the BD concept for scattering processes. 

In a very rough picture of the collision between two heavy ions in the vicinity of the Coulomb barrier, we can separate the incident flux of particles in a transmitted flux through the barrier, leading to fusion, and in a reflected flux in various elastic, inelastic and few-nucleon transfer reaction channels forming overall the so-called quasi-elastic scattering yield. 

For medium-mass systems in that energy range other reaction channels may be considered of negligible importance, hence flux conservation would tell us that any effect influencing fusion yields (e.g. coupled channels effects) would affect quasi-elastic yields in a parallel way.  
We simplify this illustration by considering the case of a single barrier (no coupled channels) and a head-on collision ($\theta$ = 180$^o$, angular momentum $l\hbar$=0). The transmission coefficient $T(E)$ and the reflection coefficient $R(E)$ have to sum up to one

\begin{equation}
T(E) + R(E) =1
\label{TandR}
\end{equation}

 where $R(E)$ and $T(E)$ can be expressed as 

\begin{equation}
R(E) = \frac{d\sigma_{qe}}{d\sigma_R}(E)
\label{express}
\end{equation}

\begin{equation}
T(E) = \frac{1}{\pi R_b^2}\frac{d}{dE}[E\sigma_f(E)]
\label{express}
\end{equation}

A further differentiation with respect to the energy gives the normalised fusion barrier distribution (see Eq.~\ref{cent})

\begin{equation}
D_f(E) = \frac{dT}{dE} = \frac{1}{\pi R_b^2}\frac{d^2}{dE^2}[E\sigma_f(E)]=\frac{1}{\pi R_b^2}B(E)
\label{dTdE_1}
\end{equation}

Classically, this is a $\delta$-function. It was however shown~\cite{rowBD} that in a quantum-mechanical description this equation is still valid, but the $\delta$-function is replaced by a 
a gaussian-like function of width $\simeq2-3$ MeV (see Eq.\ref{Wong_2}), i.e.

\begin{equation}
D_f(E)= \frac{2\pi}{\hbar\omega}\frac{e^x}{(1+e^x)^2}
\label{dTdE}
\end{equation}

where $x$=$2\pi (E-V_b)/\hbar\omega$ and, by combining the equations 

\begin{equation}
D_f(E) = \frac{dT}{dE} = -\frac{dR}{dE}=-\frac{d}{dE}\Big[\frac{d\sigma_{qe}}{d\sigma_R} (E)\Big] \equiv D_{qe}(E) 
\label{dRdE}
\end{equation}

When several barriers are involved as a consequence of channel couplings, one can show that, for a particular angle $\theta$,

\begin{equation}
\frac{d\sigma_{qe}}{d\sigma_R}(E) = \sum\limits_{m}w_m\frac{d\sigma^{m}_{el}}{d\sigma_R}(E)
\label{multiple}
\end{equation}

that is, the total quasi-elastic cross section is a weighted sum of eigenchannel elastic cross sections $d\sigma^m_{el}/d\sigma_R(E)$ associated to the various barriers~\cite{Andres}, with the same weights $w_m$ appearing in Eq.~\ref{sigmaCC}. We can differentiate this further, and obtain

\begin{equation}
D_{qe}(E) = -\frac{d}{dE}\Big[\frac{d\sigma_{qe}}{d\sigma_R} (E)\Big] \label{Dqe}
\end{equation}
\begin{equation}
\nonumber
= -\sum\limits_{m}w_m\frac{d}{dE}(E)\Big[\frac{d\sigma^{m}_{el}}{d\sigma_R}(E)\Big] = \sum\limits_{m}w_mD^m_{el}(E).
\end{equation}

\begin{figure}[h]
\centering
\resizebox{0.40\textwidth}{!}{\includegraphics{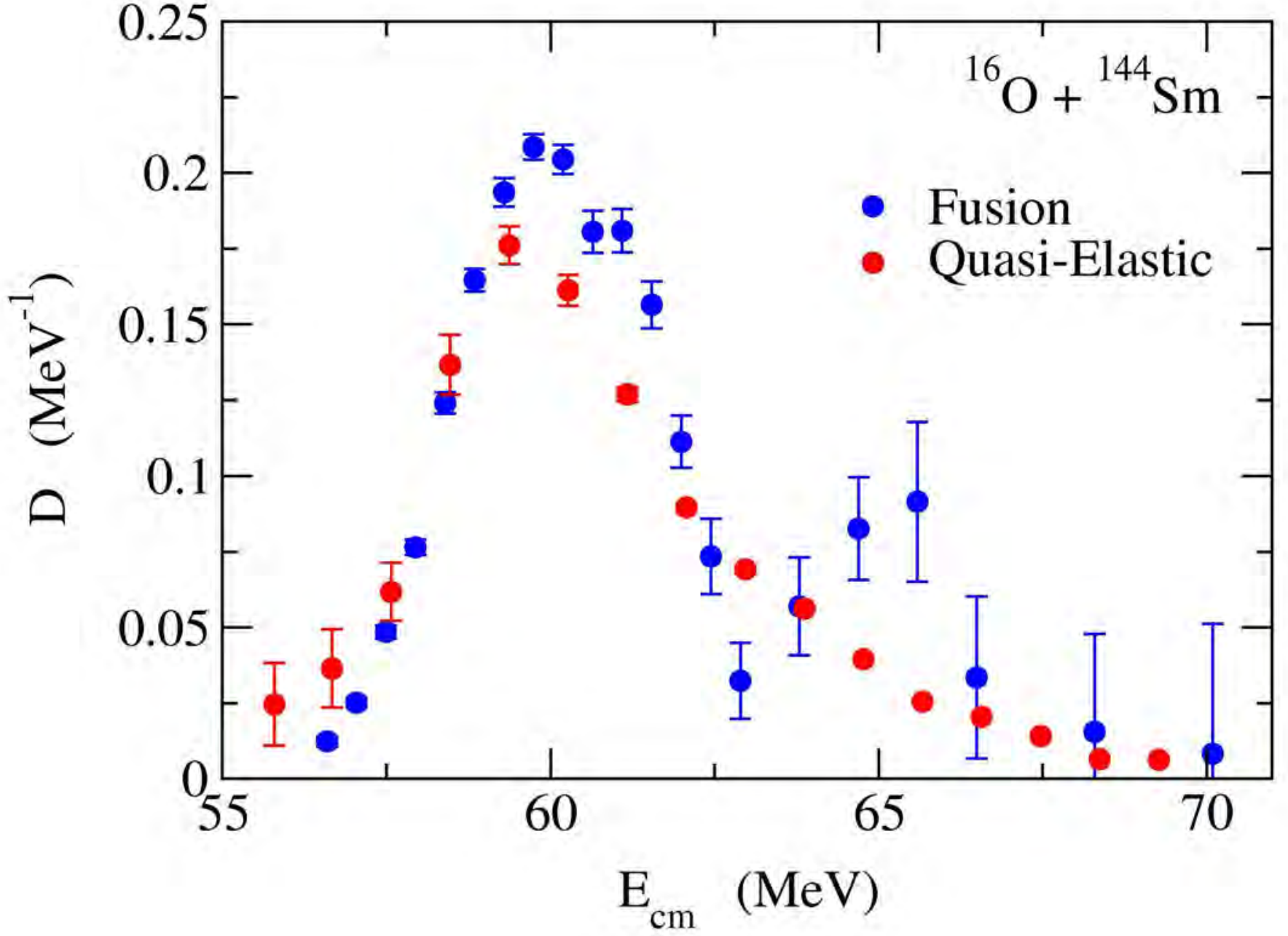}}
\caption{Representations of barrier distributions derived from fusion ($D_{f}(E)$, blue dots) and backangle quasi-elastic scattering ($D_{qe}(E)$, red dots) excitation functions of $^{16}$O + $^{144}$Sm~\cite{TimmersQE}.  
$D_{f}(E)$ shows two separate peaks, but the higher energy one disappears in the $D_{qe}(E)$ representation.
}
\label{BDqe_Timmers}     
\end{figure}

Measuring quasi-elastic scattering cross sections at $\theta = 180^o$ is usually a difficult experimental task, however, it turns out that this condition may be relaxed to some extent, that is, $D_{qe}(E)$ is a representation of the barrier distribution keeping essentially the structure of $D_f(E)$ even when $\theta$ is a backward angle only ``reasonably" close to $180^o$. We are going to see some examples of this. 
When analysing sets of data not rigorously taken at $\theta = 180^o$, it is necessary, to be able to perform a meaningful comparison with $D_f(E)$, to correct the energy by the centrifugal energy $E_{cent}$ which is given by 

\begin{equation}
E_{cent} = E_{cm}\frac{cosec(\theta/2) -1}{cosec(\theta/2) +1}
\label{multiple}
\end{equation}

when considering Rutherford orbits. The main advantage of using the $D_{qe}(E)$ representation of the barrier distribution is that it yields much smaller experimental uncertainties above the Coulomb barrier than $D_f(E)$. 
Indeed, we have seen (see Eq.~\ref{errapprox3points}) that the error associated with the extraction of the second derivative of the fusion excitation function increases with the energy and, more importantly, with the fusion cross section. In most cases, even reducing the statistical uncertainties to the level of $1\%$ is not enough to obtain the shape of the BD with reasonable accuracy at high energy.
\par
The error associated with $D_{qe}(E)$, on the contrary, while being relatively large at low energy, becomes smaller above the barrier, where $d\sigma_{qe}/d\sigma_R(E)$ drops rapidly. In fact, one can easily show~\cite{TimmersPhD} that

\begin{equation}
\Delta(D_{qe}(E)) \simeq \frac{\delta\sqrt2}{\Delta E}\Big[\frac{d\sigma_{qe}}{d\sigma_R} (E)\Big]
\label{error_Back}
\end{equation}

where, by approximating the energy dependence of $d\sigma_{qe}/d\sigma_R$ with a two point-difference formula, $\delta$ is the average of the absolute uncertainties of  the two cross sections and $\Delta$E is the energy step.
One might say that the two representations $D_{qe}(E)$ and $D_f(E)$ are complementary to each other from the point of view of the experimental accuracies. This is shown in Fig.~\ref{BDqe_Timmers} where,  on the other hand, 
an additional feature can be noted, that is, $D_f(E)$ has two distinct peaks while the higher-energy barrier disappears in $D_{qe}(E)$. This effect was early attributed~\cite{TimmersQE} to transfer channels, and was observed in various other cases later on. It seems that the sensitivity of quasi-elastic excitation functions to the real fusion barrier distribution $D_f(E)$ is reduced above the barrier.

 Both  barrier distributions from the quasi-elastic $D_{qe}(E)$ and fusion data  $D_f(E)$ were extracted for the two systems  $^{40}$Ca + $^{90,96}$Zr~\cite{Timmers}. The results  confirmed that the two representations can be complementary to each other for medium mass systems.
\par
\begin{figure}[h]
\centering
\resizebox{0.35\textwidth}{!}{\includegraphics{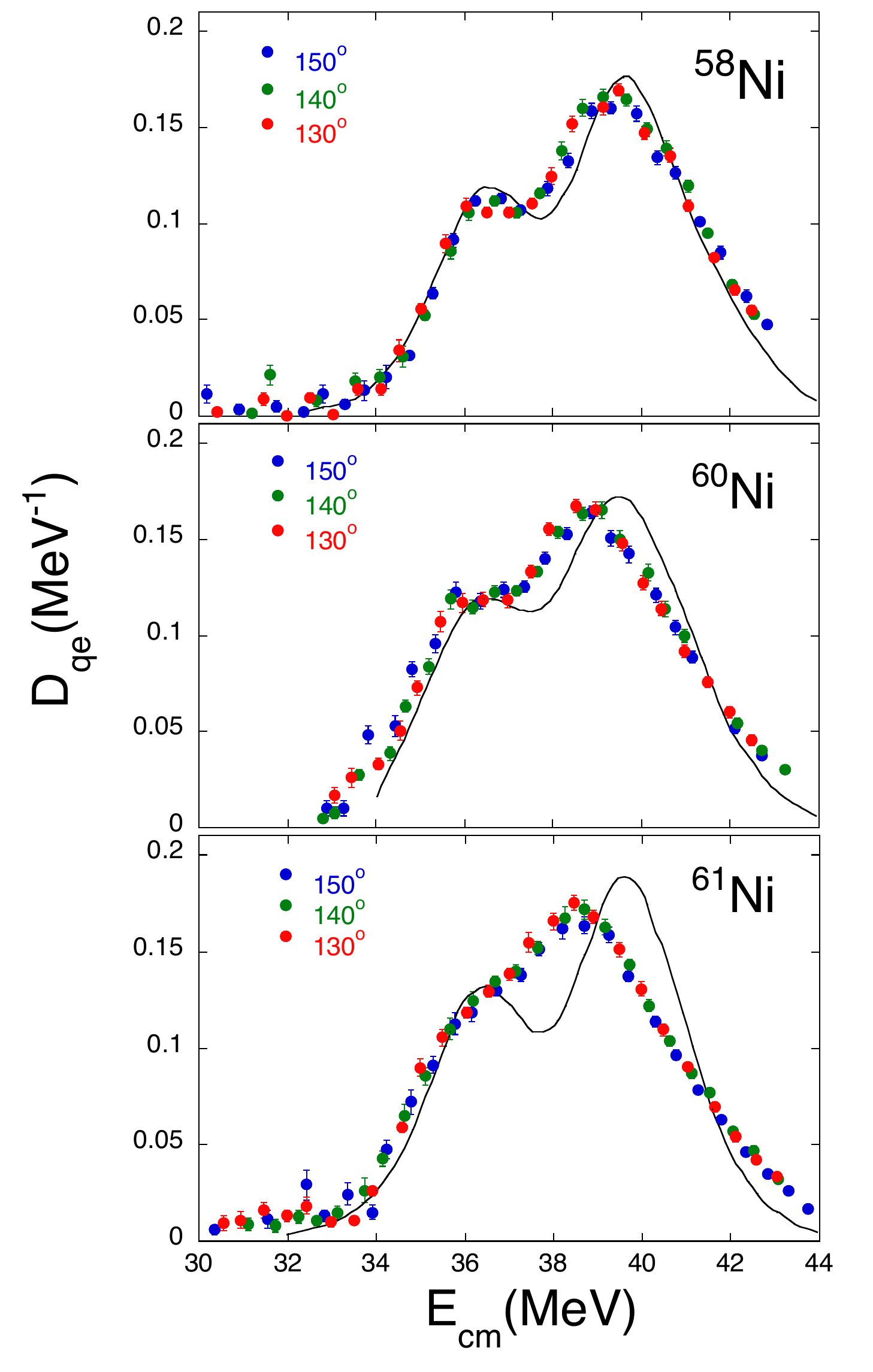}}
\caption{Barrier distributions for  $^{20}$Ne + $^{58,60,61}$Ni derived using  the quasi-elastic  excitation functions~\cite{Piase}. The solid lines are the CC calculations  obtained by
means of the CCQEL code. The distributions are very similar for the two even mass Ni isotopes, while a certain degree of smoothing can be notice for  $^{61}$Ni, see text. Figure redrawn from Ref.~\cite{Piase}.}
\label{PiaseNeNi}     
\end{figure}

Following previous results on   $^{20}$Ne + $^{90,92}$Zr~\cite{PiaseZr}
 $D_{qe}(E)$ was used by the Warsaw group~\cite{Piase} to evidence 
the influence of weak (non-collective) reaction channels on barrier height distributions and, consequently, on fusion dynamics.
By measuring the backward quasi-elastic scattering of $^{20}$Ne + $^{58,60,61}$Ni they observed a two-peak structure  in the extracted BD for the two targets $^{58,60}$Ni, which almost completely disappears for $^{61}$Ni. This is shown in Fig.~\ref{PiaseNeNi} where the data taken at three different  backward angles are reported together with the calculations performed with the  CCQEL code.
The difference between the even- and the odd-mass target supports the hypothesis that  weak  couplings
to several non-collective single-particle states cause the  smoothing of the $D_{qe}(E)$ shape of the odd Ni isotope.

 \begin{figure}[h]
\centering
\resizebox{0.45\textwidth}{!}{\includegraphics{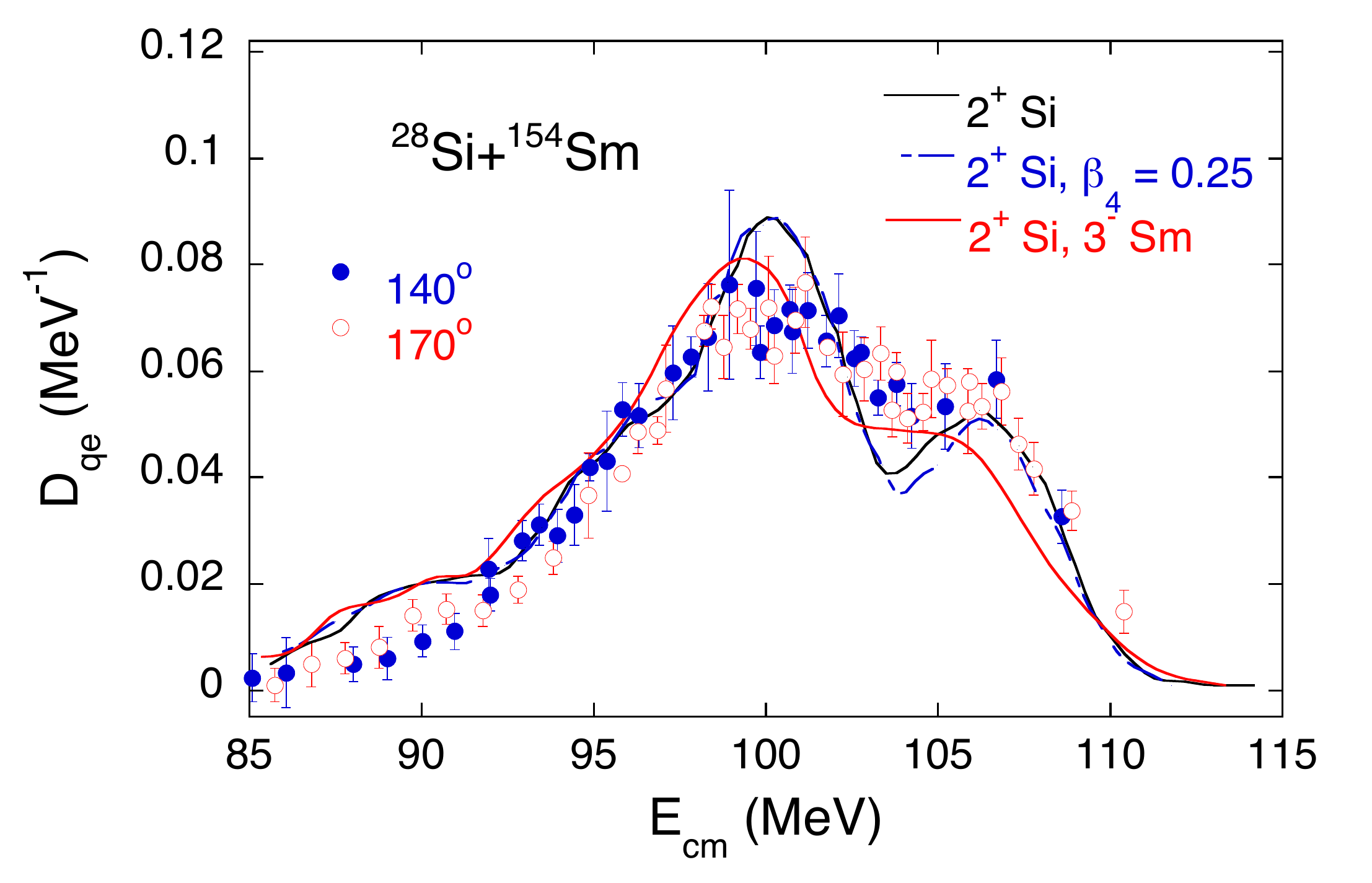}}
\caption{Barrier
distribution for the $^{28}$Si  + $^{154}$Sm system resulting from measurements of quasi elastic scattering  cross sections at backward angles~\cite{Gurp}. The lines are CC calculations using  vibrational (solid line) and rotational (dotted-dash line) couplings for the first 2$^+$ state of $^{28}$Si, using $\beta_4$=0.25  for the  hexadecapole deformation. The octupole vibration of $^{154}$Sm causes a smoothing effect shown by the red line. Figure redrawn from Ref.~\cite{Gurp}.}
\label{28154}     
\end{figure}

An interesting study was recently performed by Jia et al.~\cite{JiaBD} on the feasibility of deducing the hexadecapole deformation $\beta_4$ from backward quasi-elastic scattering. They investigated  the systems 
$^{16}$O  + $^{152}$Sm, $^{170}$Er and $^{174}$Yb near the Coulomb barrier. The extracted BD were analysed to obtain $\beta_4$ values by means of CC calculations, that are in reasonable agreement with available results.
Barrier  distributions  $D_{qe}(E)$ were also used in Ref.~\cite{Gurp}, 
to study the effect of couplings  in the $^{28}$Si  + $^{154}$Sm reaction, showing a high degree of sensitivity to 
projectile excitations. In particular, those authors show that a good fit  with CC calculations requires a strong positive hexadecapole deformation of $^{28}$Si. Fig.~\ref{28154} illustrates the result of this experiment.
 \par

\begin{figure}[h]
\centering
\resizebox{0.30\textwidth}{!}{\includegraphics{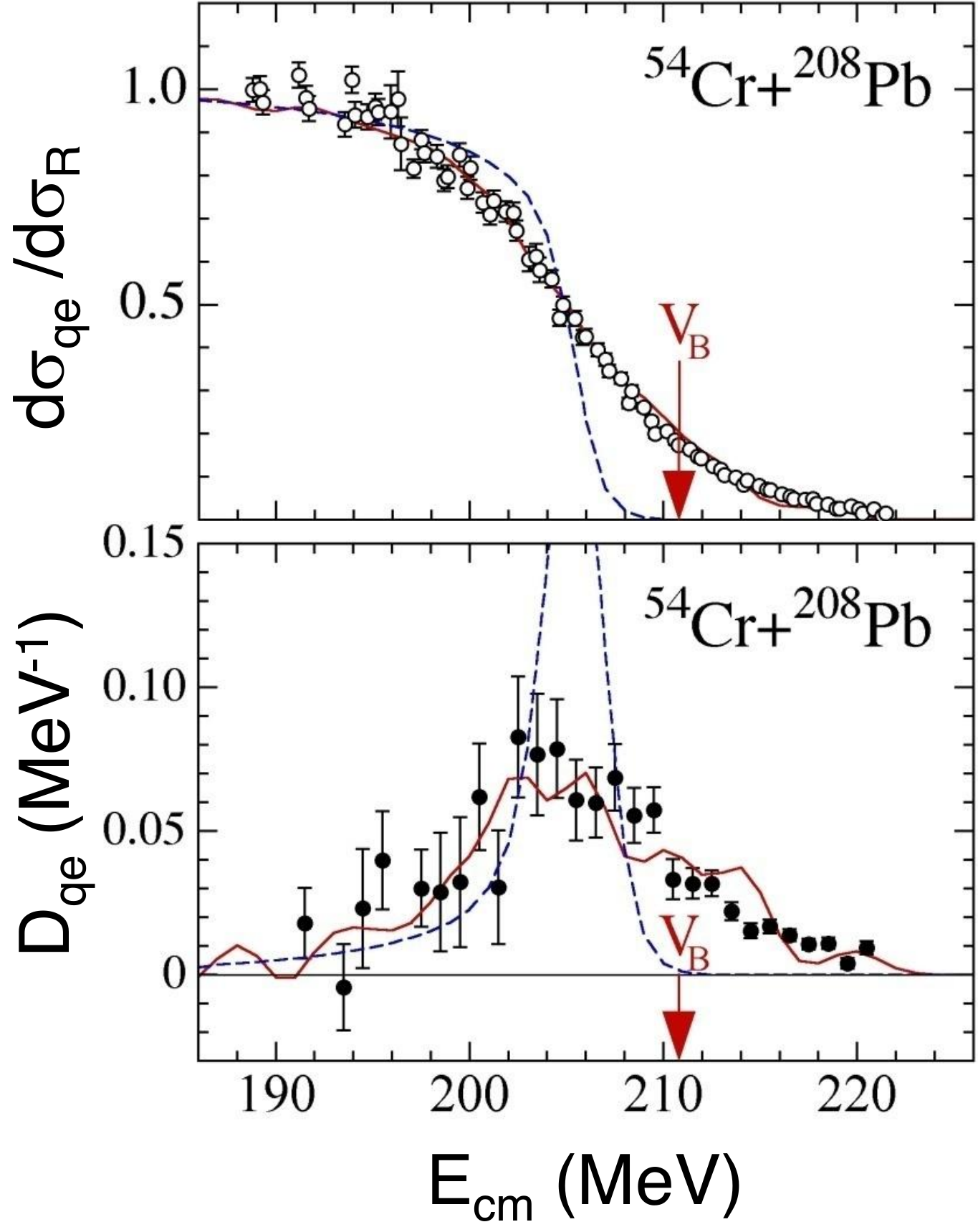}}
\caption{Excitation function of quasi-elastic scattering of $^{54}$Cr  + $^{208}$Pb~\cite{Mits} (upper panel), and extracted barrier distribution. The arrow indicates the position of the Bass barrier~\cite{Bass}. The dashed line is the calculation in the no-coupling limit, and the full line is the CC calculation taking into account the quadrupole  excitation for $^{54}$Cr and the octupole state of $^{208}$Pb.}
\label{CrPb}     
\end{figure}

Some years ago, the method of obtaining the BD from the QE excitation function was applied in an experiment on heavy systems  performed by Mitsuoka et al.~\cite{Mits} in relation to cold fusion reactions for the production of superheavy elements Z=104,106, 108, 110 and 112, at the Tandem booster facility of the Japan Atomic Energy Agency.
The results for the system  $^{54}$Cr  + $^{208}$Pb are reported  in  Fig.~\ref{CrPb} where one can notice that CC calculations  reproduce the shape of the BD rather well, and that the centroid of the distribution is lower by  $\simeq$6 MeV  with respect to the Bass  barrier. Mitsuoka et al. considered these  results and the analogous ones obtained for other heavy systems  very relevant for  the synthesis of superheavy
elements, where the barrier height  is a critical quantity  and the choice of the  bombarding energy is consequently crucial.
 \par
Barrier distributions from QE excitation functions were measured also for the very heavy system $^{86}$Kr  + $^{208}$Pb (eventually leading to the Z=118 superheavy element by cold fusion) by Ntshangase et al.~\cite{Ntsh}. That experiment supported  the concept of a barrier distribution, produced by strong entrance-channel couplings, even in such an extreme case where fusion is very nearly absent.

The data of Ref.~\cite{Mits} were analysed by Pollarolo~\cite{Polla_QE} using the semiclassical model including both surface excitations and nucleon exchange between the two nuclei. The excitation functions  and the BD  are successfully reproduced and it appears that the contribution of transfer channels is relevant at all measured energies.
However both Pollarolo and Zagrebaev~\cite{Zagre}  point out that 
for such systems  fusion and quasielastic scattering do not exhaust
most of the total reaction cross section, because many other reaction channels like deep inelastic or quasi fission, are dominant. Therefore one cannot expect the equivalence between $D_{qe}$ and $D_{f}$ and the meaning of  $D_{qe}$ is, rather, a reaction threshold distribution.
Consequently, one should take some care in using the results of similar experiment on very heavy systems 
when planning  challenging measurements for the synthesis of superheavy elements.

\section{Influence of transfer channels on fusion}
\label{InfTransf}

\subsection{General considerations}
\label{general}

The effect of couplings to nucleon transfer channels is  certainly  important in the near and sub-barrier  fusion process for  several systems following the first  suggestion of Broglia et al.~\cite{Bro83} that two-neutron transfer with $Q>$0 should enhance sub-barrier fusion. However,  identifying more quantitatively  that effect has often been elusive, when deduced from comparing with CC calculations  that are affected by unavoidable approximations.  One might say that the importance of transfer couplings is clear only in the cases where the experimental evidence is already unambiguous, as e.g. is the case of the results of the early experiments of Beckerman et al.~\cite{BeckNiNi} on the various Ni+Ni systems.

Transfer reactions have an obvious peculiarity with respect to inelastic excitations, that is, the Q-value of the reaction may be negative or {\it positive}. The CC model indicates that coupling to $Q>$0 reaction channels produces a trend of the sub-barrier excitation function much different from what is expected by coupling low-energy collective modes of the colliding nuclei with $Q<$0. Qualitatively speaking, the excitation function decreases more slowly with the energy when  $Q>$0. 

This is due to the basic different shape of the barrier distribution produced by the two kind of couplings.  When considering in the CC model, for simplicity, a situation with only one inelastic excitation to be coupled in (see Refs.~\cite{Dasso,Dasso2}), one obtains a barrier distribution possessing a peak  {\it above} the energy of the unperturbed barrier, and the main barrier is shifted to lower energies.
A typical example of such a situation is reported in Fig.~\ref{BDqe_Timmers} for the system $^{16}$O + $^{144}$Sm~\cite{TimmersQE}. The BD has two peaks, and the higher one (having lower intensity) is originated from the coupling to the lowest 2$^+$ excitation of $^{144}$Sm at E$_x$= 1.660 MeV.
\begin{figure}[h]
\centering
\resizebox{0.40\textwidth}{!}{\includegraphics{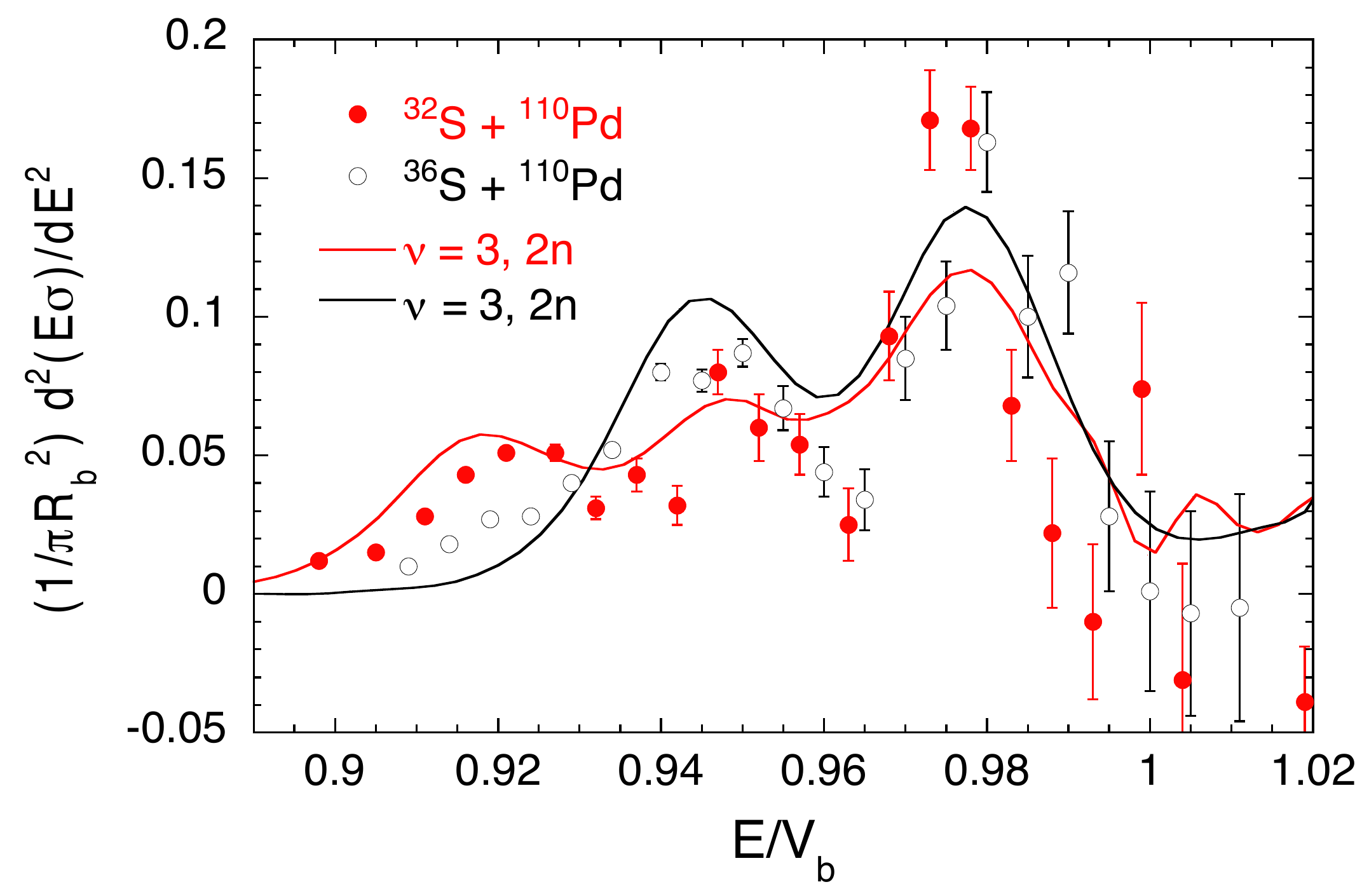}}
\caption{Fusion barrier distributions of $^{32,36}$S+$^{110}$Pd~\cite{3236110} in a reduced energy scale~\cite{Akyuz}. The curves are CC calculations including the 2$^+$ and 3$^-$ of $^{32,36}$S, three phonons of the collective quadrupole mode in $^{110}$Pd and the two neutron (2n) pick-up channel.
The small peak at $E/V_b\simeq$ 0.92 was attributed to the effect of coupling to the  transfer channel in $^{32}$S+$^{110}$Pd having $Q>$0.
}
\label{S+110}     
\end{figure}
\par
The opposite situation shows up by coupling to a $Q>$0 reaction channel, that produces  a structure {\it below} the unperturbed barrier which, in turn, is shifted to slightly higher energies. In real situations, inelastic excitations of different type may contribute, and two-, three, and possibly multiple phonon excitations have to be considered, so that the observed BD often have a complex structure. An interesting example of such a situation was revealed by the  
investigation of $^{32,36}$S+$^{110}$Pd~\cite{3236110}. In this case the dominant influence of the strong quadrupole surface vibrations of $^{110}$Pd leaves space to a significant sub-barrier cross section enhancement due to the 2-neutron pick-up channel in $^{32}$S+$^{110}$Pd having $Q>$0, that produces as well a clear structure in the barrier distribution at very low energies. This is shown in Fig.~\ref{S+110} reporting the BD extracted for the two systems. 

They are very similar to each other with at least two analogous peaks near and slightly below the AW~\cite{Akyuz} model barrier, due to the excitation of up to three phonons of the collective quadrupole mode in $^{110}$Pd (common to both systems). A smaller (but very clear) peak is observed in $^{32}$S+$^{110}$Pd only, and it was attributed to the coupling to the 2n pick-up channel in this system, that has $Q_{g.s.}$= +5.1 MeV.
\begin{figure}[h]
\centering
\resizebox{0.45\textwidth}{!}{\includegraphics{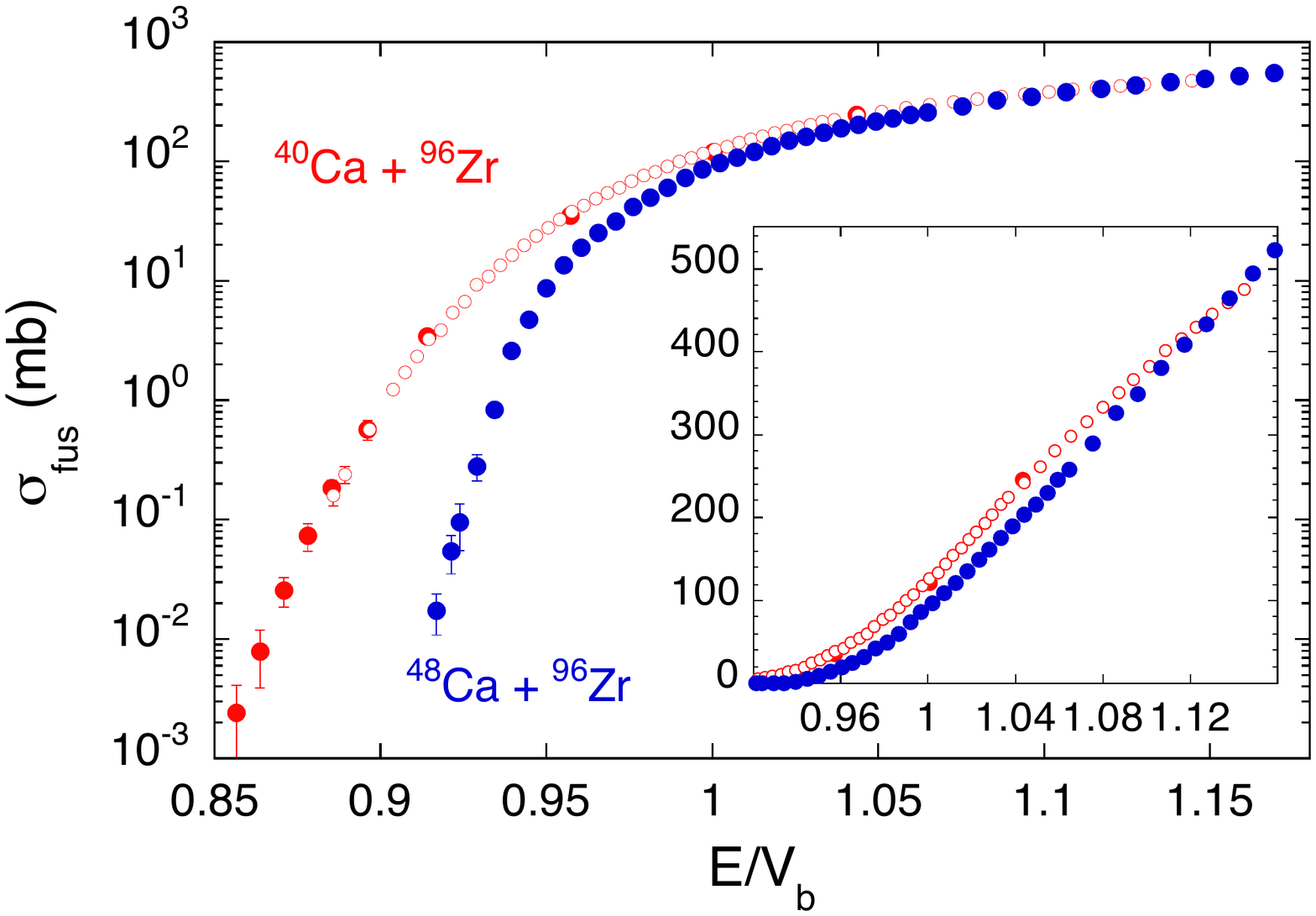}}
\caption{Fusion excitation functions of $^{40}$Ca + $^{96}$Zr ( full dots from Ref.~\cite{4096}, open symbols from Ref.~\cite{Timmers}) and $^{48}$Ca + $^{96}$Zr~\cite{48+Zr}. The insert shows the high energy part of the excitation functions in a linear cross section scale. 
The  energy scale is relative to the Coulomb barrier V$_b$ obtained with the  AW potential~\cite{Akyuz}. }
\label{40_48+96}     
\end{figure}

What appears to be important, for the influence of transfer reactions on fusion, is not the ground state Q-value, but the ``effective" Q-value.
The effective Q value $Q_{eff}$ for transfer reactions is defined in
terms of the ground state Q-value $Q_{gs}$ as $Q_{eff}=Q_{gs}+\Delta V_{CB}$~\cite{Bro83}
where $\Delta V_{CB}$ is the difference in the height of the Coulomb barrier
in the entrance and in the exit channels. 
It is usually observed that transfer couplings with $Q_{eff}>$0 can produce large enhancements of the fusion cross sctions below the barrier. Instead, couplings to transfer channels with large $Q_{eff}<$0 have a weaker effect essentially leading to an adiabatic renormalisation of the potential~\cite{PTP}.
For neutron transfer channels, $\Delta V_{CB}$ is a small quantity, so that the effective Q-value is close to the ground state Q-value. The same is not true for transfer reactions of charged particles, however, couplings to such reactions are usually, but not systematically, of little importance for fusion cross sections.
\par
A problem may arise when trying to identify (or predict) the influence of transfer channels on fusion. This is because nucleon transfer effects should show up most clearly at energies well below the barrier, and that is the energy range where fusion hindrance (see Sect.~\ref{Hindrance}) is expected to show up. Therefore, low-energy fusion cross sections are determined by the competing  effects of hindrance and enhancement, and disentangling the two opposite contributions may be problematic.

\subsection{Medium-mass and medium-light systems}
\label{medium-light}

The case of $^{40}$Ca + $^{96}$Zr~\cite{4096} is very significant for the investigation of the effects of transfer couplings on fusion.
The sub-barrier  excitation function  has been recently measured down to cross sections $\simeq$2.4$\mu$b, i.e. two orders of magnitude smaller than obtained in a previous experiment~\cite{Timmers}, where the low-energy fusion of this system was found to be greatly enhanced with respect to $^{40}$Ca + $^{90}$Zr, and the need of coupling to transfer channels was suggested. Indeed, $^{40}$Ca + $^{96}$Zr has several neutron pick-up transfer channels with large and positive Q-values.
The comparison with $^{48}$Ca + $^{96}$Zr, where no $Q>$0 transfer channels are available, is illuminating (Fig.~\ref{40_48+96}). The sub-barrier cross sections of this system drop very steeply, while the excitation function of $^{40}$Ca + $^{96}$Zr decreases slowly (and smoothly). Fig.~\ref{BD_all}  reported in Sect.~\ref{extract} shows that the two barrier distributions have quite different behavior, the one of $^{40}$Ca + $^{96}$Zr extending much more toward low energies. A low-energy tail is also present in $^{40}$Ca + $^{94}$Zr
where the situation of Q-values for transfer channels is analogous.

In the insert of Fig.~\ref{40_48+96}, one also notices that far above the barrier the excitation function for $^{40}$Ca + $^{96}$Zr increases more slowly than for $^{48}$Ca + $^{96}$Zr, and the cross sections tend to become smaller.
This is a further typical effect of strong couplings to transfer channels~\cite{Rehm,4096}.
 
\begin{figure}[h]
\centering
\resizebox{0.45\textwidth}{!}{\includegraphics{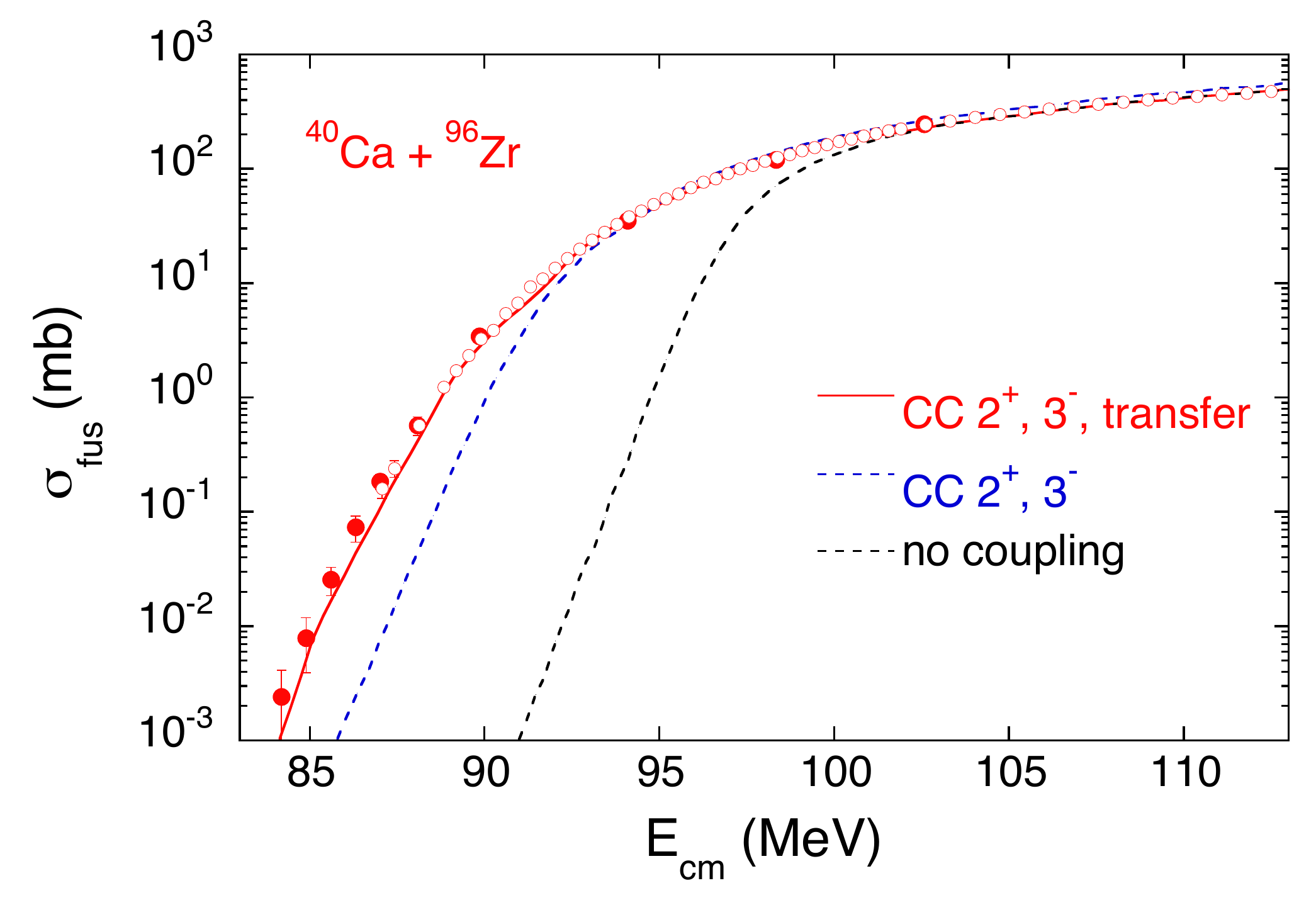}}
\caption{Fusion cross sections for $^{40}$Ca + $^{96}$Zr~\cite{4096,Timmers} are compared with CC calculations~\cite{Esb_4096} using the 
WS potential. The red line reproducing the data includes couplings to the one- and two-nucleon transfer channels besides the inelastic excitation 2$^+$ and 3$^-$ of both nuclei (additionally, the relatively strong 5$^-$ excitation of $^{40}$Ca is taken into account). Figure modified and redrawn from Ref.~\cite{Esb_4096}.}
\label{40+96_Esbensen}     
\end{figure}

The recent CC analysis~\cite{Esb_4096} of the excitation function is shown in Fig.~\ref{40+96_Esbensen}. It includes explicitly one- and two-nucleon $Q>$0 transfer channels, with coupling strengths calibrated
to reproduce the measured neutron transfer data. Such transfer couplings bring significant cross section enhancements, even at the level of a few $\mu$b.
One obtains  an excellent account of the fusion data~\cite{esbcazr}, and a large contribution to the enhancement is due also
to proton stripping  channels having positive Q-values.

No indication of hindrance (see Sect.~\ref{Hindrance}) shows up, because a standard WS potential allows to fit the low-energy data (and, indeed, the logarithmic derivative stays well below the $L_{CS}$ limit).
 Locating the hindrance threshold, if any, in $^{40}$Ca + $^{96}$Zr would require challenging measurements of cross sections in the sub-$\mu$b range.

It is surprising that the  agreement with the data is obtained using the WS  potential  and ignoring the repulsive
part characteristic of the M3Y + repulsion interaction. The suggested interpretation in Ref.~\cite{esbcazr} was that, since 
the Q-values for nucleon
transfer are large and positive, the valence nucleons can 
flow more freely from one nucleus to the other 
without being hindered by Pauli blocking.

An alternative  quantal CC approach was employed by Scamps and Hagino~\cite{Scamps} to reproduce  simultaneously   the fusion cross sections and the transfer probabilities for the $^{40}$Ca + $^{96}$Zr system.
They  used phenomenological transfer  form factors  allowing to fit  the experimental  transfer data, and used them for the calculation of sub-barrier fusion cross sections  by considering  also the collective excitations of the two nuclei.
The low energy fusion excitation function is still underestimated, even if the shape of the  BD is rather nicely reproduced.

\par
\begin{figure}[h]
\centering
\resizebox{0.45\textwidth}{!}{\includegraphics{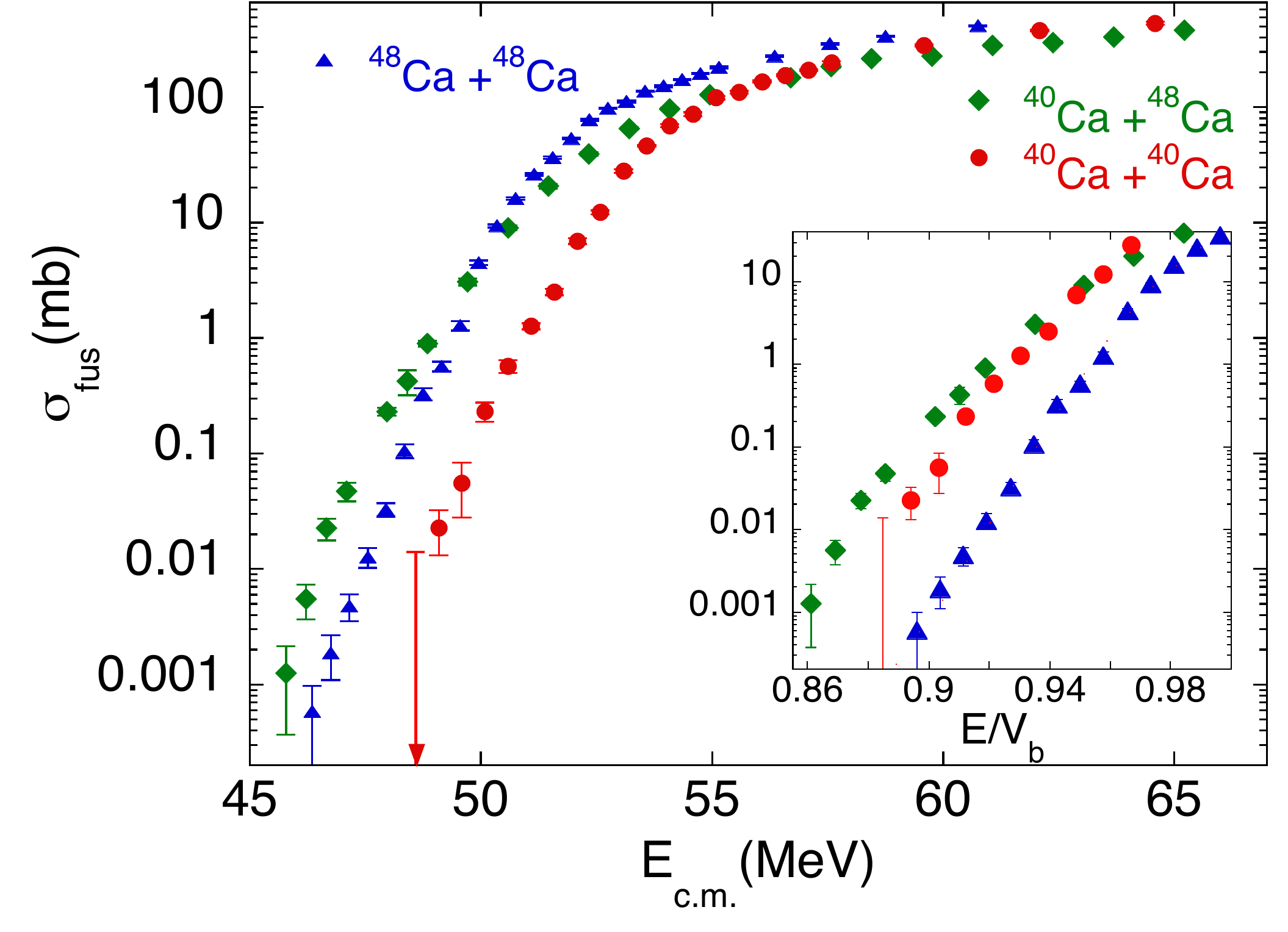}}
\caption{Fusion excitation functions of Ca + Ca systems~\cite{4040}. The inset shows the low-energy cross sections, where
the energy scale is normalised to the barrier resulting from the AW potential.}
\label{Calci}     
\end{figure}

The $^{40}$Ca+$^{96}$Zr fusion data were also analysed by V. V. Sargsyan et al. (see~\cite{Sargsyan} and Refs. therein) on the basis of the quantum-diffusion approach to barrier penetration. This model uses a double folding potential, and the influence of fluctuations and dissipation are taken into  account. This results in a quite  good fit, particularly for the previous data~\cite{Timmers}.  However, the new lowest energy points  tend to be  underestimated. More recently, the role of neutron-pair transfer reactions in sub-barrier capture processes was emphasized~\cite{Sargsyan3}, and it was suggested that the transfer of two neutrons influences fusion through the change of deformation of the colliding nuclei.

\begin{figure}[h]
\centering
\resizebox{0.48\textwidth}{!}{\includegraphics{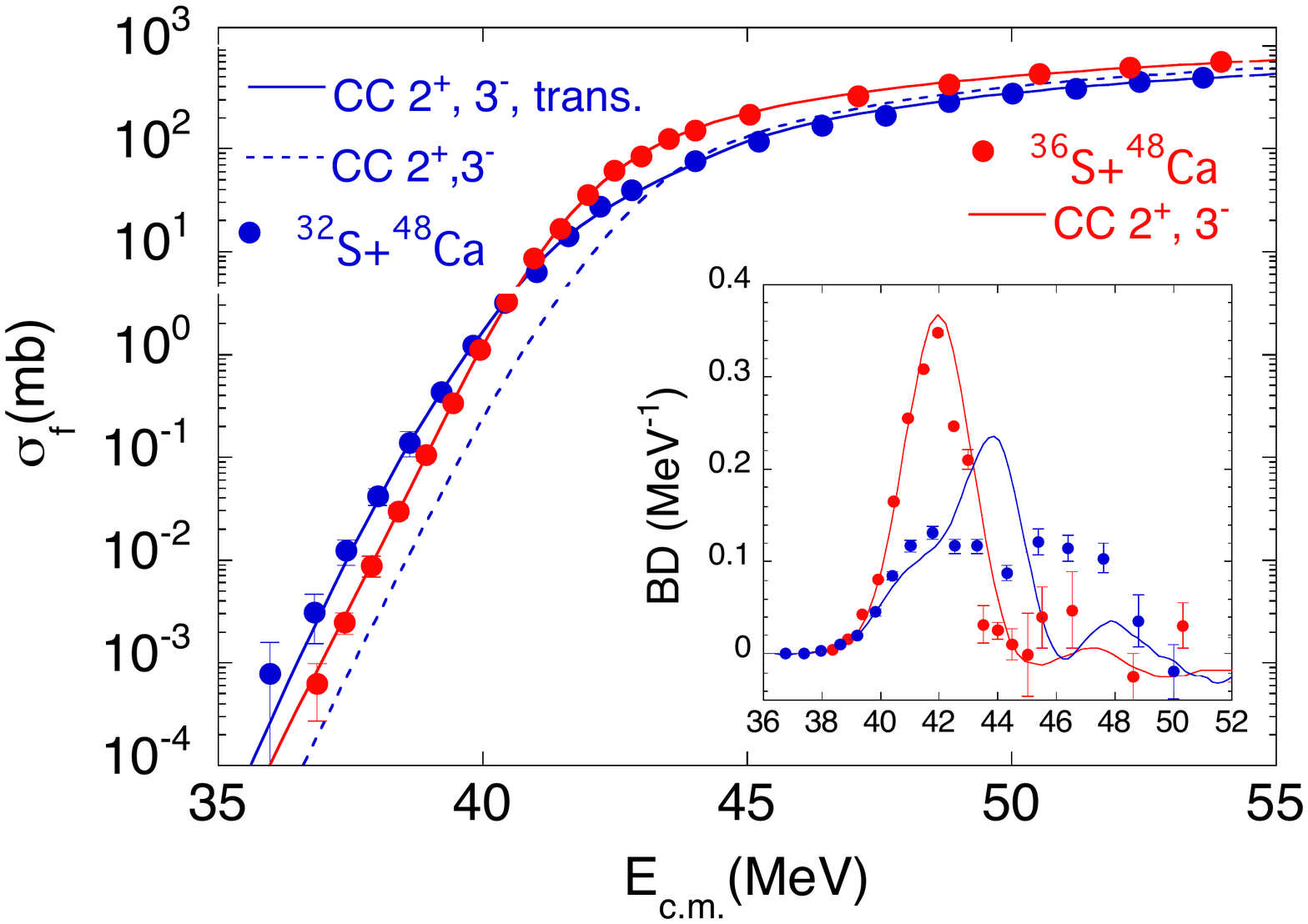}}
\caption{Fusion excitation functions of $^{32,36}$S+$^{48}$Ca~\cite{3248}
compared to each other and to the CC calculations described in the text. The inset shows the two barrier distributions compared to the same calculations.}
\label{323648}     
\end{figure}

Systematic trends that  have been attributed to transfer couplings have been observed in various other cases.
The low-energy cross sections of $^{40}$Ca + $^{48}$Ca~\cite{4040,4048} exceed the $^{48}$Ca+$^{48}$Ca data and they are suppressed compared to the $^{40}$Ca+$^{40}$Ca data at high energy. This is illustrated in Fig.~\ref{Calci} and is a characteristic feature originating from strong $Q>0$ transfer couplings. In fact, it is possible to account for the fusion cross sections of $^{40}$Ca+$^{48}$Ca 
by including couplings to one- and two-nucleon transfer channels with positive $Q$-values and by adjusting the strength
and the effective $Q$-value of the 2n-transfer.

The same enhancement/suppression effect is observed for the pair $^{32,36}$S+$^{48}$Ca~\cite{3248} (Fig.~\ref{323648}).
The effect is due to the broad barrier distributions produced by those couplings (see insert of Fig.~\ref{323648}).
Indeed, the measured barrier distribution for $^{36}$S+$^{48}$Ca  consists of a strong narrow
peak and a much weaker structure at higher energies.
On the contrary, the barrier distribution extracted from the $^{32}$S+$^{48}$Ca data is very wide and has two main peaks on either side of the unperturbed barrier. This peculiar shape is not well reproduced by calculations, even if the width of the distribution can only be reproduced by including transfer couplings in the calculations. We feel that a more realistic and detailed treatment of transfer couplings (two-nucleon transfer, in particular), might reduce the disagreement.

Further recent examples of strong effects due to transfer couplings have been  deduced  from the comparison of the systems $^{32}$S+$^{90,96}$Zr~\cite{Huanqiao} and $^{32}$S+$^{94}$Zr~\cite{Jia14}, that were investigated in Beijing using the  electrostatic beam deflector of the China Institute of Atomic Energy. 
Moreover,  an overall analysis of several systems where positive Q-value transfer channels  are available, was recently performed by the same group~\cite{Jia}.
They have provided a confirmation for the  transfer coupling effects on sub-barrier heavy-ion fusion in a number of cases.

The data on $^{32}$S+$^{90,96}$Zr were analysed using the 
``empirical coupled channels model'' (ECC)~\cite{ZagModel} where  the barrier distribution
arising from surface excitations is obtained by diagonalization and not by explicitly solving the coupled equations.
The ECC model with neutron rearrangement has been successfully used for reproducing and predicting cross sections of various other cases where transfer coupling plays an important role.
That model has been recently implemented  with 
a more realistic CC calculation of the fusion probability~\cite{Karpov}. 
This further step (Quantum Coupled Channels + Empirical Neutron Rearrangement  model) allowed them to obtain 
a good overall agreement with the shape of the barrier
distribution, as well as with the sub-barrier excitation functions,  for several S, Ca + Zr and Ni + Mo systems.
An interesting observation  is that  the parameters used to fit  the transfer probabilities (where existing) were different from those needed to correctly describe fusion cross sections.

\begin{figure}[h]
\centering
\resizebox{0.40\textwidth}{!}{\includegraphics{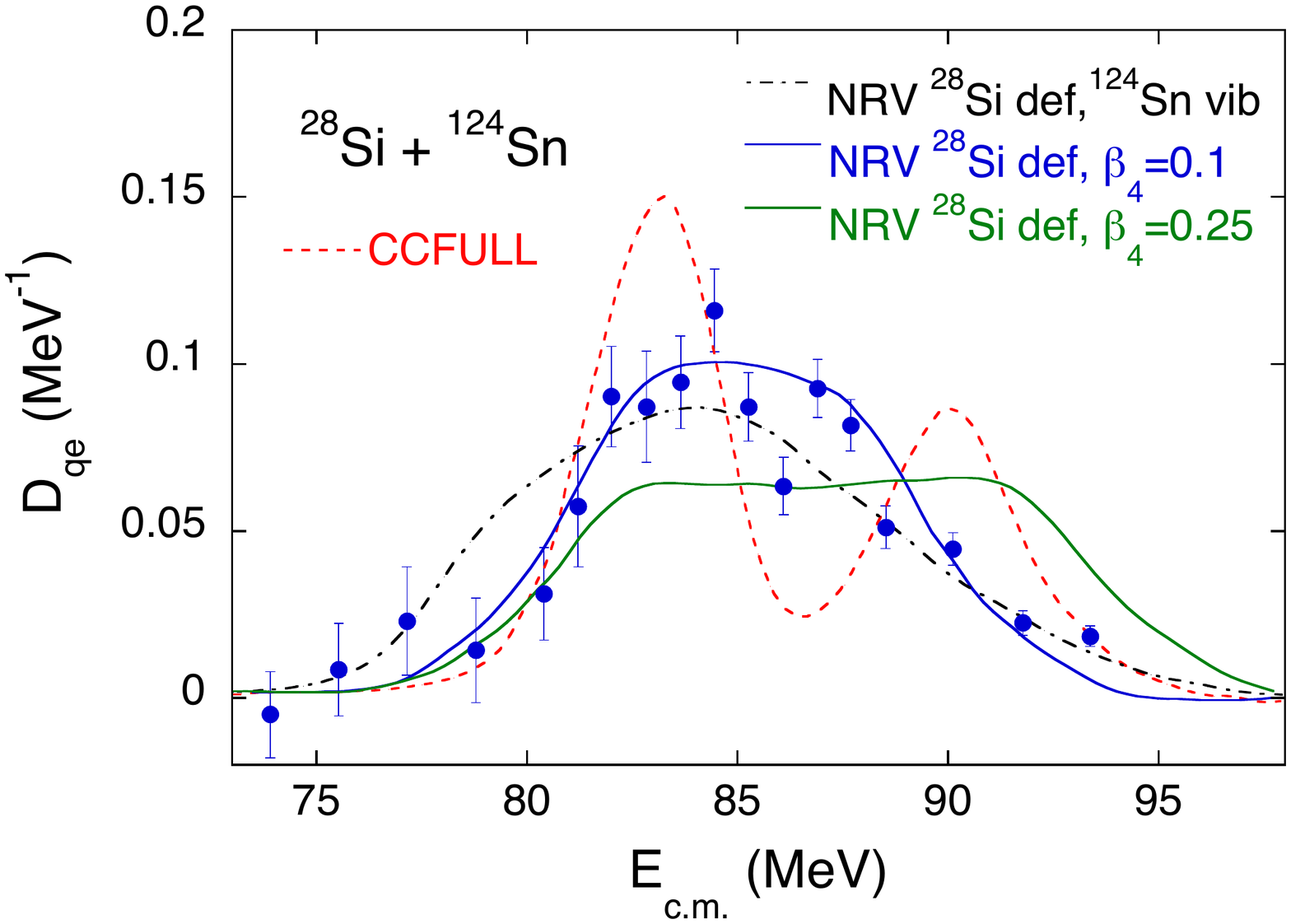}}
\caption{Barrier distribution extracted from backangle quasi-elastic excitation functions for $^{28}$Si+$^{124}$Sn~\cite{Danu}, compared to the results of CCFULL and NRV~\cite{Karpov} calculations. See text for more details, figure redrawn from Ref.~\cite{Danu}.}
\label{Danudanu}     
\end{figure}

That model (also called the NRV model because it is available on the Nuclear Reaction Video webpage~\cite{NRV}) was used in the analysis of the data obtained by Danu et al.~\cite{Danu} by measuring the large-angle quasi elastic scattering excitation functions of $^{28,30}$Si+$^{124}$Sn, and extracting the barrier distributions $D_{qe}$. In particular, the case with the $^{28}$Si projectile is attractive because several multi-nucleon transfer channels with Q$>$0 are available. The barrier distribution is shown in Fig.~\ref{Danudanu}. The CCFULL calculation (red line) includes the rotational excitation of $^{28}$Si and the vibrational modes of $^{124}$Sn, as well as a Q$>$0 transfer coupling, but shows two separate peaks not observed in the experiment.
\par
\begin{figure}[h]
\centering
\resizebox{0.45\textwidth}{!}{\includegraphics{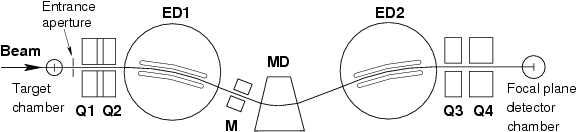}}
\caption{The HIRA spectrometer~\cite{HIRA} is based on  a symmetric electrostatic dipole-magnetic dipole-electrostatic dipole (ED-MD-ED) configuration with two quadrupole doublets placed before and after the two  electrostatic dipoles.}
\label{HIRA}     
\end{figure}

The blue line is the result of the NRV calculation with the same rotational coupling in $^{28}$Si ($\beta_2$ = -0.408 and $\beta_4$ = 0.1), and several multi-nucleon transfer channels, but no coupling for $^{124}$Sn. This gives a good data fit, however adding the vibrational modes of $^{124}$Sn significantly deteriorates the agreement (black dash-dotted line). The situation is not completely clear, and a complementary measurement of the fusion excitation function would greatly help in a case like this. Deriving the real barrier distribution $D_{f}$ would allow to exclude that the disagreement with CCFULL is due to the loss of resolution one often observes when considering $D_{qe}$ (see Sect.~\ref{BDqe}).
\par
\begin{figure}[h]
\centering
\resizebox{0.45\textwidth}{!}{\includegraphics{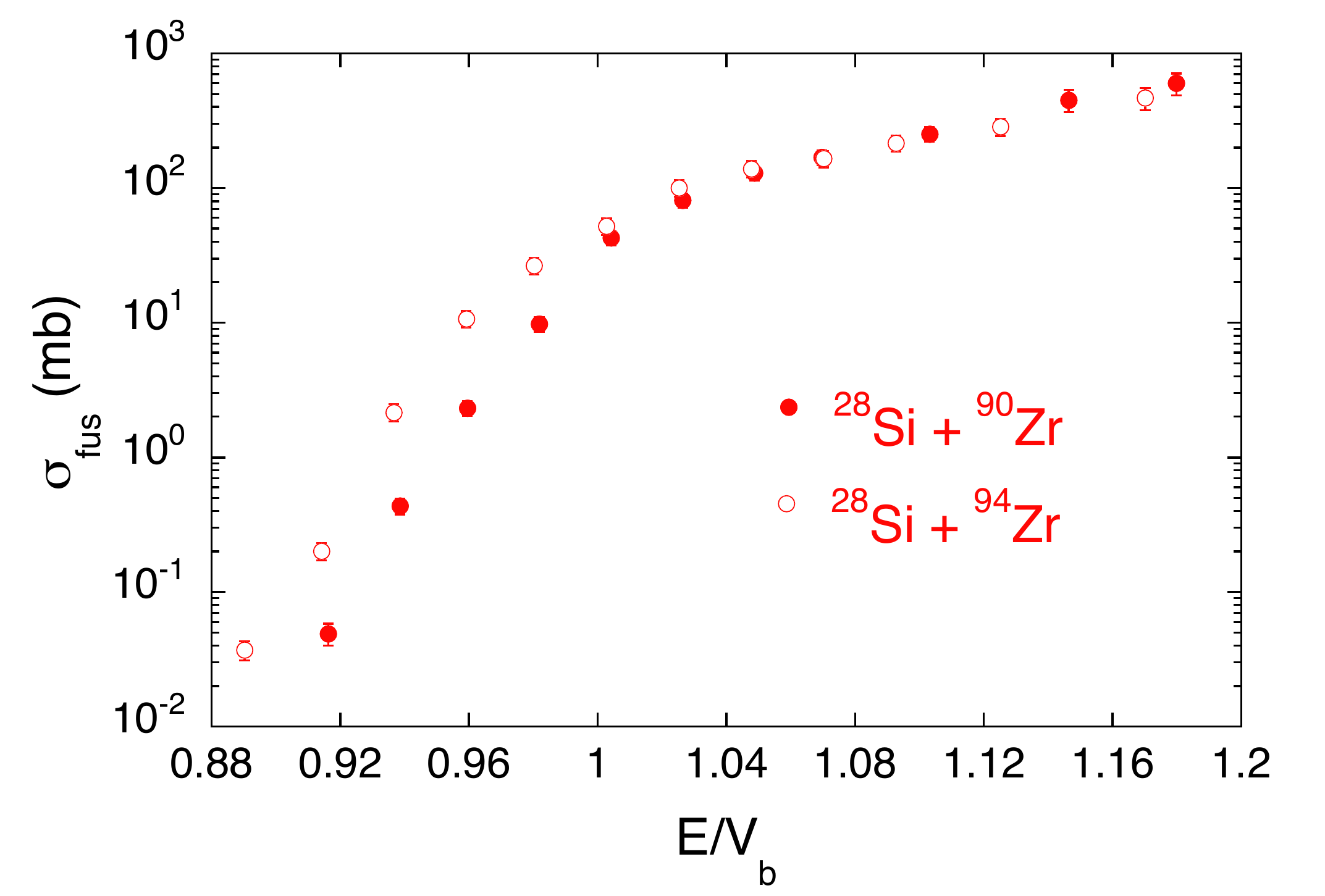}}
\caption{Comparative  cross section plot for  $^{28}$Si+$^{90,94}$Zr~\cite{Kalkal} 
in a reduced energy scale using the barrier produced by the AW potential.}
\label{kalkal}     
\end{figure}

Several studies concerning the influence of transfer on fusion were performed in New Delhi using the heavy ion reaction analyser
HIRA (see Fig.~\ref{HIRA}) that is a  recoil mass spectrometer (RMS) with a high rejection factor for the primary beam.
It can operate in the  beam direction thanks to  a high  rejection factor (of the order of 10$^{13}$), and 
the  reaction products are identified at  the focal plane with very good mass resolution ($\simeq$1/300). HIRA has a variable acceptance 1-10 msr, energy and mass acceptances $\pm$20$\%$ and $\pm$5$\%$, respectively. It is equipped with a sliding seal scattering chamber allowing a rotation up to 25$^\circ$. Alternatively, a $\gamma$-ray array can be installed around the target by using a small Al scattering chamber.

Kalkal and coworkers have evidenced the large difference in sub-barrier cross sections existing in the two systems  $^{28}$Si+$^{90,94}$Zr~\cite{Kalkal} where  transfer channels with  Q$>$0 exist only in $^{28}$Si+$^{94}$Zr. This is shown in Fig.~\ref{kalkal} and the relative enhancement of one system with respect to the other is attributed to the influence of one and multinucleon transfer even if the CC calculations reported in the original article  are not able to treat correctly such couplings. 

On the contrary, no clear isotopic effects were observed in the systems $^{48}$Ti+$^{58,60,64}$Ni~\cite{Vin} in the measured energy range.
Further investigations might evidence a  possible role of transfer couplings at lower energies. In that experimental work the mean angular momentum of the compound nuclei was also measured using the 4N to 3N evaporation residue ratio, following the method previously used in Ref.~\cite{Dasgupta_l}

 For $^{40}$Ca + $^{58,64}$Ni~\cite{Bourgin,Bourgin0}, 
 the authors used the CCFULL code and  
 the bare nucleus-nucleus potential was computed with the
frozen Hartree-Fock method and coupling parameters were taken
from known nuclear structure data. 
These calculations give a centroid of the barrier distribution that is in agreement with the TDHF fusion threshold
for $^{40}$Ca+$^{58}$Ni. In  this case the octupole and quadrupole excitations are dominant and 
neutron transfer is weak. 
For $^{40}$Ca+$^{64}$Ni the TDHF barrier  is lower than predicted by the analogous  CC results. This  could be due
to  the large  neutron transfer probabilities  evidenced in the  TDHF calculations. 

\subsection{Heavier systems}
\label{Heavier}

\begin{figure}[h]
\centering
\resizebox{0.44\textwidth}{!}{\includegraphics{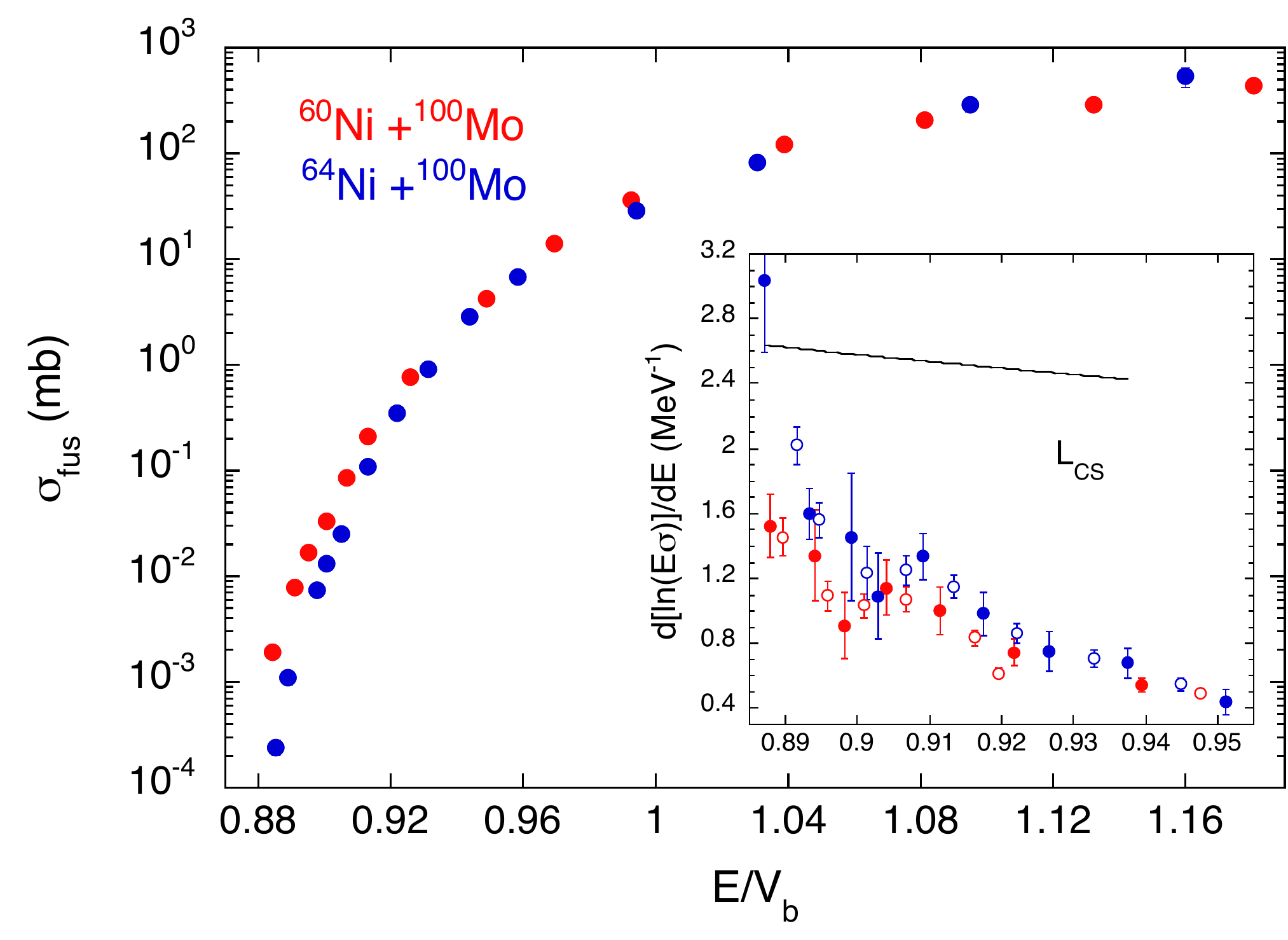}}
\resizebox{0.44\textwidth}{!}{\includegraphics{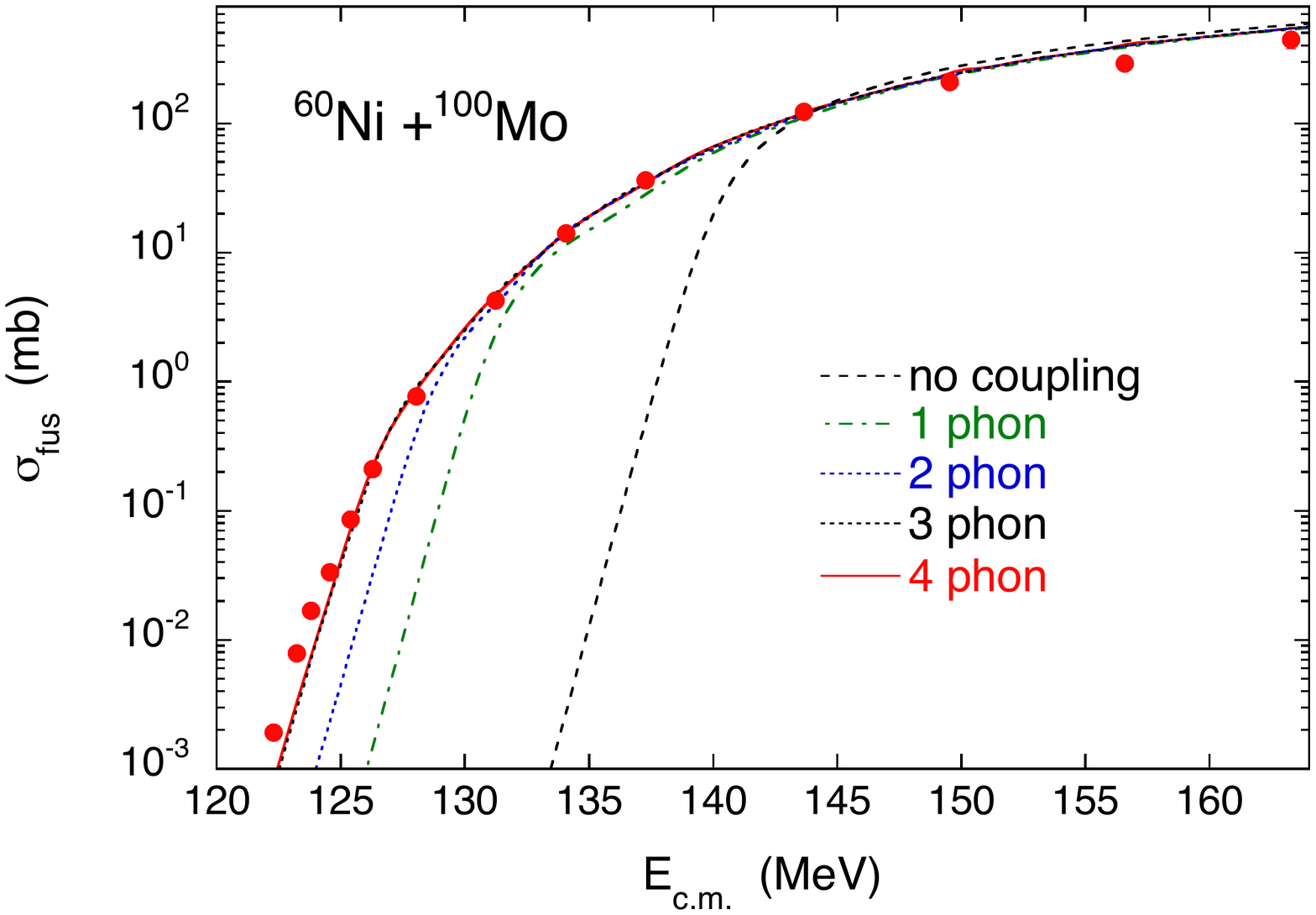}}
\caption{(upper panel) Fusion excitation functions of $^{60,64}$Ni + $^{100}$Mo~\cite{60100,64100} in an energy scale relative to the Coulomb barrier, as obtained from the AW potential~\cite{Akyuz}. The inset shows the logarithmic slopes d[ln(E$\sigma_f$)]/dE, derived from two (solid dots) or three successive points (empty dots) of the excitation function. The plotted statistical uncertainties for the cross sections are smaller than the symbol size in most cases. (lower panel) The fusion excitation function of $^{60}$Ni + $^{100}$Mo compared to several CC calculations (see text).
}
\label{60-64+100}     
\end{figure}

When moving to  heavier  and soft systems, sub-barrier fusion is  strongly affected by multi-phonon excitations, and the effect of coupling to transfer channels with Q$>$0 becomes relatively less important so that it can only be observed at very low energy. This is indeed what we can deduce from the comparison of the fusion excitation functions for  $^{60,64}$Ni + $^{100}$Mo~\cite{60100,64100} (see Fig.~\ref{60-64+100}, upper panel). The two excitation functions are very similar to each other and only differ at very low energies. This suggests that Q$>$0  transfer couplings (only existing in $^{60}$Ni + $^{100}$Mo) play a marginal role.

No evidence for hindrance is found for $^{60}$Ni + $^{100}$Mo. This is also indicated by the trend of the logarithmic slopes reported in the inset of that figure, showing that the $L_{CS}$ value is only overcome by $^{64}$Ni + $^{100}$Mo. Hindrance might appear for the lighter system only below the measured energy range, but this is obviously a simple conjecture. Please see, in this respect, the discussion in Sect.~\ref{Hindrance}.

CCFULL calculations are reported in the lower panel of Fig.~\ref{60-64+100} for $^{60}$Ni + $^{100}$Mo. Multi-phonon excitations are predominant in the cross section enhancement near and below the barrier. In particular, the strong quadrupole mode of $^{100}$Mo gives an increasing contribution up to the level of the third phonon, accounting for most of the observed enhancement. The additional contribution of the fourth phonon is negligible as can be seen in the figure. It appears that the calculation has converged 

Advances in the
theoretical approach within the CC model would be very welcome for such heavy systems, because couplings should be taken into account to all orders, avoiding at the same time the harmonic approximation of the vibrational modes. However, it is unfortunate that multi-phonon vibrations lack experimental information in most cases.

Along the same line,  
measurements of fusion excitation functions of the two systems
$^{58,64}$Ni + $^{124}$Sn have been recently extended~\cite{Jiang_15} to cross sections in the range 1-10$\mu$b.
Detailed CC calculations  reasonably
agree with the new measurements, indicating that the largest enhancement of the cross sections is due to the coupling
of inelastic excitations, as expected for these heavy systems. 
\par
The case of $^{58}$Ni + $^{124}$Sn shows a larger influence from the coupling to transfer reactions, as qualitatively expected since transfer channels with large and positive Q-values are available. For  $^{64}$Ni + $^{124}$Sn the contribution from transfer
is weaker, due to the smaller Q-values, but not negligible.
For both systems the influence of transfer shows up mainly in the cross
section region below $\simeq$ 1 mb. 
\par

\begin{figure}[h]
\centering
\resizebox{0.45\textwidth}{!}{\includegraphics{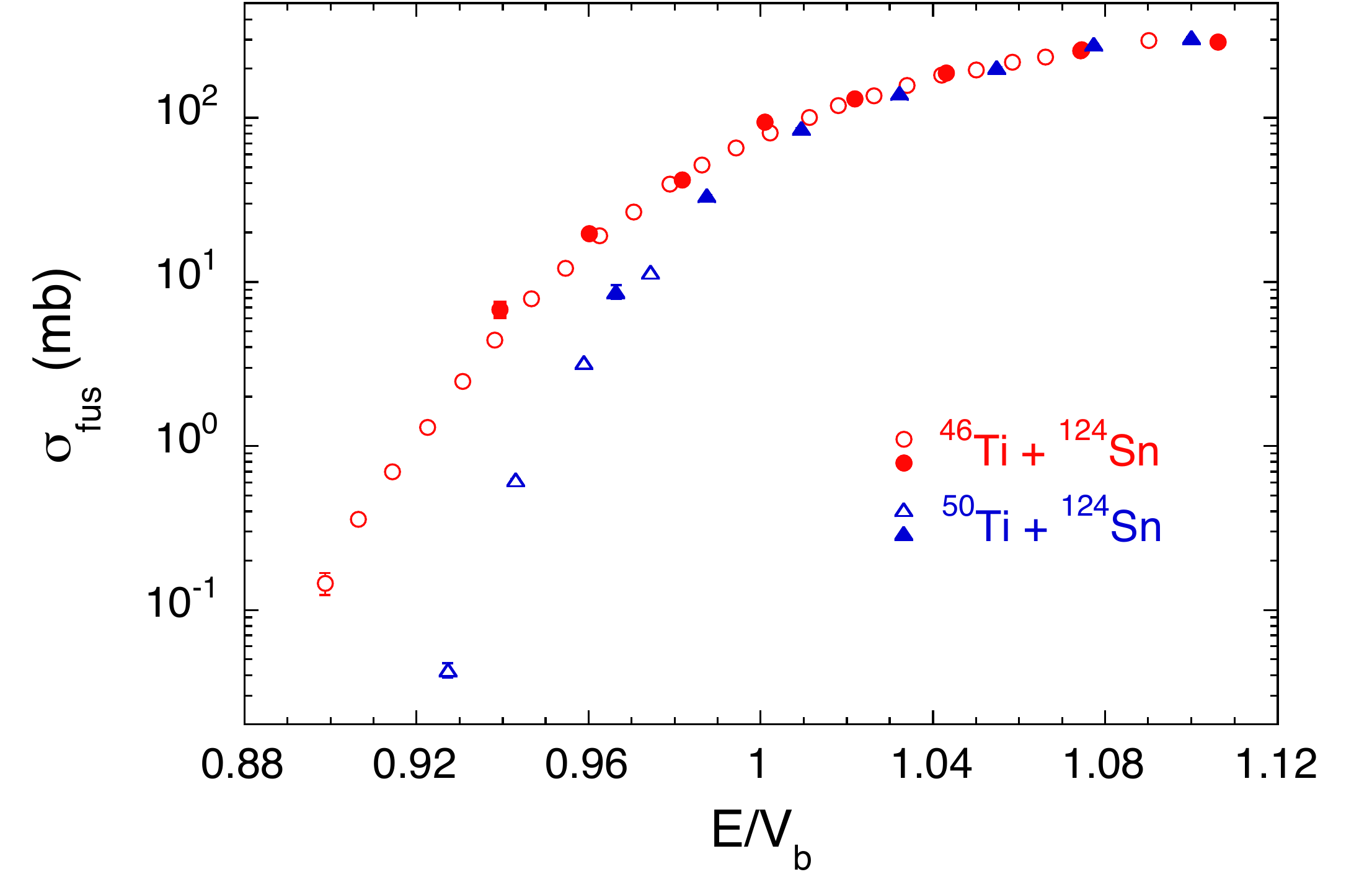}}
\caption{Comparison of  fusion excitation functions for $^{46,50}$Ti + $^{124}$Sn~\cite{Liang16}. The energy scale is obtained 
dividing the energy by the barrier height produced  by the AW potential. The full symbols are results of the measurements performed at HRIBF in inverse kinematics while the open ones come from the measurements of ANU in direct kinematics. }
\label{4650+124}     
\end{figure}

This may explain why in the experiments
of the Oak Ridge group~\cite{Liang03,Koh11,Liang07} a negligible contribution from transfer was identified in the investigation of $^{58,64}$Ni + $^{132}$Sn using the radioactive beam $^{132}$Sn (see Sect.~\ref{exoticheavy}), despite the very large and positive Q-values existing in $^{58}$Ni + $^{132}$Sn. As a matter of fact, the lowest cross sections measured in Refs.~\cite{Liang03,Koh11} were limited to $\geq$1 mb, due to the relatively low intensity of the exotic beam.

\par
We have already seen that for lighter systems involving e.g. $^{40}$Ca, transfer couplings become significant already at much higher cross sections (see e.g. Fig.~\ref{40_48+96}). By the way, this is also the result of the study of  $^{40}$Ca + $^{124}$Sn~\cite{40124}, where a very large enhancement was found. 

In order to clear up the source of such specific differences,
the near-barrier excitation functions were measured for the two systems $^{46,50}$Ti + $^{124}$Sn~\cite{Liang16}. The experiment using the 
$^{46}$Ti beam was performed with high statistical accuracy and small energy steps, so to allow extracting the barrier distribution. The results for the two systems are compared in Fig.~\ref{4650+124}, showing a large relative enhancement that has been attributed to the effect of coupling to transfer channels with $Q>0$. The barrier distribution of $^{46}$Ti + $^{124}$Sn shows a peak below the uncoupled barrier, in analogy with the case of $^{40}$Ca + $^{124}$Sn where, however, that peak is shifted further below the uncoupled barrier. This has been attributed to the very strong octupole mode of $^{40}$Ca, that produces, together with the transfer couplings, the huge fusion enhancement of  $^{40}$Ca + $^{124}$Sn.

The experiments on $^{46,50}$Ti + $^{124}$Sn were performed following a series of measurements using the heavy radioactive beams of Holifield Radioactive
Ion Beam Facility (HRIBF) at Oak Ridge National Laboratory. These measurements will be described in some detail in Sect.~\ref{radioactive}.

\begin{figure}[h]
\centering
\resizebox{0.45\textwidth}{!}{\includegraphics{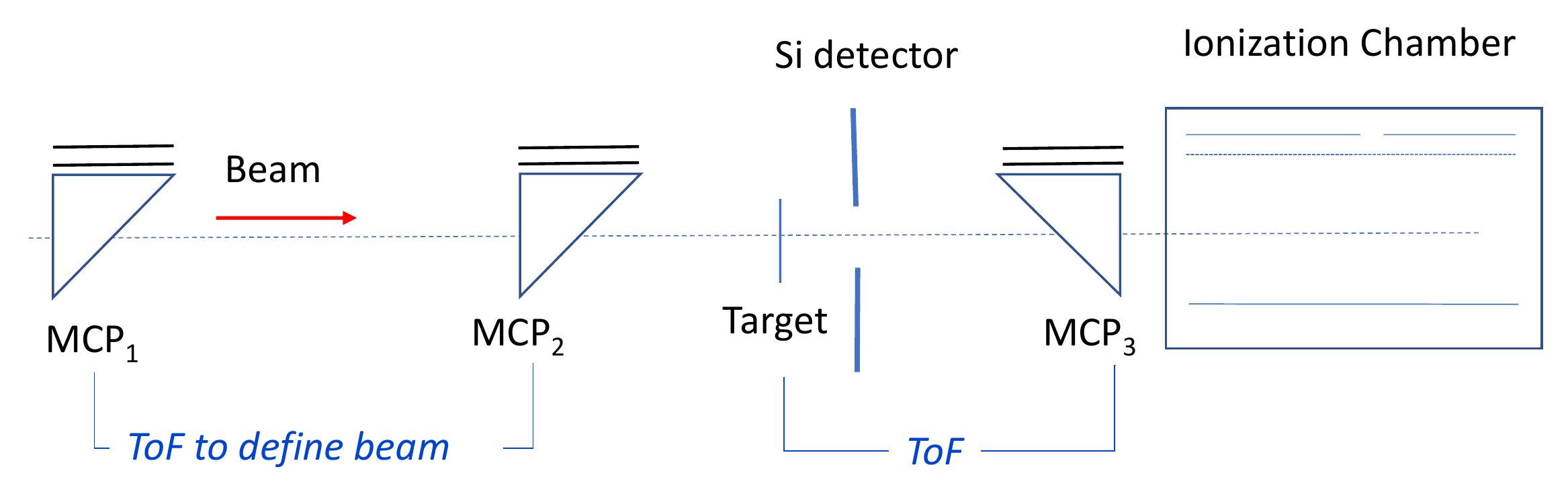}}
\caption{Set-up developed at ORNL for measurements of ER cross
sections with  low-intensity beams in inverse kinematics~\cite{Liang07}. The detector set-up is placed in the same direction as the incoming beam. The two first micro-channel plate detectors (MCP$_1$ and MCP$_2$) are used to define the beam condition and direction. The ToF measured using MCP$_3$ can discriminate the ER from the transmitted beam. The final ionization chamber gives the $\Delta E-E$ signals and therefore the Z identification. Figure redrawn and modified from Ref.~\cite{Liang07}.}
\label{ORNL}     
\end{figure}

The first part of the measurements on $^{46,50}$Ti + $^{124}$Sn were performed at the HRIBF, using a $^{124}$Sn beam on titanium targets in inverse kinematics. The ER were identified in a TOF-$\Delta$E-E  telescope using microchannel-plate detectors and a
multi-anode ionization chamber~\cite{Shapira} (see Fig.~\ref{ORNL}).

\begin{figure}[h]
\centering
\resizebox{0.45\textwidth}{!}{\includegraphics{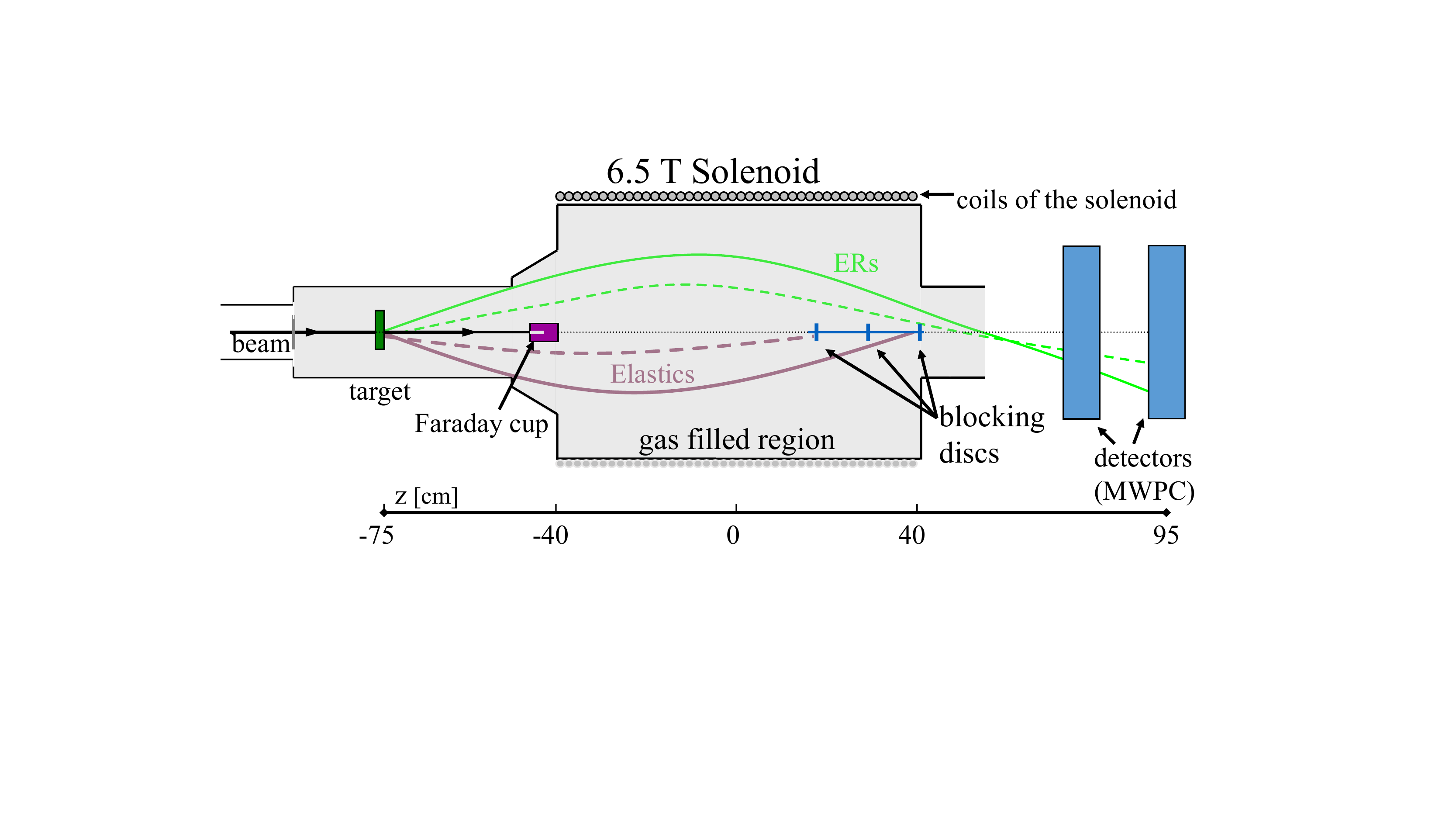}}
\caption{Layout of SOLITAIRE, based on a gas-filled 6.5T superconducting solenoid~\cite{Soli}.}
\label{SOLI}     
\end{figure}

The second part of the experiment followed at the Australian
National University (ANU) in direct kinematics (beams of $^{46,50}$Ti) using the superconducting solenoid SOLITAIRE.
This solenoidal fusion product separator is schematically shown in Fig.~\ref{SOLI}. It has a very  high efficiency for the detection of the evaporation residues, since it covers a large fraction of their angular distribution (from 0.45$^\circ$ to 9.5$^\circ$).

We want to point out that 
a systematic survey of the excitation functions of several systems has been recently performed~\cite{Jiang_transfer} with the purpose of obtaining a common signature of transfer effects. By analyzing the data with the Wong formula~\cite{Wong} they observed  that 
 the slopes of the fusion excitation functions and, consequently, the sub-barrier enhancement of fusion cross sections, are actually correlated, for both light {\it and}  heavy systems, with the strength of the total 
 neutron transfer cross sections. This suggests that transfer couplings should in any case be considered in CC calculations of near- and sub-barrier fusion reactions.

\section{Hindrance far below the barrier}
\label{Hindrance}

\subsection{Early evidence and significant cases}
\label{subhindrance}
Below the lowest barrier produced by channel couplings, one would expect that the energy-weighted excitation functions  $E\sigma$ display a simple exponential falloff with decreasing energy~\cite{Wong}. This is not always true, as shown for the first time for the system $^{60}$Ni + $^{89}$Y by Jiang et al.~\cite{Jiang0}. It was found that, at deep sub-barrier energies, the cross section decreases very rapidly, so that the excitation function is much steeper than expected. This was named a hindrance effect.

\begin{figure}[h]
\centering
\resizebox{0.45\textwidth}{!}{\includegraphics{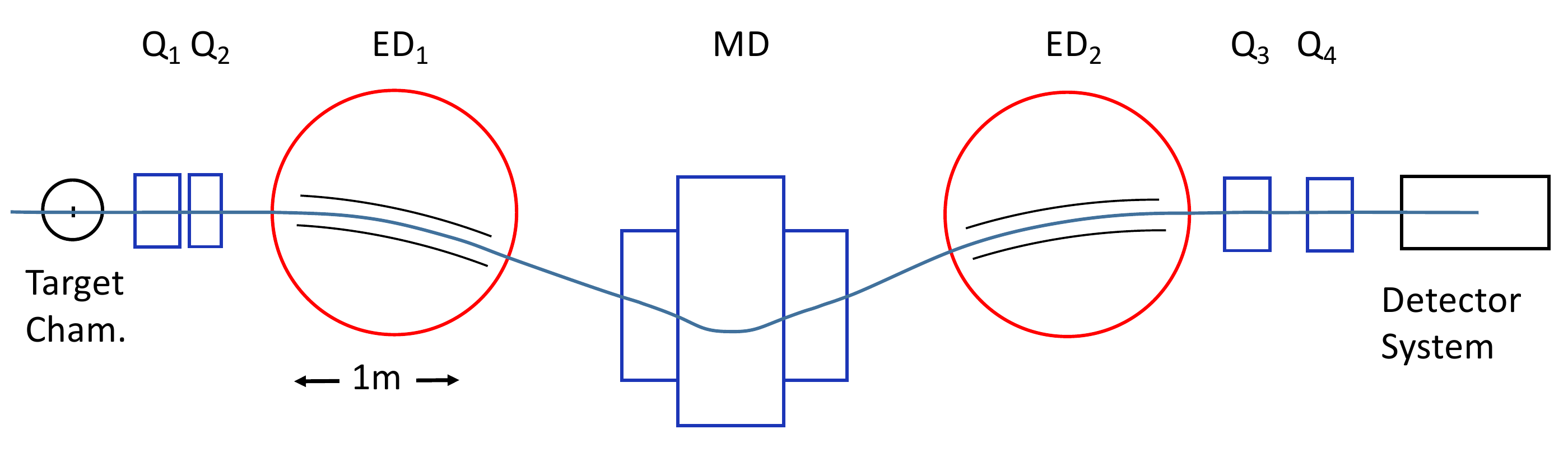}}
\caption{Schematic layout (top view) of the FMA at ATLAS. ED$_1$ and ED$_2$ are electric dipoles, and MD is a
magnetic dipole. Q$_1$,Q$_2$,Q$_3$ and Q$_4$ are magnetic  quadrupoles. The target chamber  and the detector system  are also indicated. Figure redrawn and modified from Ref.~\cite{Jiang05}.}
\label{FMA}     
\end{figure}

The cross section was measured down to 1.6$\mu$b (an upper limit of 95 nb was established for the very lowest measured energy) using the Fragment Mass Analyzer (FMA)~\cite{Davids} installed at the Argonne superconducting linear accelerator ATLAS. A scheme of the spectrometer is shown in Fig.~\ref{FMA}. Its large momentum
and angular acceptances (10$\%$ and  $\theta_{lab}\leq$2.3$^\circ$, respectively) result in a high detection efficiency
for the evaporation residues (50-70$\%$ for each charge state), while achieving an excellent beam suppression
factor of about 4$\times$10$^{17}$~\cite{Jiang05}.
\par
\begin{figure}[h]
\centering
\resizebox{0.40\textwidth}{!}{\includegraphics{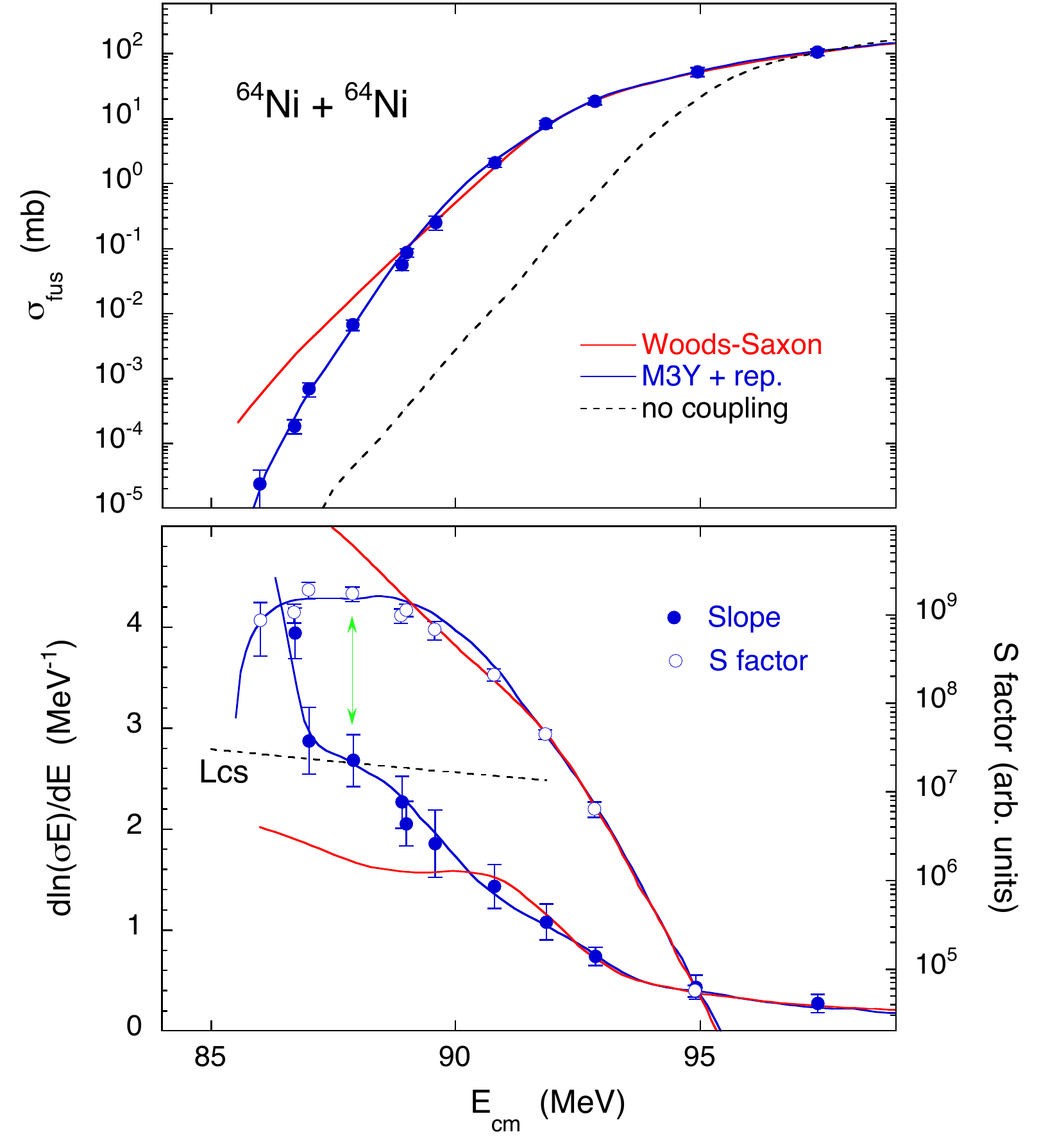}}
\caption{Fusion excitation function, logarithmic slope and astrophysical $S$ factor of $^{64}$Ni + $^{64}$Ni~\cite{Jiang04}.}
\label{6464}     
\end{figure}

Further measurements far below the barrier for other systems evidenced that the slope of the excitation function keeps increasing with decreasing energy. A remarkable case in this sense is $^{64}$Ni + $^{64}$Ni~\cite{Jiang04} whose behavior is shown in Fig.~\ref{6464}.
The fusion excitation function was measured down to $\simeq$20 nb (upper panel). One notices that the cross sections far below the barrier
are much lower than expected on the basis of standard CC calculations employing the Woods-Saxon potential (red line). 

A good data fit in that energy region is obtained by using the M3Y+ repulsion potential  having a  shallow pocket in the entrance channel (blue line)~\cite{Misi}.
This model will be described in some detail in Section~\ref{models}. As a reference, the figure also shows the result of the ``no coupling" calculation (dashed line), that is, the excitation function obtained by pure tunnelling through a one-dimensional barrier.

One has strong evidence that the low-energy hindrance effect is a general phenomenon for heavy-ion fusion, however it shows up with varying intensities and distinct features in different systems. In order to emphasise the appearance and the characteristics of hindrance it is often very useful to represent the data by means of  two further quantities, i.e. the logarithmic slope $L(E)$ 
of the excitation function and the astrophysical $S$ factor $S(E)$~\cite{BBFH}. 
They are defined as 
\begin{equation}
L(E)=d[ln(E\sigma)]/dE=\frac{1} {E\sigma} \frac{d(E\sigma)} {dE}
\label{slope}
\end{equation}

\begin{equation}
S(E)=E\sigma(E)exp(2\pi\eta)
\label{Esse}
\end{equation}

These two quantities extracted from the measured cross sections for the system $^{64}$Ni + $^{64}$Ni are plotted in the lower panel of Fig.~\ref{6464}.
One sees that the slope $L(E)$ is continuously increasing below the barrier (which is around 94 MeV) and that the $S$ factor develops a maximum 
vs. energy close to the point where the slope reaches the value 
$$L(E)=\frac{\pi\eta} {E}$$ which is conventionally called $L_{CS}$ ($\eta$ is the Sommerfeld parameter $Z_{1}Z_{2}e^2/\hbar v$  and $v$ is
the beam velocity). Indeed the two quantities $L(E)$ and $S(E)$ are algebraically related~\cite{Jiang04_2},
since it is easy to show that the energy derivative of $S(E)$ is 

\begin{equation}
\frac{dS} {dE}=S(E)\Big[L(E)-\frac{\pi\eta} {E}\Big]
\label{S_deriv}
\end{equation}
and therefore vanishes when the slope equals $\pi\eta/E$.

Plotting the $S$ factor is a useful and straightforward way of representing the trend of the excitation function in the energy region below the barrier. Indeed, S is directly extracted from the cross sections, whereas the logarithmic slope $L(E)$ and the barrier distribution $B(E)$ are $derivatives$ of the excitation function, and are therefore subject to larger experimental errors.

\begin{figure}[h]
\centering
\resizebox{0.35\textwidth}{!}{\includegraphics{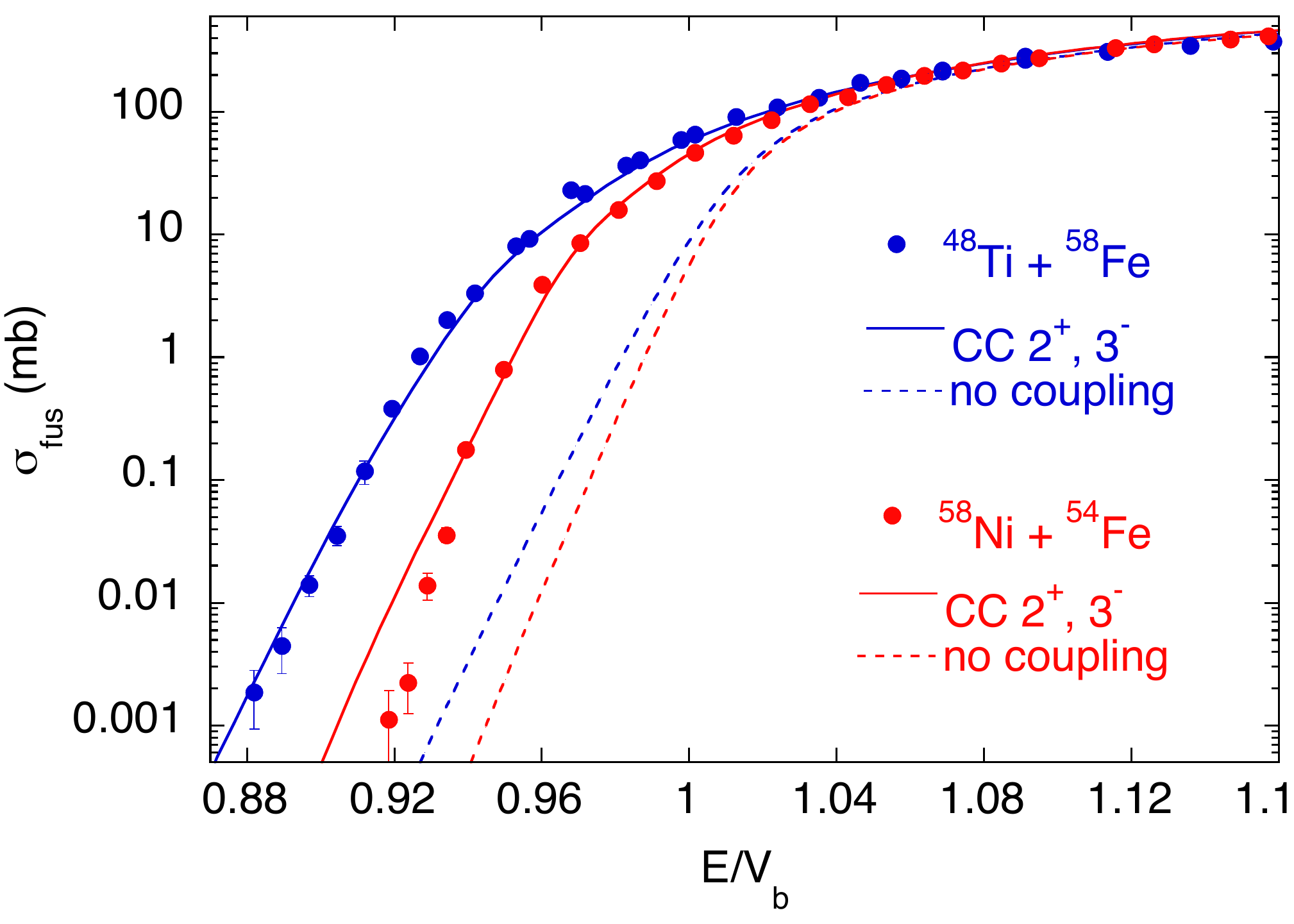}}
\resizebox{0.36\textwidth}{!}{\includegraphics{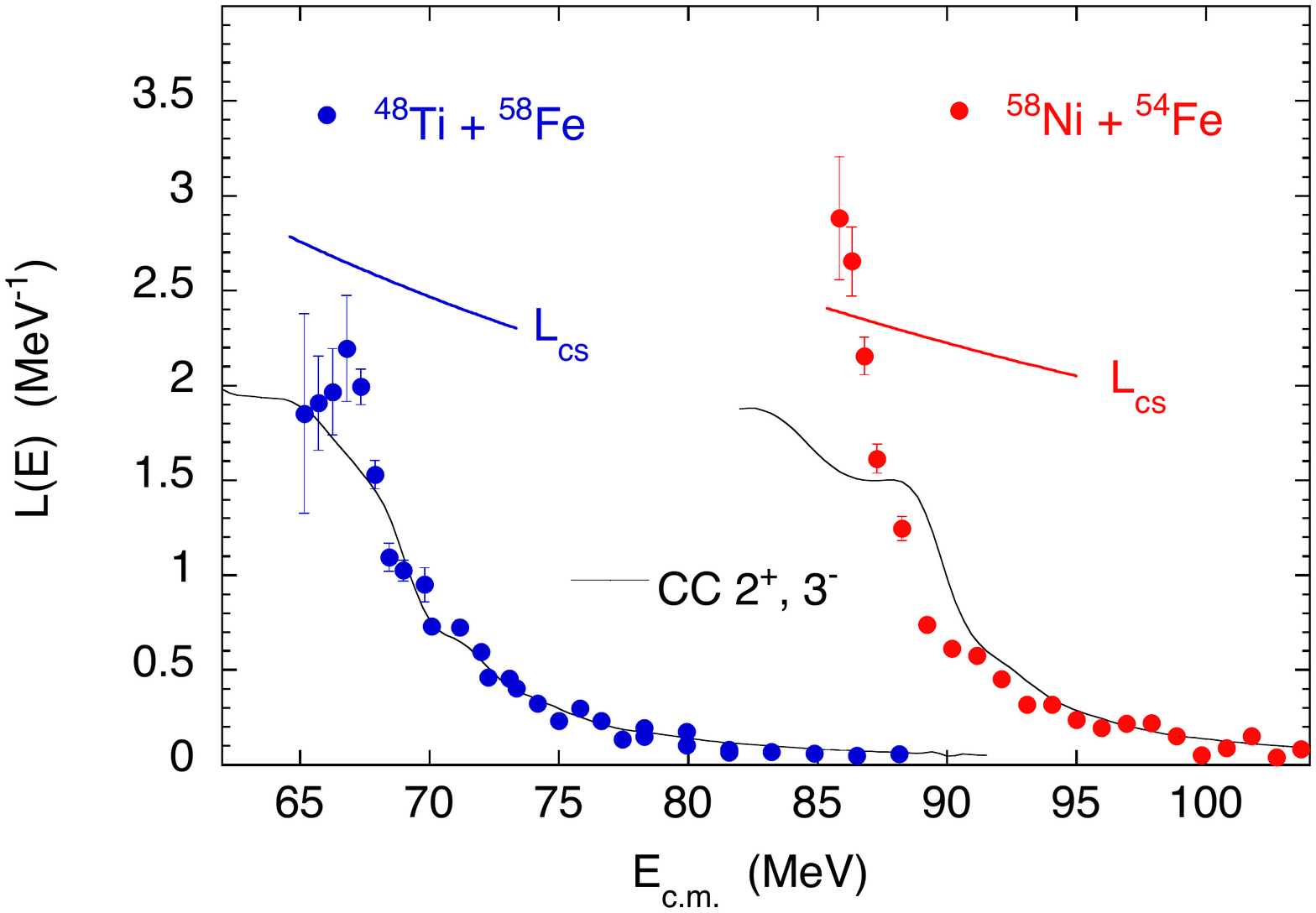}}
\caption{Fusion excitation function and logarithmic slope of $^{48}$Ti + $^{58}$Fe and $^{58}$Ni +  $^{54}$Fe~\cite{NiFedeep,48+58}.}
\label{48+58-58+54}     
\end{figure}

The green arrow in Fig.~\ref{6464} for $^{64}$Ni + $^{64}$Ni indicates the link between $L_{CS}$ and the maximum of the $S$ factor. The energy where $L(E)$=$L_{CS}$ (and the $S$ factor has a maximum) has often been 
phenomenologically taken as the threshold for the hindrance effect. However, we will see that hindrance may show up even in the absence of a maximum of $S(E)$.

\begin{figure*}[ht!]
\centering
\resizebox{0.70\textwidth}{!}{\includegraphics{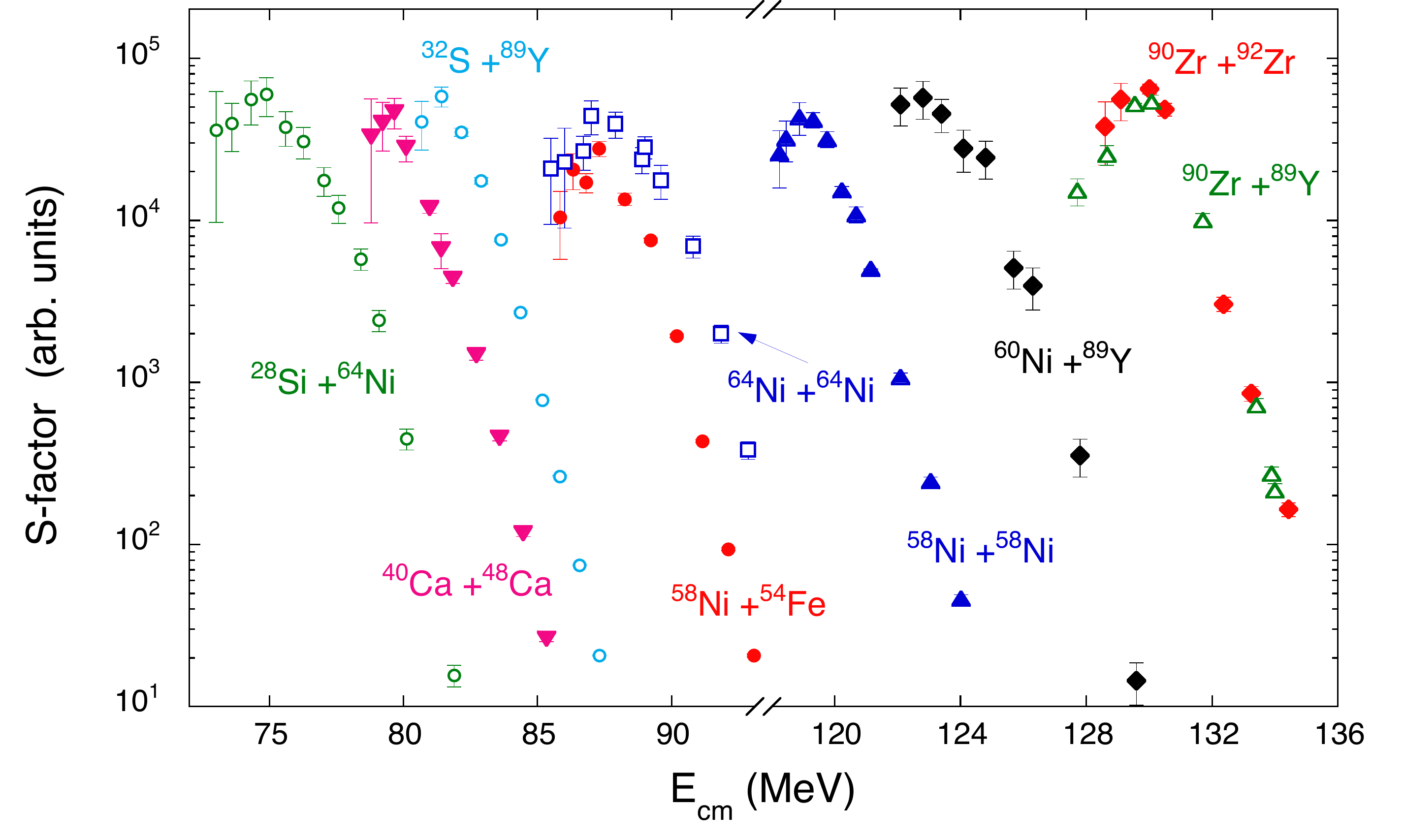}}
\caption{Astrophysical $S$ factor of several systems where a maximum shows up. They were measured at ANL ($^{28}$Si + $^{64}$Ni~\cite{Jiang_2864},   $^{64}$Ni + $^{64}$Ni~\cite{Jiang04} and $^{60}$Ni + $^{89}$Y~\cite{Jiang0}), LNL ($^{40}$Ca + $^{48}$Ca~\cite{4048} and $^{58}$Ni + $^{54}$Fe~\cite{NiFedeep}), ANU-Canberra ($^{32}$S + $^{89}$Y~\cite{3289}), MIT ($^{58}$Ni + $^{58}$Ni~\cite{BeckNiNi}) and GSI ($^{90}$Zr + $^{92}$Zr,$^{89}$Y~\cite{Keller}).  }
\label{Mag8}     
\end{figure*}

We show in Fig.~\ref{48+58-58+54} another clear example of the hindrance phenomenon, observed for  $^{58}$Ni + $^{54}$Fe~\cite{NiFedeep}. The situation is contrasted with the behaviour of $^{48}$Ti + $^{58}$Fe, whose fusion excitation function was the object of a more recent experiment~\cite{48+58}.

In the upper panel,  the two excitation functions are plotted vs. the energy with respect to the  AW Coulomb barrier~\cite{Akyuz}. This allows us to notice immediately the very steep slope of $^{58}$Ni + $^{54}$Fe fusion with respect to the other system in the sub-barrier region, and with respect to the results of standard CC calculations including the low-lying 2$^+$ and 3$^-$ excitations. This is better evidenced in the lower panel of the figure showing the logarithmic slopes $L(E)$ of the two systems. The slope of $^{58}$Ni + $^{54}$Fe keeps increasing, reaches and overcomes the value $L_{CS}$~\cite{Jiang04}. Consequently, a clear maximum of S develops with decreasing energy.

We also notice that the behaviour of $^{48}$Ti + $^{58}$Fe is remarkably different; indeed, its slope saturates below the barrier and remains much lower than $L_{CS}$. No maximum of the $S$ factor develops, so that, in other words, no fusion hindrance seems to show up in this case in the measured energy range. 
\par

The different trends originate from the dissimilar low-energy nuclear structure of the involved nuclei. Indeed, $^{48}$Ti and $^{58}$Fe are soft and have a low-lying quadrupole excitation lying at $\approx$ 800-900 keV only. Instead, $^{58}$Ni and  $^{54}$Fe have a closed shell (protons and neutrons, respectively) and are rather stiff. Therefore the barrier distribution of $^{48}$Ti + $^{58}$Fe (Fig.~\ref{BD_4858}) is wider and  extends to lower energies with respect to $^{58}$Ni + $^{54}$Fe (see Fig.~3 of Ref.~\cite{48+58}). This probably pushes the onset of hindrance to an energy below  the measured range. On the other hand, the barrier distribution of $^{58}$Ni + $^{54}$Fe is narrower, thus  allowing the hindrance to show up already at the level of $\approx$ 200 $\mu$b.
\par 

\par
\begin{figure}[h]
\centering
\resizebox{0.45\textwidth}{!}{\includegraphics{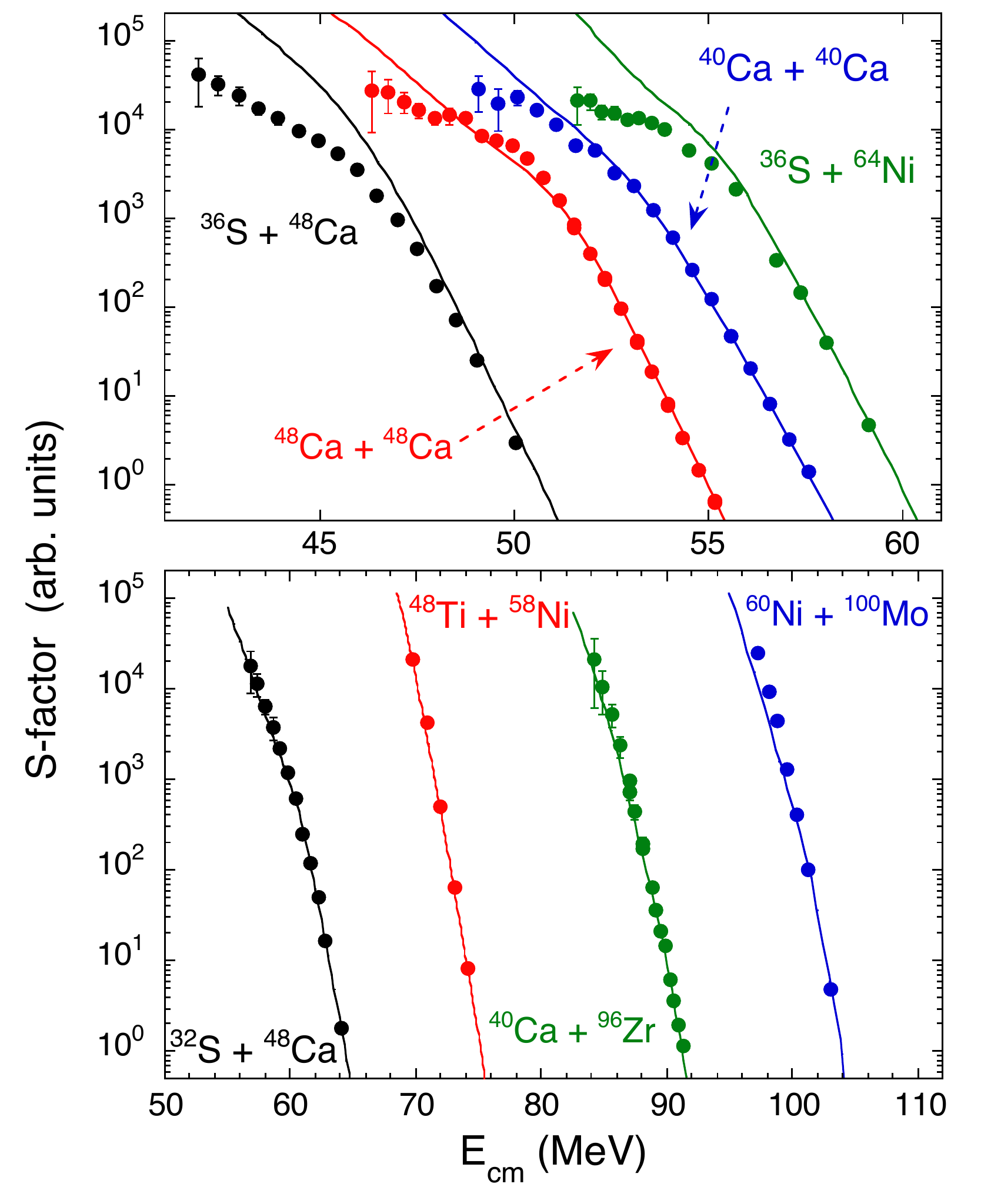}}
\caption{Astrophysical $S$ factor for several systems where no maximum shows up. In the upper panel the hindrance can be recognised by the comparison of the data with results of standard CC calculations. In the lower panel the hindrance can not even  be recognised by such comparison. 
The reported cases are $^{36}$S + $^{48}$Ca~\cite{3648}, $^{48}$Ca + $^{48}$Ca~\cite{4848}, $^{40}$Ca + $^{40}$Ca~\cite{4040} and $^{36}$S + $^{64}$Ni~\cite{3664}, $^{32}$S + $^{48}$Ca~\cite{3248}, $^{48}$Ti + $^{58}$Ni~\cite{48+58}, $^{40}$Ca + $^{96}$Zr~\cite{4096,Esb_4096} and $^{60}$Ni + $^{100}$Mo~\cite{60100}. For  $^{36}$S + $^{48}$Ca, $^{32}$S + $^{48}$Ca and $^{60}$Ni + $^{100}$Mo the energy scale has been shifted up by 5 MeV, 20 MeV and -25 MeV, respectively, for graphical convenience.}
\label{Stutti}     
\end{figure}

Several experiments 
have shown that  fusion cross sections at very low energies exhibit the hindrance effect, and consequently the $S$ factor develops a maximum vs. energy. 
A selection  of cases where the $S$ factor  shows a clear maximum vs. energy, is presented in Fig.~\ref{Mag8}. 
In particular, the systems $^{90}$Zr + $^{92}$Zr,$^{89}$Y were measured at GSI in the 80's~\cite{Keller} down to very small cross sections ($\simeq$ 120 nb and 340 nb, respectively), below the onset of hindrance. Results for  $^{90}$Zr + $^{90}$Zr were obtained in the same experiments, however the behaviour of its excitation function at very low energies is somewhat irregular, and no clear maximum of the $S$ factor can be identified.

For other systems  no maximum of  the $S$ factor can be observed in the measured energy range, so that in order to decide whether the hindrance phenomenon is present, one has to compare the experimental data 
with the results of standard CC calculations. One may observe that such calculations overpredict  the experimental cross sections at low energies and this is the signal of hindrance. 

The use of the $S$ factor representation, is very convenient also in these cases. A few examples are shown in Fig.~\ref{Stutti} (upper panel) for some medium-mass systems recently investigated at LNL.
The standard CC calculations were performed using the Aky\"uz-Winther potential~\cite{Akyuz} and including the lowest 2$^+$ and 3$^-$ collective modes. It is evident that for none of those four systems a maximum of the $S$ factor appears, however for all of them 
the $S$ factor  is largely over-predicted by the calculations below the barrier and consequently one may deduce that the hindrance effect occurs.

However, there are several other cases where even the comparison of the experimental $S$ factor with standard CC calculations does not give any indication of the hindrance phenomenon.  The lower panel of Fig.~\ref{Stutti} shows a few examples of this situation which may be a consequence of different influence of  nuclear structure and/or strong transfer couplings, that probably push the hindrance threshold 
below the lowest measured energy. In particular, the case of  $^{40}$Ca + $^{96}$Zr~\cite{4096,Esb_4096}
is a very significant example of the effect of couplings to quasi-elastic transfer channels with $Q>0$, as discussed in greater detail in Section~\ref{InfTransf}.

Some years ago a phenomenological analysis led to a purely empirical formula~\cite{Jiang06}  for the expected energy $E_s^{emp}$ of the $S$-factor maximum.  
The formula was originally developed  for  medium-mass  systems with negative fusion $Q$-value,  involving 
closed-shell nuclei  for both projectile and target (stiff systems). In its most recent version~\cite{Jiang09}  describing also lighter systems,  it  reads 
\begin{equation}
  E_s^{emp} = [0.495\zeta/(2.33+580/\zeta)]^{2/3} \ \ (\rm{MeV}) 
\label{syste}
\end{equation} 
where the quantity $2.33+580/\zeta$ represents the logarithmic slope $L_s^{emp}$ of the excitation function at the energy $E_s^{emp}$.

The energy  $E_s$ where the maximum actually shows up in different systems generally tends to decrease with respect to  $E_s^{emp}$, when the total number of ``valence nucleons" outside closed shells in the entrance 
channel, increases~\cite{64100}. We will come back to this point further below when discussing light systems.



\begin{figure}[h]
\centering
\resizebox{0.45\textwidth}{!}{\includegraphics{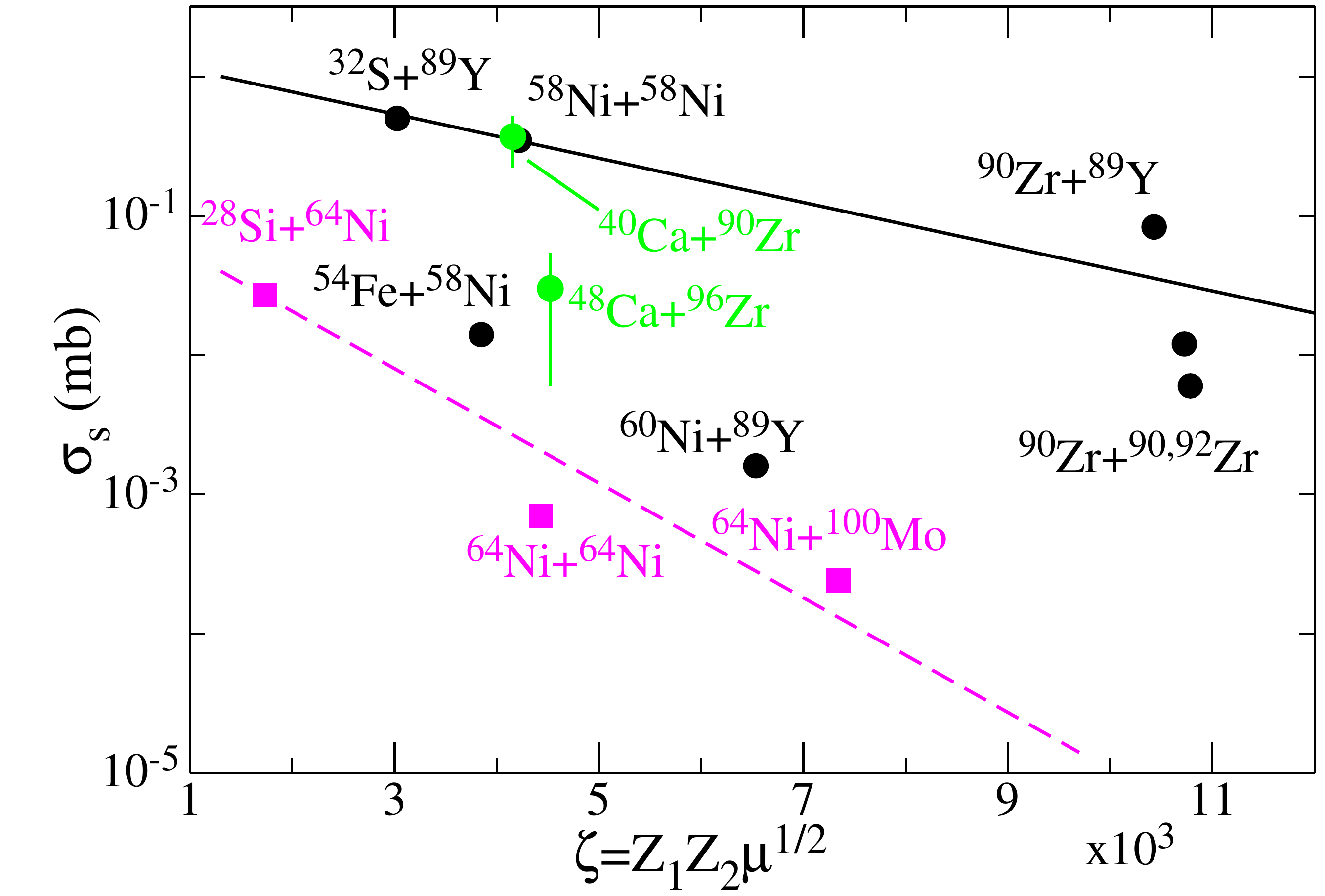}}
\caption{Fusion cross section at the $S$ factor maximum vs. system parameter $Z_1Z_2 \sqrt{\mu}$.
Black and magenta symbols were determined by measurements (stiff and soft systems respectively). Green symbols have been obtained from
extrapolations (courtesy of C.L.Jiang).}
\label{Jiang_syst_sigma}     
\end{figure}

Here we want to point out that in relation with  the trend  of  $E_s$, the cross section, $\sigma_s$ where the $S$-factor maximum  appears 
varies considerably, being much higher  (up to hundreds of $\mu$b) for stiff systems, and not exceeding $\sim$100 nb for softer cases like $^{64}$Ni + $^{64}$Ni~\cite{Jiang04} or $^{64}$Ni + $^{100}$Mo~\cite{64100}. This systematics is reported in Fig.~\ref{Jiang_syst_sigma} (see Table I in Ref.~\cite{Jiang0}).
Therefore it has been generally easier from the experimental point of view to identify the hindrance effect associated to an $S$-factor maximum, in stiff systems.

\subsubsection{The systems Si+Si}

The fusion of the symmetric system $^{28}$Si + $^{28}$Si and of the asymmetric one $^{28}$Si + $^{30}$Si offers several ``a priori" elements of interest originating from the strong oblate deformation of $^{28}$Si and from the presence of the (elastic) transfer reaction channel in the second system. 
\par
$^{28}$Si + $^{30}$Si was studied a few years ago at ANL~\cite{2830_bis}. It was the first measurement involving a light system with a positive $Q_{fus}$  with the purpose of evidencing the possible existence of hindrance, and 
the excitation function was measured down to $\simeq$40$\mu$b. 
A further experiment was performed more recently  at LNL~\cite{2830} extending the measured cross sections by around one order of magnitude. The complete data set was analysed by CC calculations using the M3Y+ repulsion potential, where one- and two-phonon
excitations as well as mutual excitations of the low-lying 2$^+$ and 3$^-$ states in both projectile and target, were included. This gives a rather poor fit to the data, and the additional influence of one- and two-neutron transfer reactions built on surface excitations, was taken into account.

\par
\begin{figure}[h] 
\centering
\resizebox{0.40\textwidth}{!}{\includegraphics{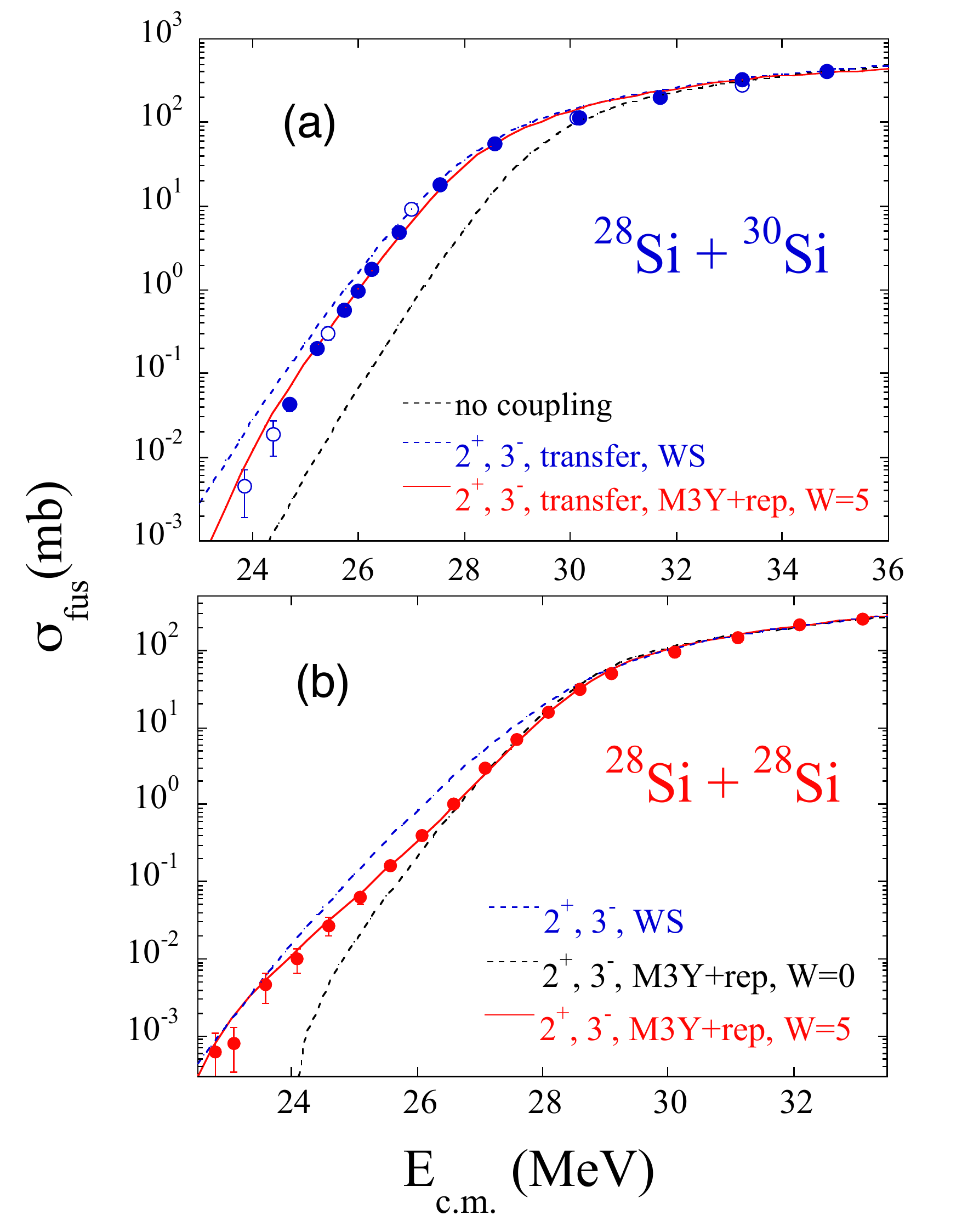}}
\caption{Measured fusion cross sections for $^{28}$Si+$^{30}$Si 
~\cite{2830_bis,2830} (a), and for 
$^{28}$Si+$^{28}$Si~\cite{2830}  (b)  are compared to CC calculations
based on M3Y+ repulsion and Woods-Saxon  potentials. Transfer couplings are necessary to fit the cross sections of  $^{28}$Si+$^{30}$Si (see text).
The data fit for both systems required  an imaginary potential with parameters 
$a_w$=0.2 fm and $W_0$=5 MeV (red  curves).
The calculation without any imaginary potential (W=0, black curve) strongly underestimates the low  energy data of 
$^{28}$Si+$^{28}$Si. 
The blue dashed curves are obtained using a standard WS potential (see Ref.~\cite{2830})..
}
\label{282830}     
\end{figure}

This provides a good fit of the data (see Fig.~\ref{282830}(a)), while the analogous calculation using a standard WS potential over-predicts the low-energy cross sections. Thus we observe hindrance in this system having a positive Q-value for fusion $Q_{fus}$= + 13.4 MeV, supporting the previous (weak) evidence of the effect coming from the trend of the $S$ factor where a clear maximum is not observed.
\par
The fusion of $^{28}$Si + $^{28}$Si was recently investigated down to very low cross sections $\sigma_f\simeq$ 600 nb. The excitation function is  reported in Fig.~\ref{282830}(b) together with the results of calculations analogous to those described here above for the asymmetric system $^{28}$Si + $^{30}$Si. When using the M3Y+ repulsion potential, the data are nicely reproduced without the need of transfer couplings (Q-values are all negative), while the cross sections are overestimated just below the barrier if a WS potential is employed. This indicates the presence of hindrance but this effect disappears at the lowest energies. 

The low-energy
data of both systems  are reproduced only by applying also a weak, short-ranged imaginary potential. This  probably simulates the effect of the oblate deformation of $^{28}$Si causing  the pocket minimum  in the different reaction
channel potentials to be located at different radial distances. Since  the  incoming-wave boundary
conditions (IWBC) are imposed at the minimum of the entrance channel potential, 
the fusion in the 2$^+$ channel is cut off at an energy that is higher than the minimum
of the 2$^+$ channel potential (see Fig. 4 of Ref.~\cite{2830}).



Very recently the measurement of the excitation function of  $^{30}$Si+$^{30}$Si 
has been performed~\cite{3030}. The nucleus $^{30}$Si is essentially spherical and preliminary CC calculations not using any imaginary potential
seem to reproduce the data. In the measured energy range this system does not show any indication of the hindrance effect, because no maximum of $S$ factor is observed and the excitation function is nicely reproduced by simply using a standard Woods-Saxon potential. 
These evidences support the hypothesis that the oblate deformation of $^{28}$Si is the reason 
why the fusion excitation function of $^{28}$Si+$^{28}$Si has a somewhat unusual shape.

\subsection{Models of fusion hindrance}
\label{models}

The discovery of the fusion hindrance phenomenon triggered a widespread discussion about the underlying physics. It was pointed out 
~\cite{Nanni} that deep sub-barrier fusion cross sections may be sensitive to the shape of the nuclear potential in the
inner side of the Coulomb barrier. In other words, one can use the results of those
measurements far below the barrier to investigate the radial dependence of the ion-ion potential at extremely close distances.

From a phenomenological point of view,  the experimental data indicate that a thicker barrier is needed, with respect to what standard potentials (like e.g. the AW potential) produce.
 Misicu and Esbensen~\cite{Misi,Esbe07} observed  that the incompressibility of nuclear matter may lead to a repulsive core in the potential at short distances.

Consequently, in the sudden approach, they proposed to use the 
double folding potential M3Y with  an additional  repulsive core in the ion-ion interaction (M3Y + repulsion), that indeed 
produces a shallow pocket inside the Coulomb barrier. 
This sudden approach using the M3Y +  repulsion potential  has been quite successful in reproducing the hindrance behaviour in 
several cases~\cite{BBB}.
\par
Very recently, it has been suggested by Simenel et al.~\cite{Sime17} that the Pauli exclusion principle gives rise to a short range repulsion in the ion-ion potential.
They have used  the density-constrained frozen Hartree-Fock method to calculate the bare potential including the Pauli repulsion exactly.
The resulting potential has a shallow pocket just as in the phenomenological approach  of  the M3Y+ repulsion interaction, and 
 the consequence is that tunnelling probability is reduced, thus allowing to reproduce rather well low energy experimental data.
 Their microscopic calculations do not predict repulsive effects due to the incompressibility of nuclear matter as in~\cite{Misi,Esbe07}, 
 which is then suggested to simply simulate the effect of Pauli repulsion.
\par
\begin{figure}[h]
\centering
\resizebox{0.40\textwidth}{!}{\includegraphics{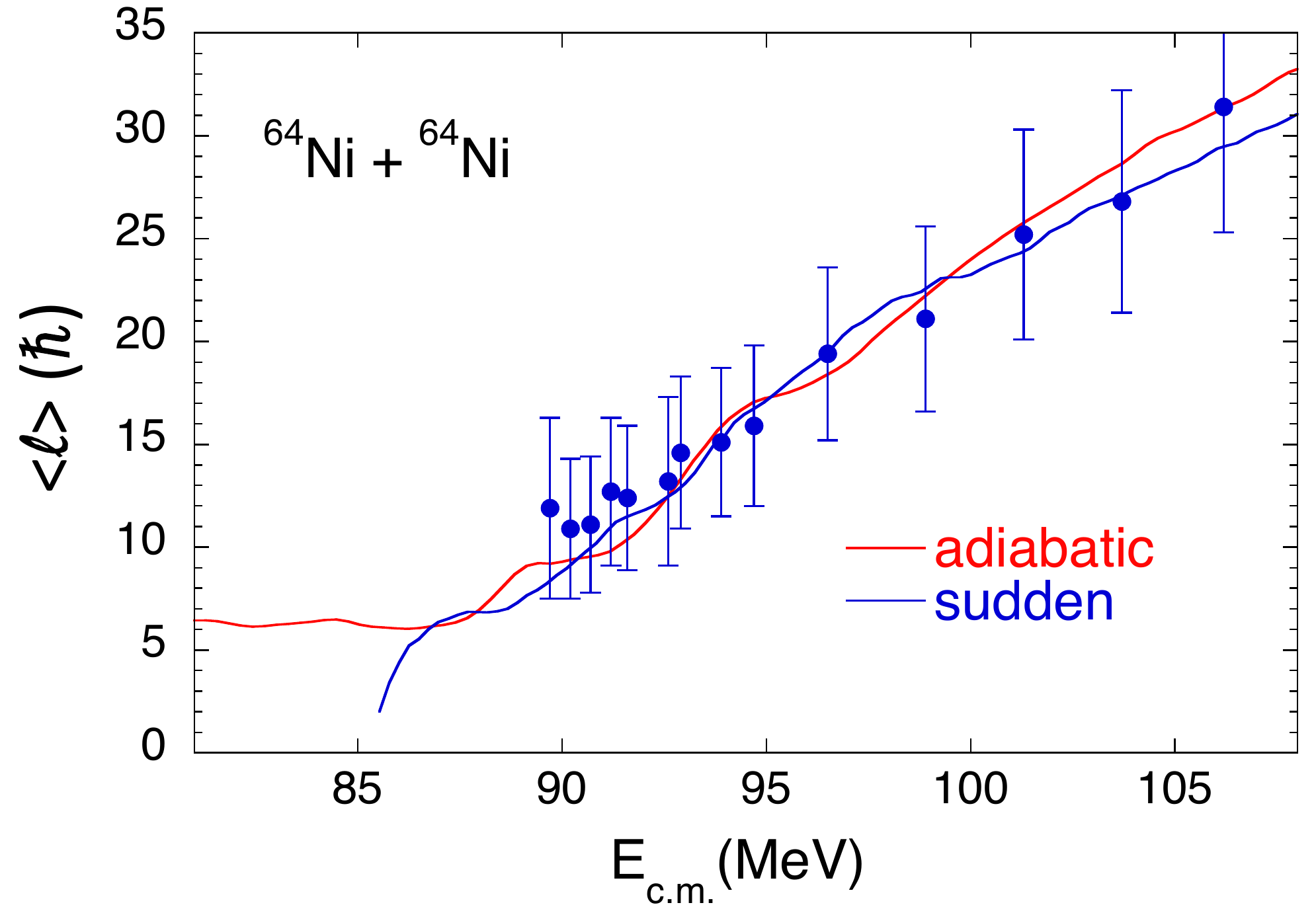}}
\caption{Average angular momentum of compound nucleus vs. incident energy for $^{64}$Ni + $^{64}$Ni. The data are from~\cite{Dieter} and the calculations from~\cite{Ichi15}.  The results of the sudden model were performed using the M3Y+rep. potential. }
\label{laseconda}     
\end{figure}

As an alternative to sudden models, an adiabatic approach was  proposed by Ichikawa and Hagino~\cite{Ichi07}, 
where neck formation between the colliding nuclei is considered and the interaction evolves from a two-body potential to a one-body potential.
In the overlap region at ion-ion distances smaller than the touching point, 
a damping of the coupling form factors was introduced in Ref.~\cite{Ichi07}. It was suggested  that in the RPA method~\cite{IchiRPA},  this can originate  from the damping of quantum vibrations of  target and  projectile,  near the touching point.
\par

The fusion cross sections, $S$ factors and logarithmic derivatives calculated within this adiabatic model using the Yukawa-plus-exponential (YPE) potential 
are in very good agreement with experimental data~\cite{Ichi15} for a number of systems. The author points out that, indeed, the energy at the touching point  strongly correlates with the threshold energy observed for the hindrance phenomenon in the various cases, apart from medium-light mass systems.
\par
A way to discriminate between sudden and adiabatic  models is a comparison of the calculated average angular momentum $<l>$ of the compound nuclei, which are significantly different in the two approaches below the threshold energy where hindrance shows up.
The $<l>$ estimated with  the sudden model is strongly
suppressed at low energies, because the high
angular-momentum components in the partial waves are cut-off due to
the shallow potential pocket. On the other hand,  in the adiabatic approach, the damping factor affects $\it{each}$  partial-wave cross section, so that in this model the compound nucleus has a higher content of angular momentum, below the hindrance threshold.
\par
\begin{figure}[h]
\centering
\resizebox{0.40\textwidth}{!}{\includegraphics{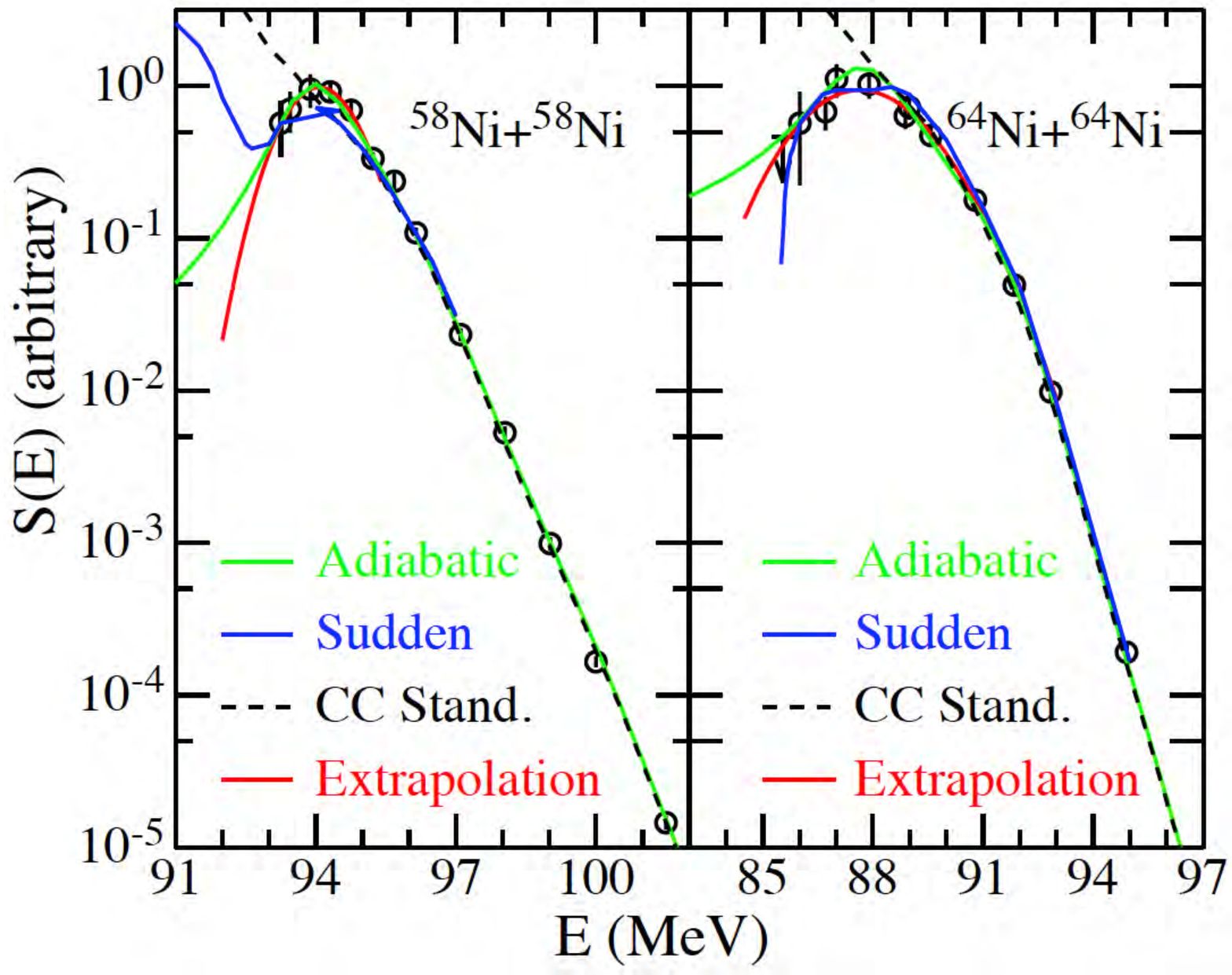}}
\caption{$S$ factor  for the systems $^{58}$Ni + $^{58}$Ni~\cite{BeckNiNi} and $^{64}$Ni + $^{64}$Ni~\cite{Jiang04}. Black dashed curves are CC calculations with standard WS potentials. The green and blue lines  are calculations from the adiabatic~\cite{Ichi15} and sudden models~\cite{Esbe07}, respectively. The red lines are empirical
extrapolations~\cite{Esbe09}.}
\label{S_AGFA}     
\end{figure}

Thus,  it would be important to measure the average angular momentum $\it{<l>}$ at
sub-barrier energies. Measurements of this kind are difficult especially at sub-barrier energies, and data have been obtained in very few cases. 
Fig.~\ref{laseconda} shows the results for $^{64}$Ni + $^{64}$Ni~\cite{Dieter}. The lowest energy where $\it{<l>}$ was measured is E$_{cm}$= 89.7 MeV where the fusion cross section is $\simeq$55 $\mu$b. That energy is still much higher than the threshold for hindrance E$_{cm}\simeq$ 87.8 MeV ($\sigma_f\simeq$ 6 $\mu$b), see Fig.~\ref{6464}, so that no conclusion can be drawn.

A further way to discriminate between the theoretical models~\cite{Esbe07,Ichi15} is provided by the trend of the  $S$ factor at very low energies. Indeed, a comparison between experimental data and theoretical calculations is shown for the two systems, $^{58}$Ni + $^{58}$Ni~\cite{BeckNiNi} and $^{64}$Ni + $^{64}$Ni~\cite{Jiang04} in Fig.~\ref{S_AGFA}. It is evident that the standard CC calculation
(black dashed line) fails below the $S$ factor maximum, whereas the data do not
extend far enough to distinguish between the sudden and the adiabatic model. 
To this end, more experimental data are clearly needed at lower energies.

It is clear that a discrimination may be experimentally attempted for systems where the hindrance threshold is relatively high, that is, for stiff systems, as shortly discussed above when commenting Fig.~\ref{Jiang_syst_sigma}.

\subsubsection{The case of $^{16}$O + $^{208}$Pb}

\begin{figure}[h]
\centering
\resizebox{0.35\textwidth}{!}{\includegraphics{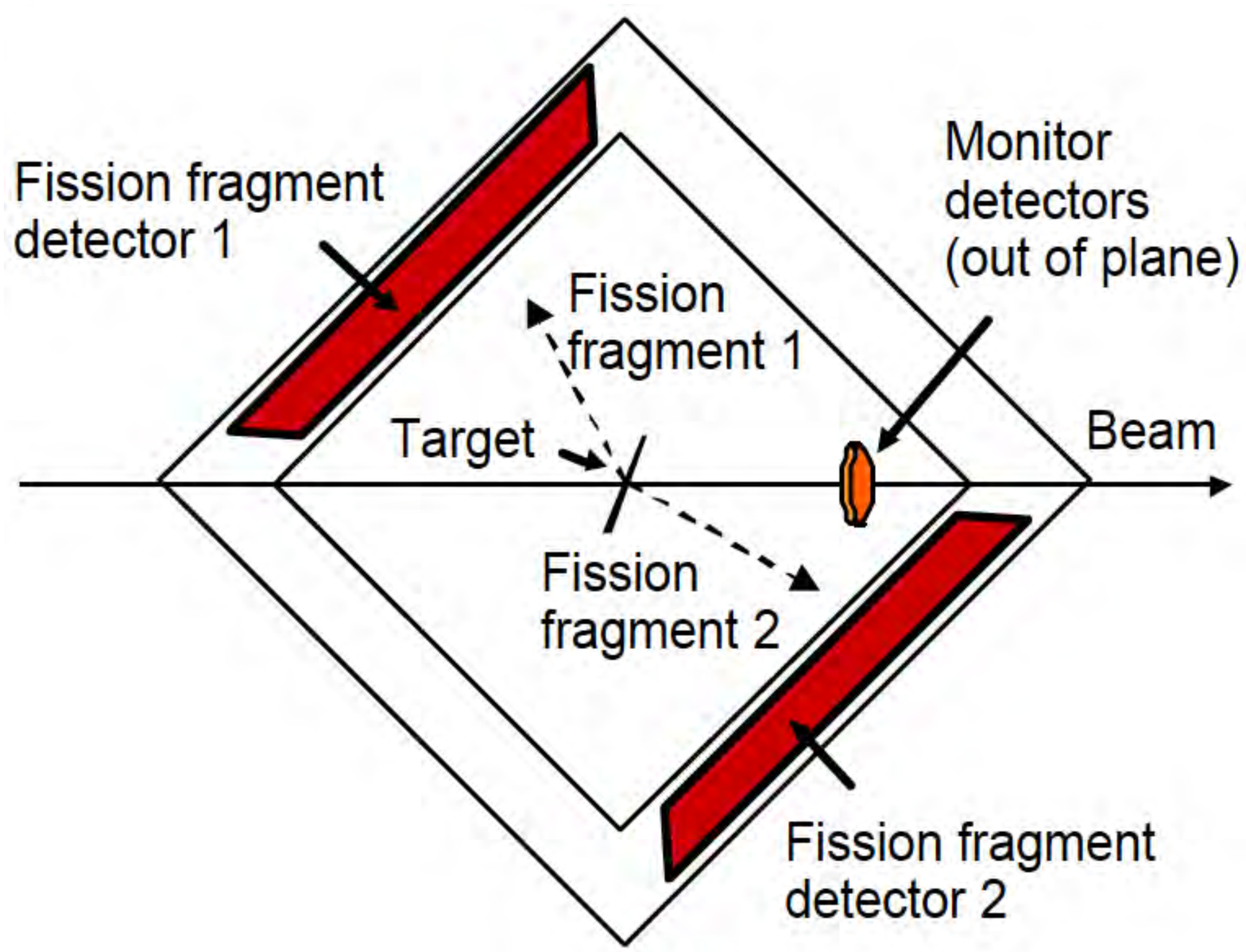}}
\caption{Set-up for the measurement of fission fragments at ANU-Canberra~\cite{Hinde96}.}
\label{Cube}     
\end{figure}

\begin{figure}[h]
\centering
\resizebox{0.36\textwidth}{!}{\includegraphics{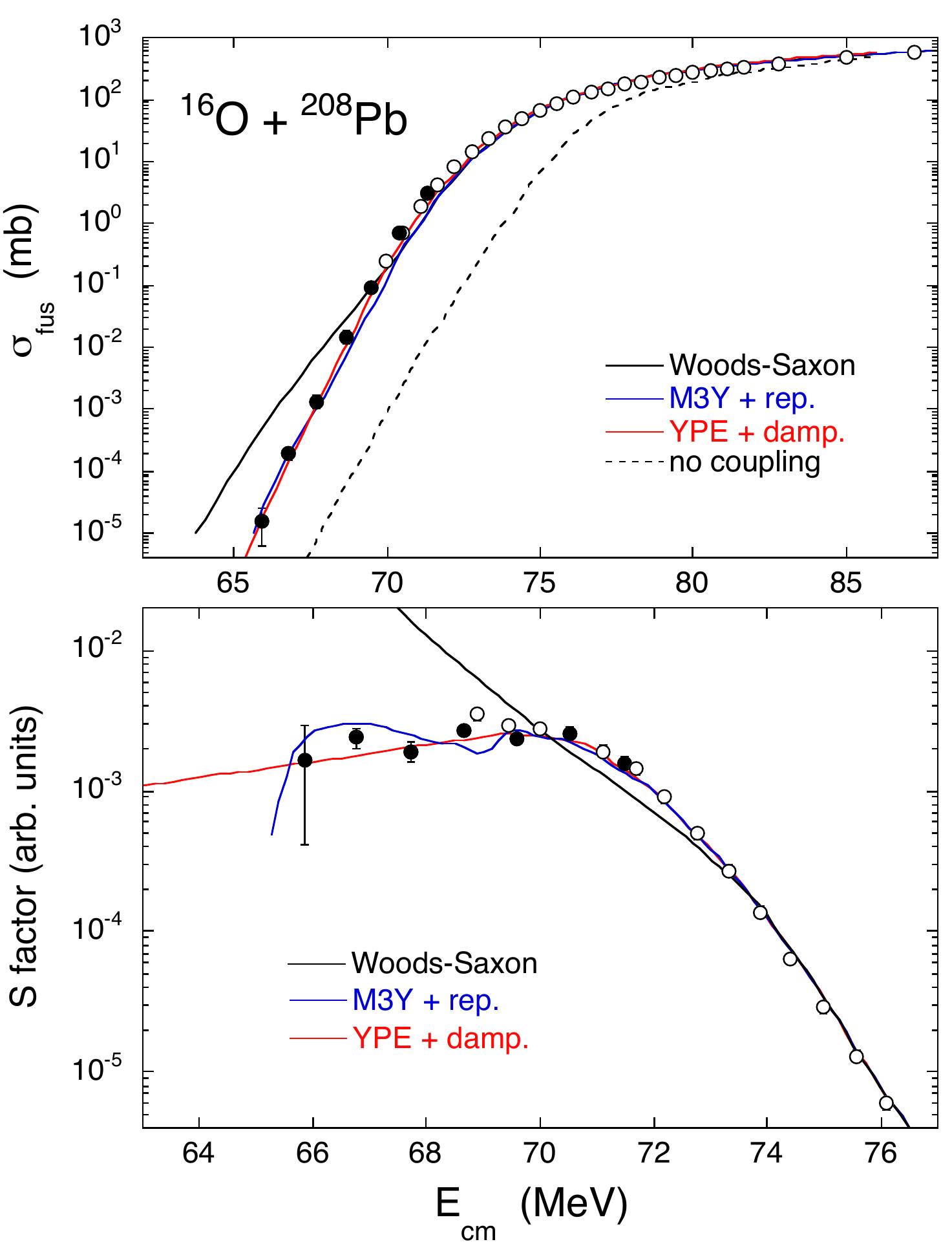}}
\caption{Fusion excitation function (upper panel) and astrophysical $S$ factor (lower panel) of $^{16}$O + $^{208}$Pb compared to several CC calculations. The full symbols refer to the more recent measurement of Ref.~\cite{Nanda07} while the open symbols come from Morton et al.~\cite{Morton}.}
\label{16208}     
\end{figure}

The fusion excitation function of $^{16}$O + $^{208}$Pb~\cite{Nanda07} was measured down to the  very low cross section $\sigma_f$ = 16$\pm$10 nb, complementing the previous data of Morton et al.~\cite{Morton}.  The ER were detected by means of a compact velocity filter and a silicon detector (see Fig.~\ref{ANU}) down to $\sim$10$^{-3}$mb, and two large position sensitive multi-wire proportional counters, schematically shown in Fig.~\ref{Cube}~\cite{Hinde96,Fiera07}, measured in coincidence  the fission fragments  down to  
 $\sigma_{fis}$ $\sim$ 10$^{-5}$mb. The complete excitation function is reported in Fig.~\ref{16208}  (upper panel), and we observe that the low-energy slope is very steep. The lower panel shows the  astrophysical $S$ factor that saturates in the same energy range. As a matter of fact,  the CC calculation using a Woods Saxon potential strongly overpredicts the  low-energy cross sections and the $S$ factor. This is a clear manifestation of the hindrance effect.
 
Dasgupta et al.~\cite{Nanda07} in their original work suggested  that the coherent 
coupled-channels model is inadequate. Indeed,  in this model  an irreversible energy dissipation starts occurring inside the barrier,  but does not influence the coherence of quantum states. On the contrary, they argued that
 a gradual onset of decoherence takes place with increasing overlap of the two nuclei, leading to hindrance of quantum tunnelling~\cite{Diaz08}. 
 
Fig.~\ref{16208} shows also the results of subsequent analyses~\cite{Esbe07,Ichi15} performed within the sudden and adiabatic models we have outlined here above. The results of the two models, as far as the excitation function is concerned, are almost indistinguishable. Discriminating between the two approaches would require measuring at even lower energies because only there the predicted $S$ factors  deviate from each other.  Such measurements are very interesting, however, they would be very challenging.

 Yao and Hagino~\cite{Yao} considered  the 
anharmonicity of the multi-octupole phonon states of $^{208}$Pb. 
They obtained  improved results for the near-barrier excitation function with respect to 
CC calculations in the harmonic-oscillator limit, and the BD is  much better reproduced.
 
\subsection{Hindrance in light systems}
 \label{Light systems}

The hindrance phenomenon was first studied in medium-heavy mass systems where the fusion Q value (Q$_{fus}$) is always negative. Under this conditions a maximum
of $S(E)$ is algebraically necessary~\cite{Jiang04_2}. Indeed, the 
cross section must be zero at $E = -Q$, i.e., at the finite energy corresponding to the ground state of 
the compound nucleus.
Consequently, $L_{CS}$(E)=$\pi\eta$/E remains finite when $E \rightarrow -Q$ while 

\begin{equation}
L(E)=\sigma^{-1}d\sigma/dE + 1/E \rightarrow + \infty
\end{equation}

in that limit.
Consequently, the $S$ factor  has necessarily a maximum at  some  energy E $>$ -- Q where 
 $L(E)$ reaches the $L_{CS}$ value. 

On the contrary, an $S$ factor maximum may not develop for systems having $Q_{fus}>0$ where, therefore, no fusion energy threshold exists. Actually in the limit of E=0   both $L(E)$ and $L_{CS}$(E) become infinite. This means that $L(E)$ may not reach $L_{CS}$ at any energy.

 Several experiments have been performed in recent years on systems with Q$_{fus}>0$ in order to investigate the possible existence of the hindrance effect.  
 It is still an open question, whether the existence  of an $S$ factor maximum at 
low energies, is a common feature of such systems.
Indeed precise measurements of very small fusion cross
sections are experimentally challenging, and most investigated cases may have 
not reached a low enough cross section level in order to 
clear up this issue.

\subsection{Consequences for astrophysics}
\label{astrophysics}

Fusion reactions of light systems are critical for a variety of stellar environments and the accurate knowledge of sub-barrier fusion cross sections is essential for valid simulations of the nucleosynthesis processes. 
Therefore it was soon realised  that the  hindrance phenomenon  may have important 
consequences on the nuclear processes occurring in astrophysical scenarios, 
if that effect is a general behaviour of heavy-ion fusion at extreme 
sub-barrier energies, including those
reactions involving light systems {\it e.g.}  fusion of carbon and oxygen nuclei 
\cite{jiang8_12}.


\begin{figure}[h]
\centering
\resizebox{0.48\textwidth}{!}{\includegraphics{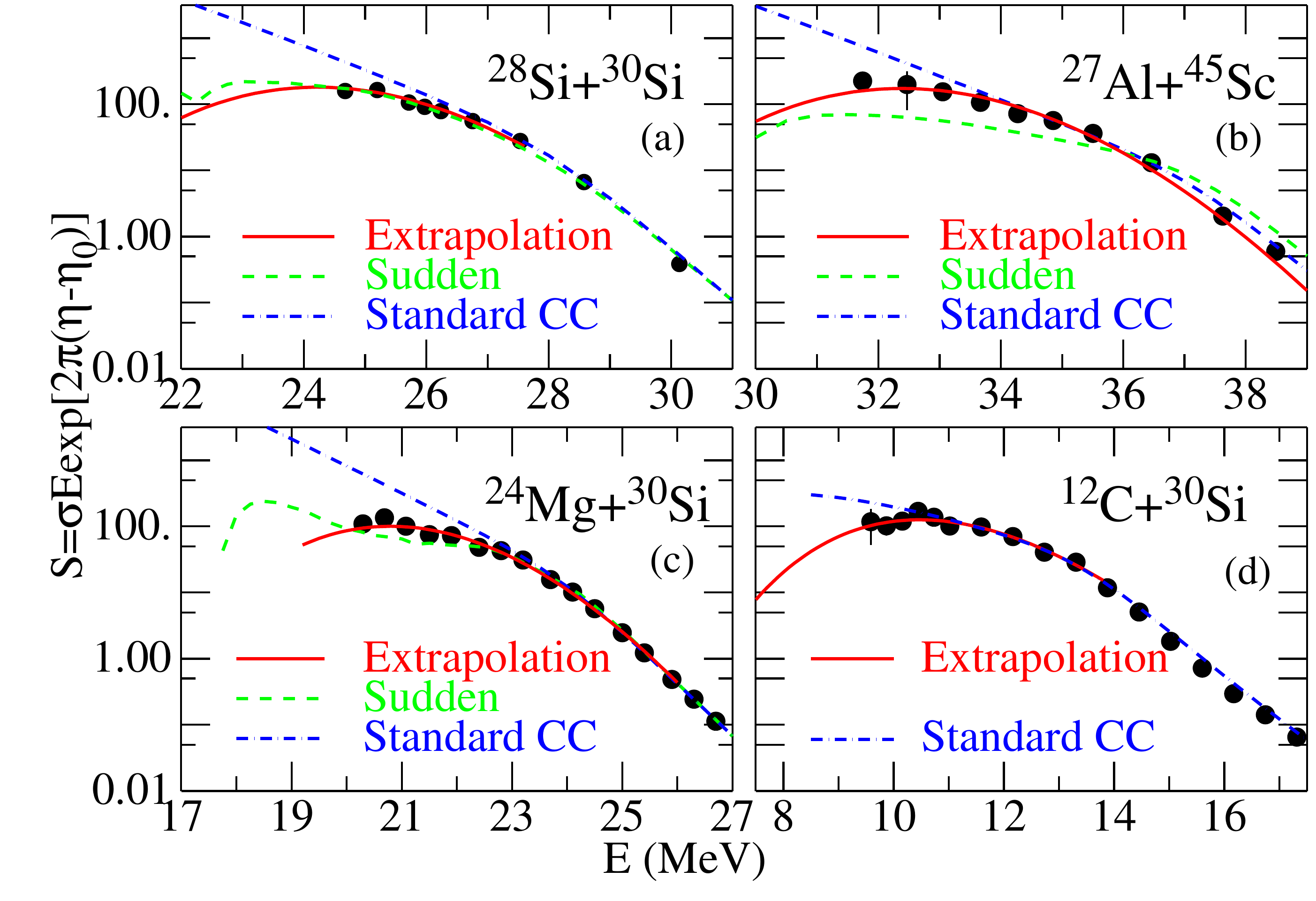}}
\caption{Astrophysical $S$ factor for four systems with positive Q-value for fusion ($Q_{fus}>0$).
The reported cases are $^{28}$Si + $^{30}$Si~\cite{2830,2830_bis}, $^{27}$Al + $^{45}$Sc~\cite{2745}, $^{24}$Mg + $^{30}$Si~\cite{2430} and $^{12}$C + $^{30}$Si~\cite{1230} (courtesy of C.L. Jiang).}
\label{Qpositivi}     
\end{figure}

There are many studies of the fusion reactions 
$^{12}$C + $^{12}$C and  
$^{16}$O + $^{16}$O. However, these measurements  often have 
large uncertainties and there are serious 
discrepancies between different experiments in the low energy range important for astrophysics.
Therefore studies of systems slightly heavier are of interest since their behaviour 
at low energy can give us guidelines for the reliable extrapolation to
$^{12}$C + $^{12}$C and similar cases towards extremely low energies. 

As a matter of fact, whether there is an $S$-factor maximum at 
very low energies for systems with a positive fusion $Q$ value
has been an experimentally challenging question for some years.
Some studies of systems with medium to light 
masses and positive $Q$ values have been recently performed at various laboratories 
 (see {\it e.g.}~\cite{2830_bis,3648,2745,4048,3248,2830,2430}), and a few cases are shown in Fig.~\ref{Qpositivi}. There 
is  some evidence for an $S$-factor maximum, but the energy range covered in these experiments 
is limited and so the possible existence of that maximum is not so clear.  

Let us first discuss the three cases
$^{28}$Si + $^{30}$Si \cite{2830,2830_bis},
$^{27}$Al + $^{45}$Sc \cite{2745} and
$^{24}$Mg + $^{30}$Si \cite{2430}.
Fusion $Q$ values are all positive for 
these systems, that is, 14.3, 9.63, and 17.89 MeV, respectively. 
Three kinds of calculations and extrapolations are included in the figure.
The blue dashed curves
are CC calculations with a standard WS potential, which
always overpredict the experimental data at low energies.

\begin{figure}[h]
\centering
\resizebox{0.48\textwidth}{!}{\includegraphics{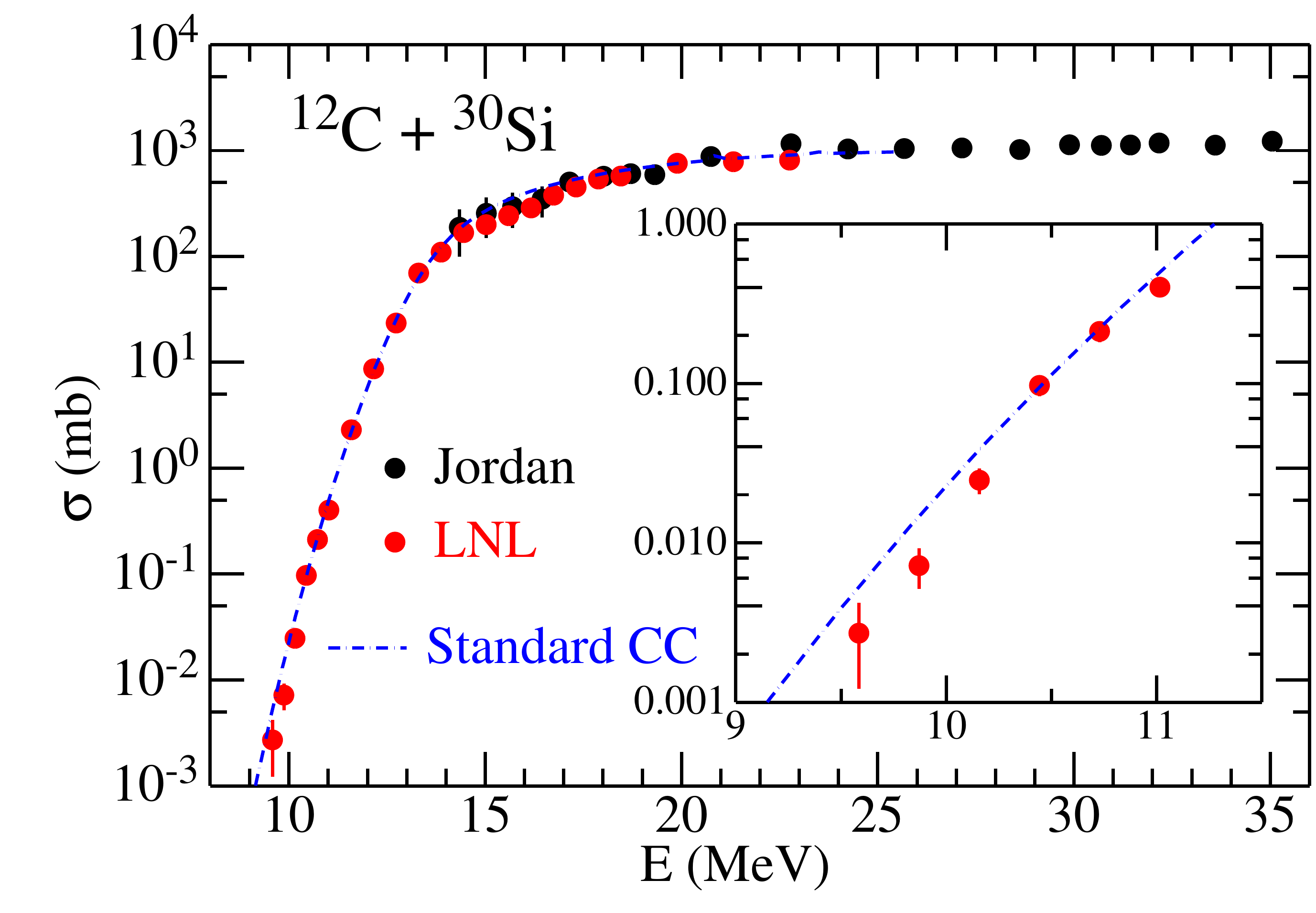}}
\caption{Excitation function for the 
system $^{12}$C + $^{30}$Si. The black and red circles are from Jordan 
{\it et al.}~\cite{jord} 
and the new measurement~\cite{1230} with the inverse-kinematics 
technique, 
respectively. 
}
\label{1230}     
\end{figure}

The green-dashed lines are CC calculations with a repulsive core
included in the potential (M3Y + rep. potential, sudden model~\cite{Misi}), 
while the red lines are from the empirical 
extrapolations developed in Ref. \cite{jiang8_12}.
For these medium-light-mass systems,  the sudden model
reproduces the experimental data quite well
as can be seen in Fig.~\ref{Qpositivi}. 

There is some evidence in all three cases for a maximum of $S$, but
$^{12}$C + $^{30}$Si (the fourth panel in the figure, with $Q$-value = +14.11 MeV) probably displays  the clearest evidence for it. Indeed, a recent
experiment with inverse-kinematics technique  has been performed~\cite{1230} for this system. The preliminary
excitation function is displayed in Fig.~\ref{1230}. The earlier experiment by Jordan {\it et al.}~\cite{jord} 
measured cross sections down to only 200 mb, while the new one reached 3~$\mu$b. 

The insert in Fig.~\ref{1230} shows 
that fusion cross sections at very low energies are clearly hinderend with respect to the standard CC calculations reported there. In fact, a rather clear $S$-factor maximum has been observed 
(see again Fig.~\ref{Qpositivi}). For this system only the standard CC calculation and the 
extrapolation are shown.
We point out that $^{12}$C + $^{30}$Si is the lightest system of the four reported in Fig.~\ref{Qpositivi}.

\begin{figure}[h] 
\centering
\resizebox{0.45\textwidth}{!}{\includegraphics{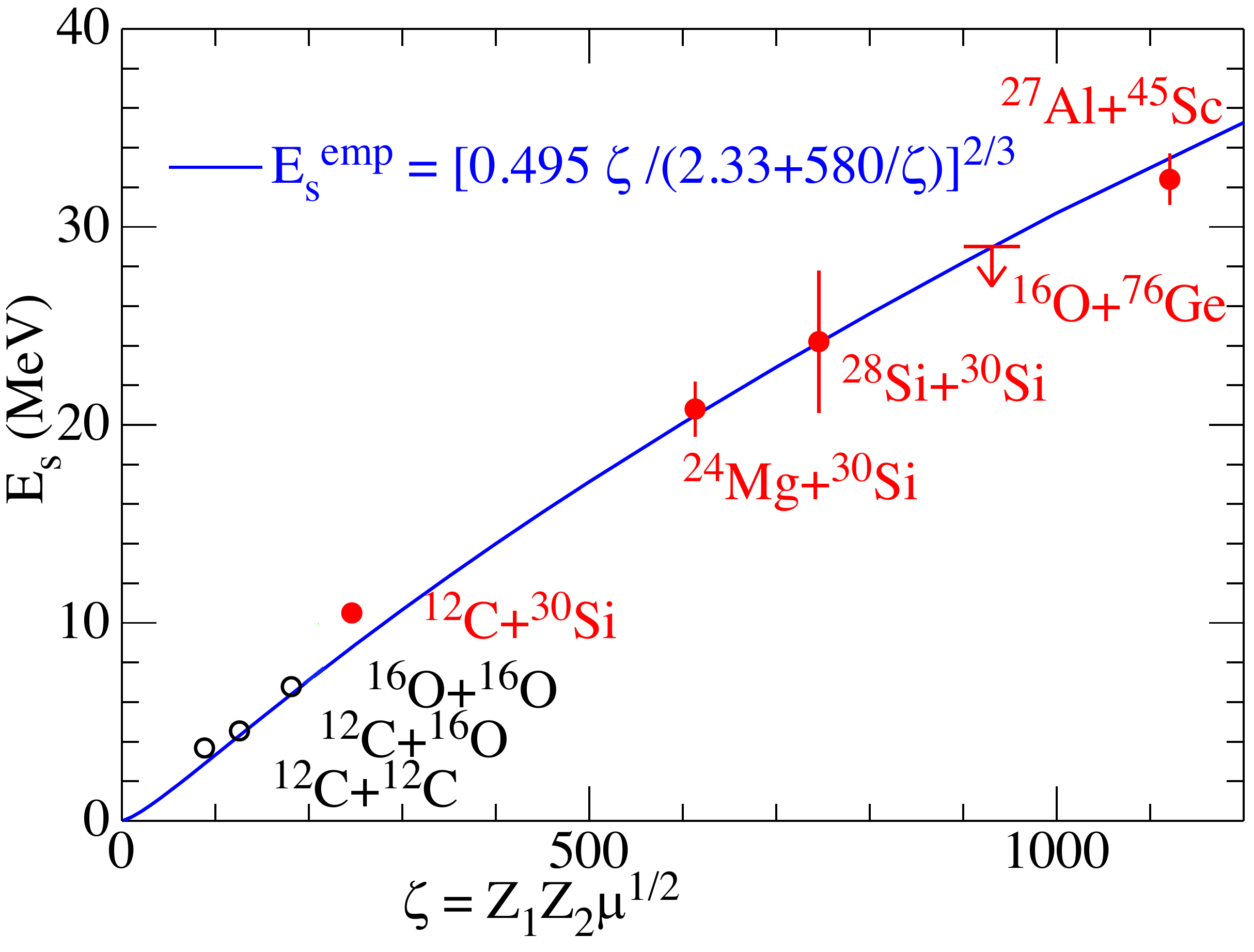}}
\caption{ Systematics of $E_s$ in several light- and medium-light mass 
systems~\cite{Jiang09}. The location for   $^{12}$C + $^{30}$Si  is very close to 
the astrophysically relevant $^{12}$C + $^{12}$C and $^{16}$O + $^{16}$O cases. 
For these systems several data sets exists, but they are sometimes contradictory and the errors large. This results in  large uncertainties in the  expected values for their hindrance threshold.
 The corresponding points  (open symbols) have therefore been obtained from extrapolations. 
} 
\label{systematics_Jiang}
\end{figure}

It is instructive to place $^{12}$C + $^{30}$Si in the phenomenological systematics cited above (see Eq.~\ref{syste}).
We see in Fig.~\ref{systematics_Jiang} that  the hindrance threshold for $^{12}$C + $^{30}$Si, having a system parameter quite close to the astrophysically relevant reactions, follows rather closely that systematics. 
This gives us reasonable confidence that a hindrance exists even for the lighter systems, around the predicted values.

\begin{figure}[h]
\vspace{-2mm}
\centering
\resizebox{0.38\textwidth}{!}{\includegraphics{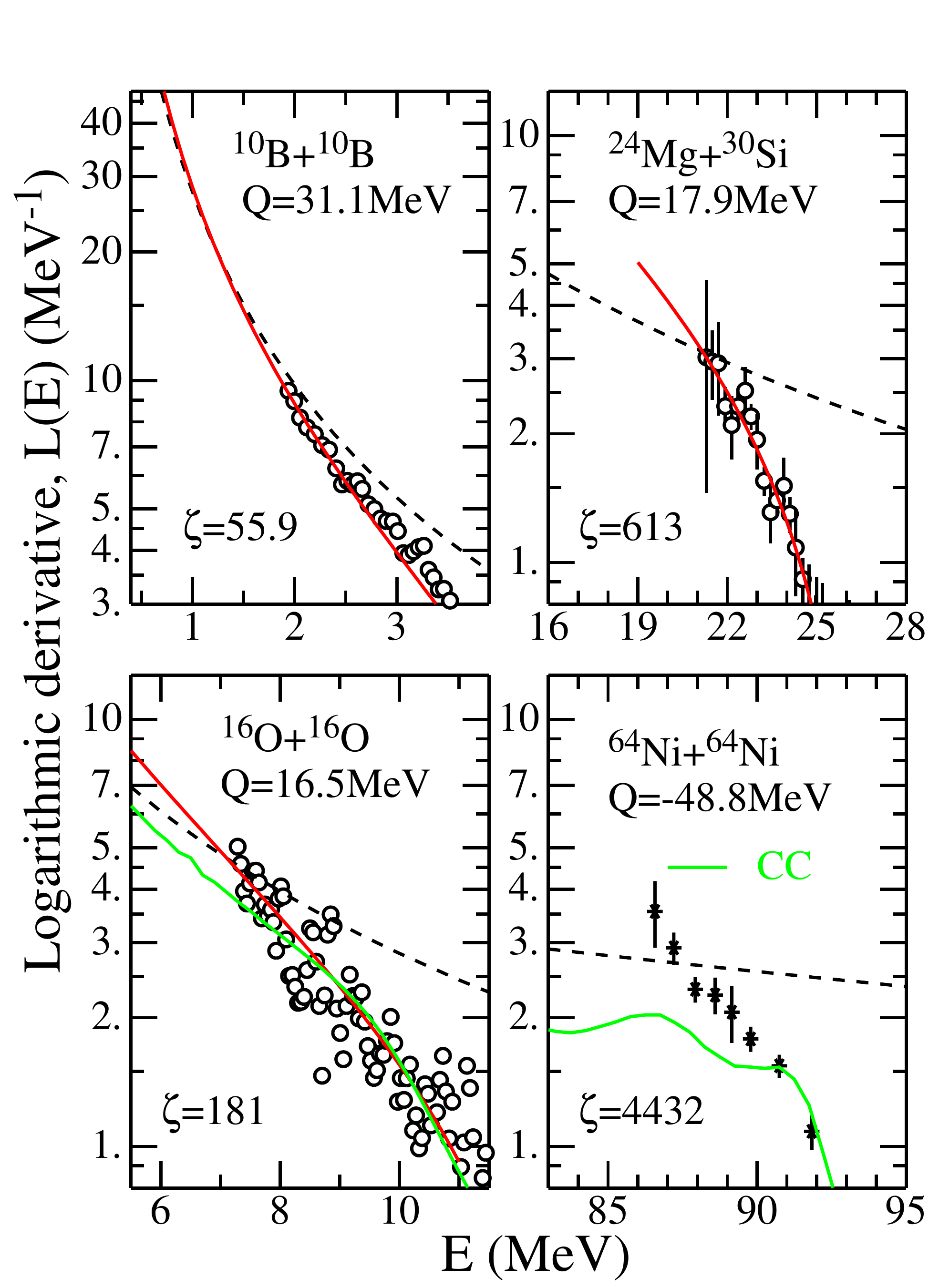}}
\caption{ Logarithmic derivative for various  heavy and light systems. In each case the Q-value for fusion is indicated,  as well as the system parameter $\zeta$= $Z_1Z_2\mu^{1/2}$. The dashed line is  $L_{CS}$(E), the red one is a simple extrapolation and the green lines are standard CC calculations (courtesy of C.L. Jiang).}
\label{9systems} 
\vspace{-3mm}    
\end{figure}

The astrophysically relevant reactions, like $^{12}$C + $^{12}$C, $^{16}$O + $^{16}$O,  systematically have Q$_{fus}>0$. 
For light systems it has been found that the slope $L(E)$ tends to become almost parallel with
$L_{CS}$(E)  and therefore the $S$ factor increases slowly with decreasing energy possibly reaching a plateau. Under such condition it is difficult to recognise an energy where $S$ has a maximum.
This situation is reported in Fig.~\ref{9systems} for various cases, both heavy and light systems. For example, we notice that the slope of $^{64}$Ni + $^{64}$Ni is very steep and crosses $L_{CS}$ with a large angle. Here the existence and the threshold of hindrance are quite well identified. On the contrary, in the light case of $^{10}$B + $^{10}$B the two curves $L(E)$ and $L_{CS}$ tend to overlap with each other at low energies.

\begin{figure}[h] 
\centering
\resizebox{0.45\textwidth}{!}{\includegraphics{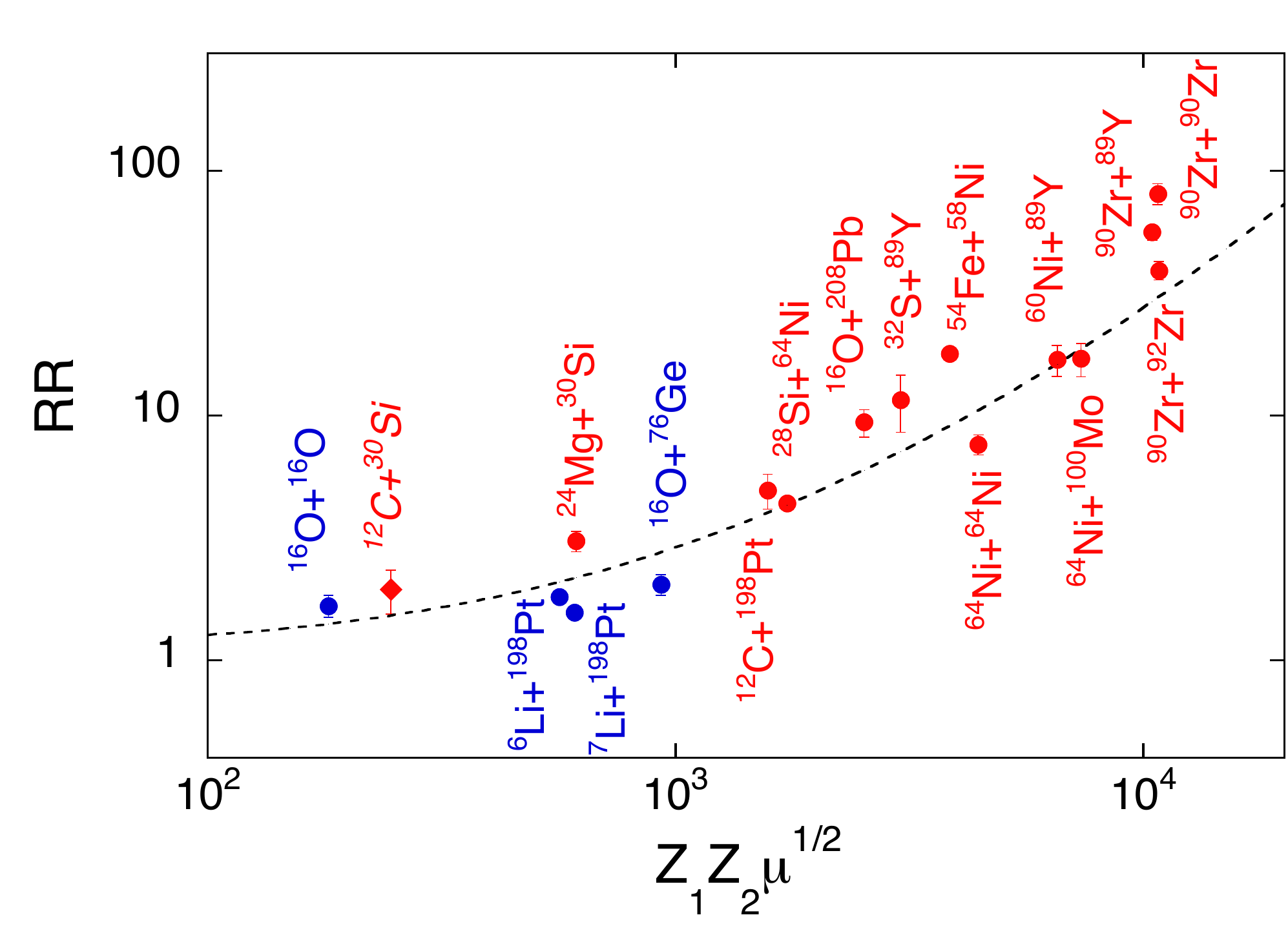}}
\caption{Ratio (RR) of energy derivatives of the slopes $L(E)$ and $L_{CS}$(E) at their crossing points, vs. the system parameter $\zeta = Z_1Z_2\mu^{1/2}$. The red dots are extracted from experimental data while the blue ones are extrapolated from data at higher energies. The red diamond is the value obtained from preliminary data on $^{12}$C + $^{30}$Si~\cite{1230}. The dashed line is the  curve resulting from the systematics discussed in Ref.~\cite{jiang07}.} 
\label{Shriva_slope}
\end{figure}

This general trend  was well represented by Jiang et al.~\cite{jiang07} introducing the quantity
\begin{equation}
RR=\frac{dL(E)/dE}{dL_{CS}(E)/dE}
\label{RR}
\end{equation}
i.e. RR is the ratio of the  energy derivatives of the slopes $L(E)$ and $L_{CS}$(E) at the crossing point (see  Fig.~\ref{Shriva_slope}). A sharp intersection point is observed for the heavier systems so that a well defined $S$ factor maximum develops. Instead, RR approaches unity for the lighter systems for which a less well defined maximum shows up. This does not imply that the fusion hindrance disappears in such cases, however, its existence may be more easily evidenced from comparing with standard CC calculations.


\begin{figure}[h] 
\centering
\resizebox{0.45\textwidth}{!}{\includegraphics{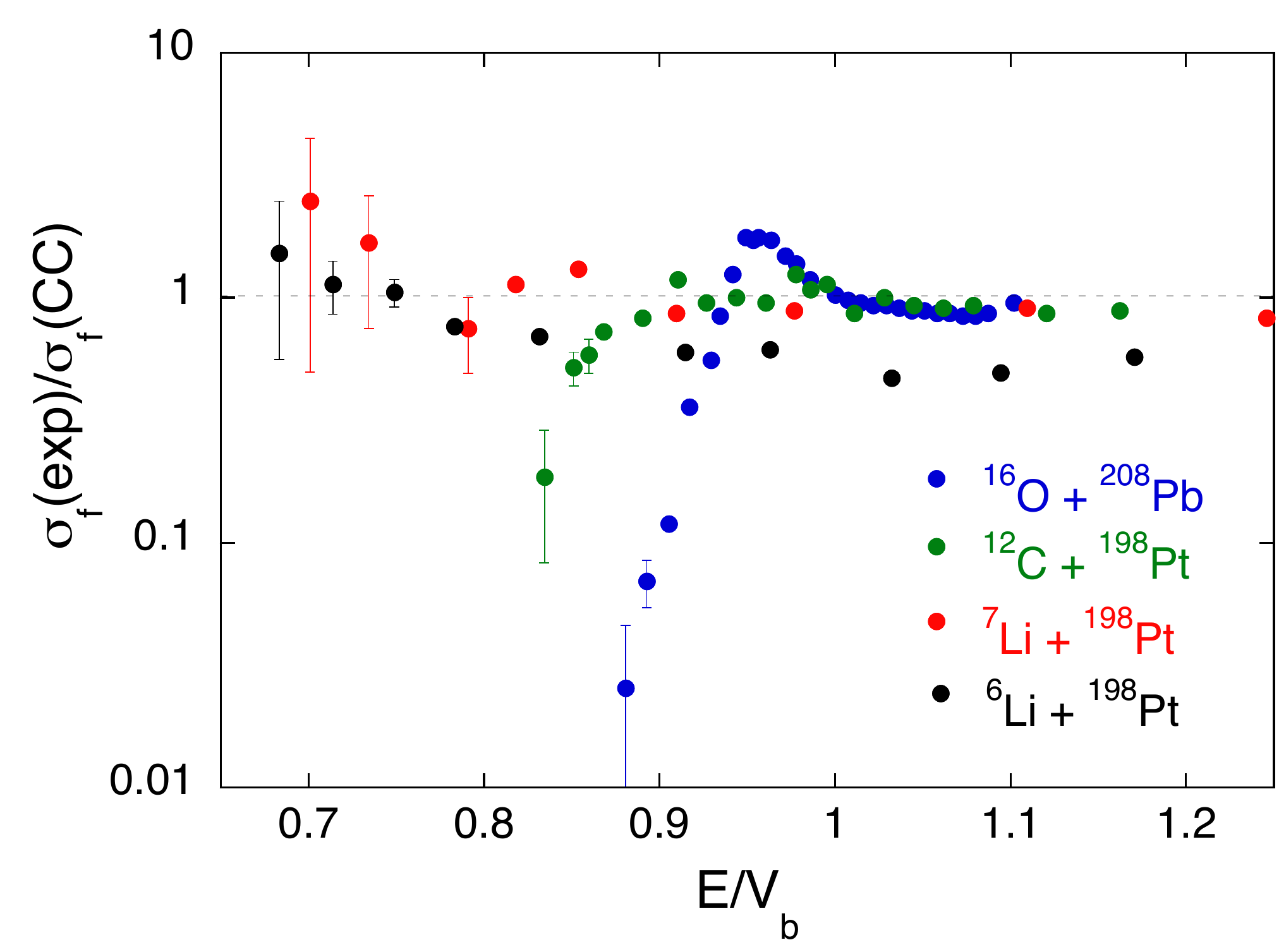}}
\caption{ Measured vs. calculated fusion cross sections as a function of the energy difference from the Coulomb barrier for $^{6,7}$Li + $^{198}$Pt~\cite{Shriva6,Shriva_PLB}, $^{12}$C + $^{198}$Pt~\cite{Shriva_PLB}, $^{16}$O + $^{208}$Pb~\cite{Nanda07,Morton} systems. The calculated values correspond to  CCFULL calculations using a standard WS potential. Figure redrawn from Ref.~\cite{Shriva_PLB}.} 
\label{systematics_Shriva}
\end{figure}
Shrivastava et al.~\cite{Shriva_PLB} made recent  studies of the way
fusion hindrance shows up and evolves with increasing mass and charge of light projectiles like $^{6,7}$Li and $^{12}$C on the  heavy target $^{198}$Pt. The case of $^{16}$O + $^{208}$Pb~\cite{Nanda07,Morton} was also re-analysed.
They measured fusion cross sections down to very small values ($\approx$100 nb) using an off-line method by detecting coincidences between characteristic X- and $\gamma$-rays (see also~\cite{Lemasson}).

Fig.~\ref{systematics_Shriva} shows that  the hindrance  effect is not observed for  $^{6,7}$Li,  but appears
and becomes more evident when going to the heavier $^{12}$C and  $^{16}$O projectiles.
 On the basis of this observed trend, it seems very important to study light and more symmetric  systems 
 where the existence of the hindrance phenomenon is still unclear and a matter of debate.
 
\section{Oscillations above the barrier}
\label{oscilla}

Oscillations were discovered  in the fusion excitation function of light heavy-ion systems like $^{12}$C + $^{12}$C, $^{12}$C + $^{16}$O and $^{16}$O + $^{16}$O~\cite{Sperr,Sperr2,Kovar,Tserruya}, in the energy region above the Coulomb barrier. Analogous oscillatory structures were observed in the elastic and inelastic data and were tentatively interpreted on the basis of quasi-molecular states in those systems.

It was suggested~\cite{Poffe} that the fusion oscillations are due to the overcoming of the centrifugal barriers due to the successive partial waves $L$ contributing to the total fusion cross section. Indeed in light systems the separation between the sequence of centrifugal barriers is large enough for the oscillations to be observable. Even the case of $^{20}$Ne + $^{20}$Ne~\cite{Poffe} seems to show oscillatory structures. In general,
$L$-dependent fusion barriers $V_B(L)$ can be parametrized as 

\begin{equation}
V_B(L)=V_{CB}+\frac{\hbar^2L(L+1)}{2\mu R_{CB}^2}
\label{partial}
\end{equation}

where $V_{CB}$ is the Coulomb barrier, $R_{CB}$ is its radius and $\mu$ is the reduced mass of the system. It follows that the energy difference between the heights of successive barriers is

\begin{equation}
\Delta V_B=V_B(L+1)-V_B(L)\approx\frac{\hbar^22(L+1)}{2\mu R_{CB}^2}
\label{endiff}
\end{equation}

\begin{figure}[h] 
\centering
\resizebox{0.45\textwidth}{!}{\includegraphics{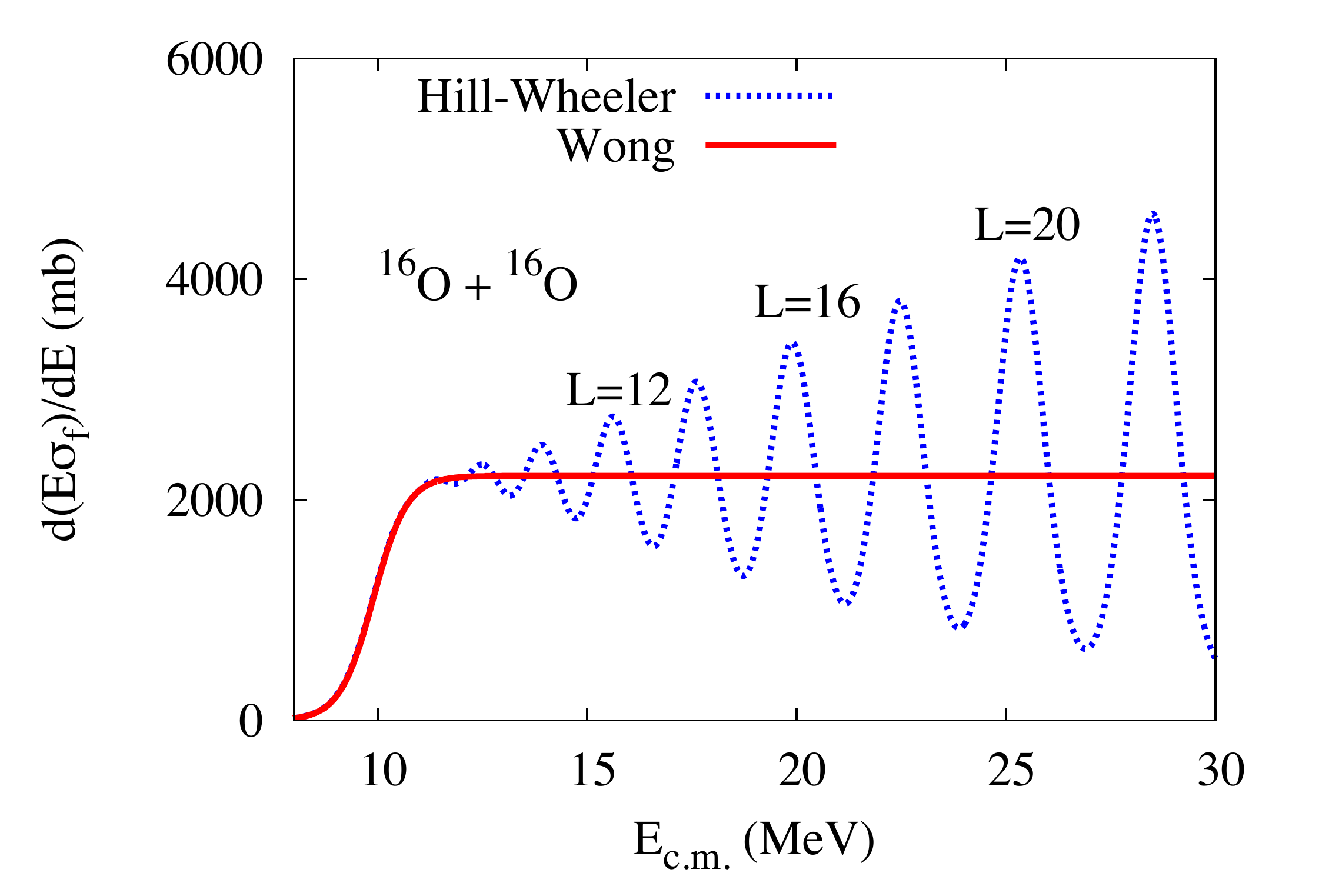}}
\caption{First derivative of the energy-weighted
fusion cross sections for $^{16}$O+$^{16}$O calculated from Hill-Wheeler's~\cite{HW}
and Wong's~\cite{Wong} formulae (see text). Figure from Ref.~\cite{Esbeosc} \copyright~American Physical Society (APS).
} 
\label{oscilla_Hen}
\end{figure}

For a symmetric system with two 0$^+$ ground state
nuclei, $\Delta V_B$ becomes twice as that because only even values of $L$ have to be considered. In the parabolic approximation of the Coulomb barrier, the width of the individual barriers is $\epsilon_o$=$\hbar\omega/2\pi$.
It is easy to estimate that for a light system like $^{16}$O + $^{16}$O $\Delta V_B\simeq$ 2.02 MeV and $2\epsilon_o\simeq$ 0.99 MeV, so that oscillations (if existing) are observable. In a case like $^{28}$Si + $^{28}$Si $\Delta V_B\simeq$ 1.52 MeV and $2\epsilon_o\simeq$ 1.13 MeV, and oscillations should still be observable. 

\begin{figure}[h] 
\centering
\resizebox{0.45\textwidth}{!}{\includegraphics{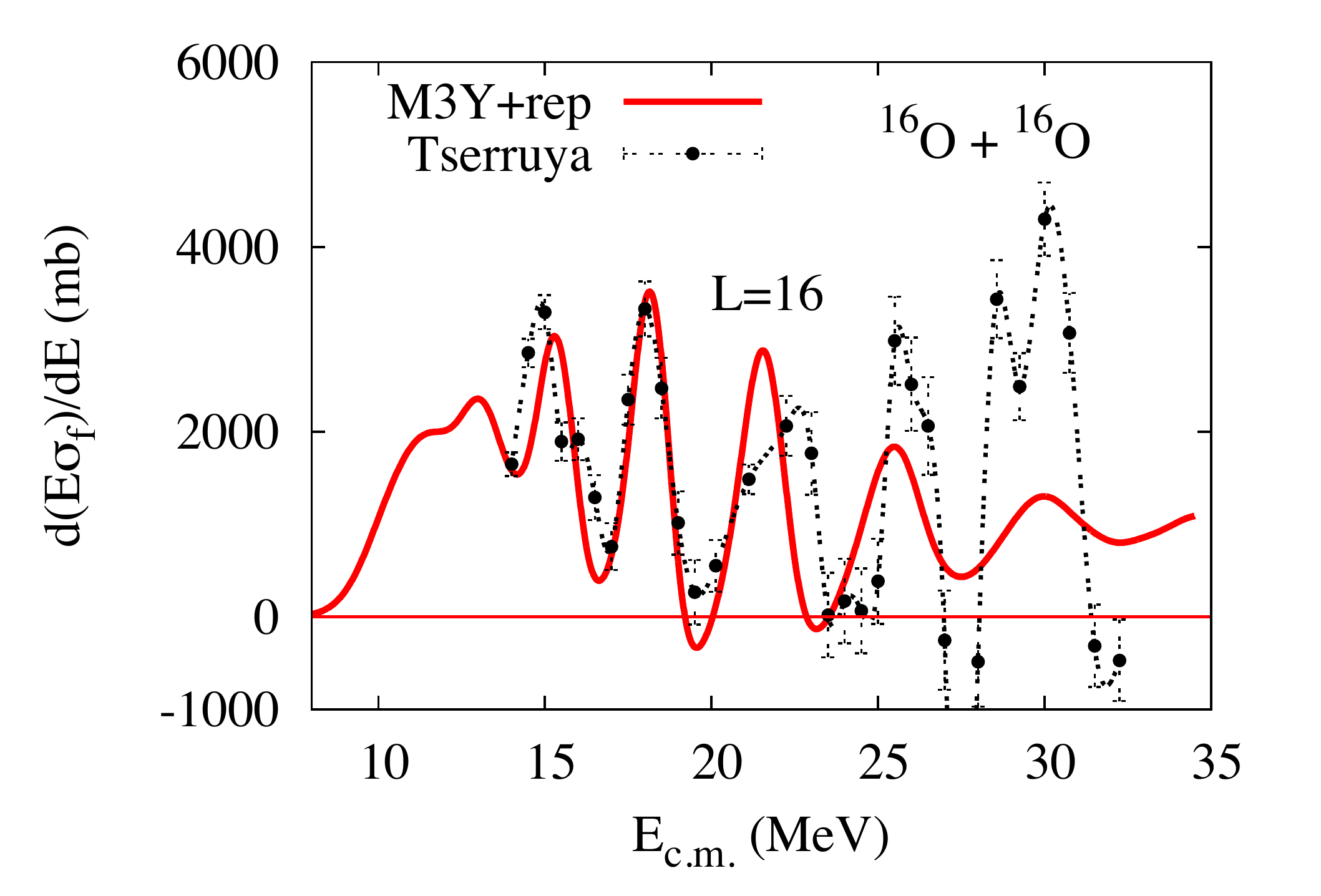}}
\caption{ Experimental data extracted from the measured excitation function of $^{16}$O+$^{16}$O~\cite{Tserruya}, compared
to CC calculations~\cite{Esben08} using  the M3Y+repulsion potential. Figure from Ref.~\cite{Esbeosc} \copyright~American Physical Society (APS).
} 
\label{oscilla_Hen2}
\end{figure}

When going to heavier systems (e.g. $^{40}$Ca + $^{40}$Ca) the distance between successive centrifugal barriers becomes comparable or smaller than 
the intrinsic energy width associated with the quantal penetration, so that oscillations of this kind turn out to be undetectable.
In heavy systems, the condition for separating the individual centrifugal barriers (see Eq.~\ref{endiff}) requires large $L$-values where however, many reaction channels open up and smear out the structures. 

In a recent paper~\cite{Esbeosc} (see also Simenel et al.~\cite{Simenel13}) this topic was re-analyzed. In Ref.~\cite{Esbeosc} Esbensen pointed out that using the first derivative of the energy-weighted excitation function $d(E\sigma_f)/dE$ makes it easier to observe oscillations. 

We consider $^{16}$O + $^{16}$O to illustrate the essential features of this phenomenon.
Fig.~\ref{oscilla_Hen} shows the calculated derivative for $^{16}$O + $^{16}$O using Wong's formula~\cite{Wong} (red line), i.e. 

\begin{equation}
\Big(\frac{d(E\sigma_f)}{dE}\Big)_W=\pi R_{CB}^2\frac{exp(x_0)}{1+exp(x_0)}
\label{endiff2}
\end{equation}

where $x_0=(E-V_{CB})/\epsilon_0$. This derivative is proportional to a Fermi function and does not show any oscillation. Above the Coulomb barrier, it approaches the value $\pi R_{CB}^2$.  The information on the individual centrifugal barriers is lost when applying  Wong's formula. The dotted line is the derivative obtained using  Hill-Wheeler's formula~\cite{HW}

\begin{equation}
\Big(\frac{d(E\sigma_f)}{dE}\Big)_{HW}=\frac{\pi\hbar^2}{2\mu}\sum\limits_{L}(2L+1)\frac{1}{\epsilon_L}\frac{exp(x_L)}{(1+exp(x_L))^2}
\label{endiff3}
\end{equation}

that almost coincides with the previous derivative below and near the barrier, and then starts oscillating above. The various peaks are the calculated locations of the individual centrifugal barriers (the $L$ = 12, 16, and 20 peaks are marked). The experimental derivative shows very clear oscillations (see Fig.~\ref{oscilla_Hen2})
that are nicely reproduced by employing an M3Y+repulsion shallow potential in CC calculations that also fit the sub-barrier excitation function of $^{16}$O + $^{16}$O. We remind that also the oscillations in the quasi-elastic data were associated to shallow ion-ion potentials~\cite{Esben08}.

\begin{figure}[h] 
\centering
\resizebox{0.45\textwidth}{!}{\includegraphics{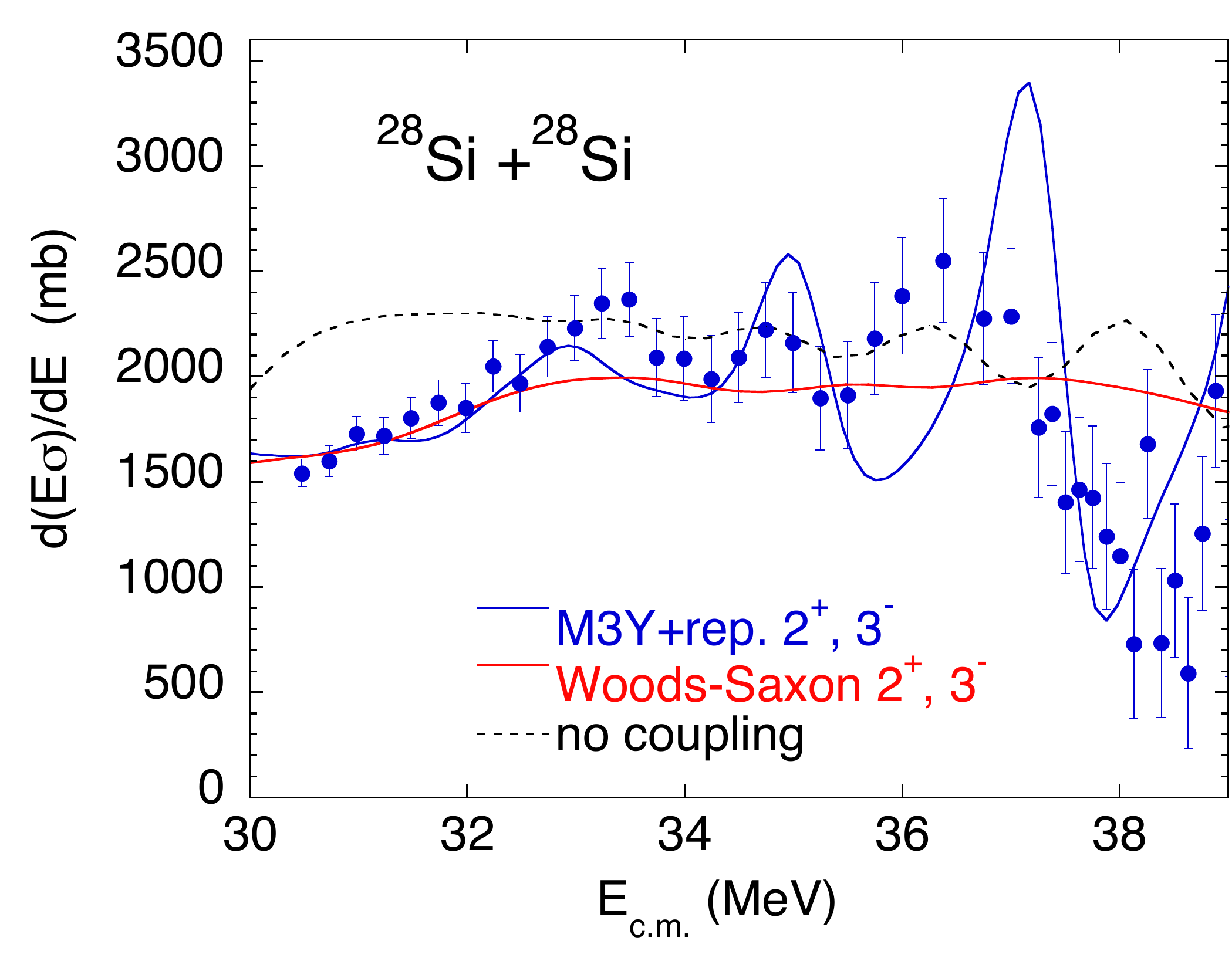}}
\caption{ The first derivative of the  energy-weighted fusion excitation function $d(E\sigma)/dE$ of $^{28}$Si+$^{28}$Si shows distinct oscillations above the barrier. The derivative was obtained as the incremental ratio
between successive points, with an energy step $\Delta$E$_{c.m.}$ = 0.75 MeV.
A comparison with CC calculations using the M3Y+rep. and the WS potentials is also shown (see text). Figure modified and redrawn from Ref.~\cite{oscilla}.} 
\label{oscilla_rev}
\end{figure}

When going to heavier systems,
these interesting results provide very useful information on the potential and on coupling effects in a wide energy range. That appears to be especially true for heavier systems where sub-barrier fusion enhancements (couplings) are stronger and hindrance may show up. Therefore, detailed measurements were performed of the above-barrier excitation function in $^{28}$Si + $^{28}$Si~\cite{oscilla},
 where previous data~\cite{Gary,Agui} did not allow any clear-cut conclusion about the existence of oscillations.
The basic result is shown in Fig.~\ref{oscilla_rev}, where the first derivative $d(E\sigma_f)/dE$ reveals three rather regular oscillations. Such oscillations $and$ the sub-barrier cross sections were reproduced  within the CC model~\cite{2830} using the same shallow M3Y+repulsion potential. 

The analogous calculation where no couplings are included (black dashed line), shows only weak oscillations. Even more flat is the result of the calculation including the 2$^+$ and 3$^-$ states of the two nuclei, but where a Woods-Saxon potential is used. The conclusion was that
in this relatively heavy system 1) a shallow potential is needed to fit the sub-barrier fusion behaviour $and$ the oscillatory structures above the barrier, and 2) the oscillations are associated to channel couplings, while in lighter cases  they are related to the overcoming of successive centrifugal barriers well spaced in energy. Indeed, it was pointed out that
in the oscillations observed for $^{28}$Si + $^{28}$Si the one-to-one relation  between each peak and the height of a centrifugal barrier is lost because of strong coupling effects. The stable oblate deformation of $^{28}$Si may play a role in this.
\par

Therefore, analogous measurements involving $^{30}$Si which is a spherical nucleus, are in progress.

\section{Radioactive beams and new set-ups}
\label{radioactive}

Several detailed experiments and theoretical investigations have been performed in recent years using light weakly-bound radioactive beams, by exploiting the fast developing techniques and the new facilities that are being installed  in several countries. Despite the great interest and far-reaching implications of such studies, they are outside the scope of the present review, as anticipated in the Introduction, therefore we send the reader to Refs.~\cite{BBB,Liang,Canto,Kolata16} for detailed reports of the results obtained  with those light exotic beams. 
Here we limit ourselves to describing  a few measurements where medium-heavy or heavy exotic beams were used.

We want to mention the early experiments performed at the National Superconducting Cyclotron Laboratory
(NSCL) at Michigan State University, where beams of $^{38}$S were produced by projectile fragmentation, and sent onto heavy targets ($^{208}$Pb and $^{181}$Ta) for studying capture-fission reactions at energies around and above the Coulomb barrier~\cite{Love2006,Zyro}.
Comparing with previous studies of $^{32}$S + $^{208}$Pb allowed to conclude that the interaction barrier for the system with the very neutron-rich projectile is substantially lower ($\Delta V_b\simeq16\pm10$ MeV), and that, consequently, this might have significant implications for the synthesis of very heavy nuclei using exotic radioactive beams.

We have introduced in Sect.~\ref{BDqe} the concept of the fusion barrier distribution extracted from measurements of the backward-angle quasi-elastic scattering D$_{qe}$. Such experiments, as shown in several significant cases, have proved to be very useful in the analysis of fusion dynamics as a complementary tool to detailed measurements of fusion cross sections.

We want to point out that the method of  extracting the barrier distribution from quasi-elastic data might be very convenient when performing  fusion investigations by means of exotic beams, because of the relatively low intensity presently available with such beams. This is especially true when going  into the near- and sub-barrier energy range where fusion cross sections  decrease by orders of magnitude, while quasi elastic scattering yields increase substantially   
and approach the Rutherford values. Moreover, directly  detecting  fusion evaporation residues (ER)  at forward angles requires in any case rejecting the beam, and separating the ER from beam-like particles with an efficiency as close to 100$\%$ as possible.

It should be clear that this method is valid for light or medium heavy systems where the reaction channels like deep inelastic or quasi-fission can  safely be neglected near and below the barrier. For heavier cases one has to take into account the remarks of Pollarolo and Zagrebaev~\cite{Polla_QE,Zagre}, so that D$_{qe}$ is not a good representation of the fusion barrier distribution any more
and one should take care in the interpretation of back scattering experiments in this sense.

\subsection{Recent results with heavy exotic beams }
\label{exoticheavy}
 
A few years ago the Holifield Radioactive
Ion Beam Facility (HRIBF) at Oak Ridge National Laboratory developed very neutron rich exotic nuclei like $^{132}$Sn and  $^{130,134}$Te using the ISOL technique.  They were employed in a series of experiments with lighter targets as $^{58,64}$Ni and $^{40,48}$Ca for the investigation of the effect of transfer couplings on sub-barrier fusion in cases where the neutron excess is very large and consequently the transfer Q-values are very positive even for 12-14 transferred neutrons.

\begin{figure}[h] 
\centering
\resizebox{0.45\textwidth}{!}{\includegraphics{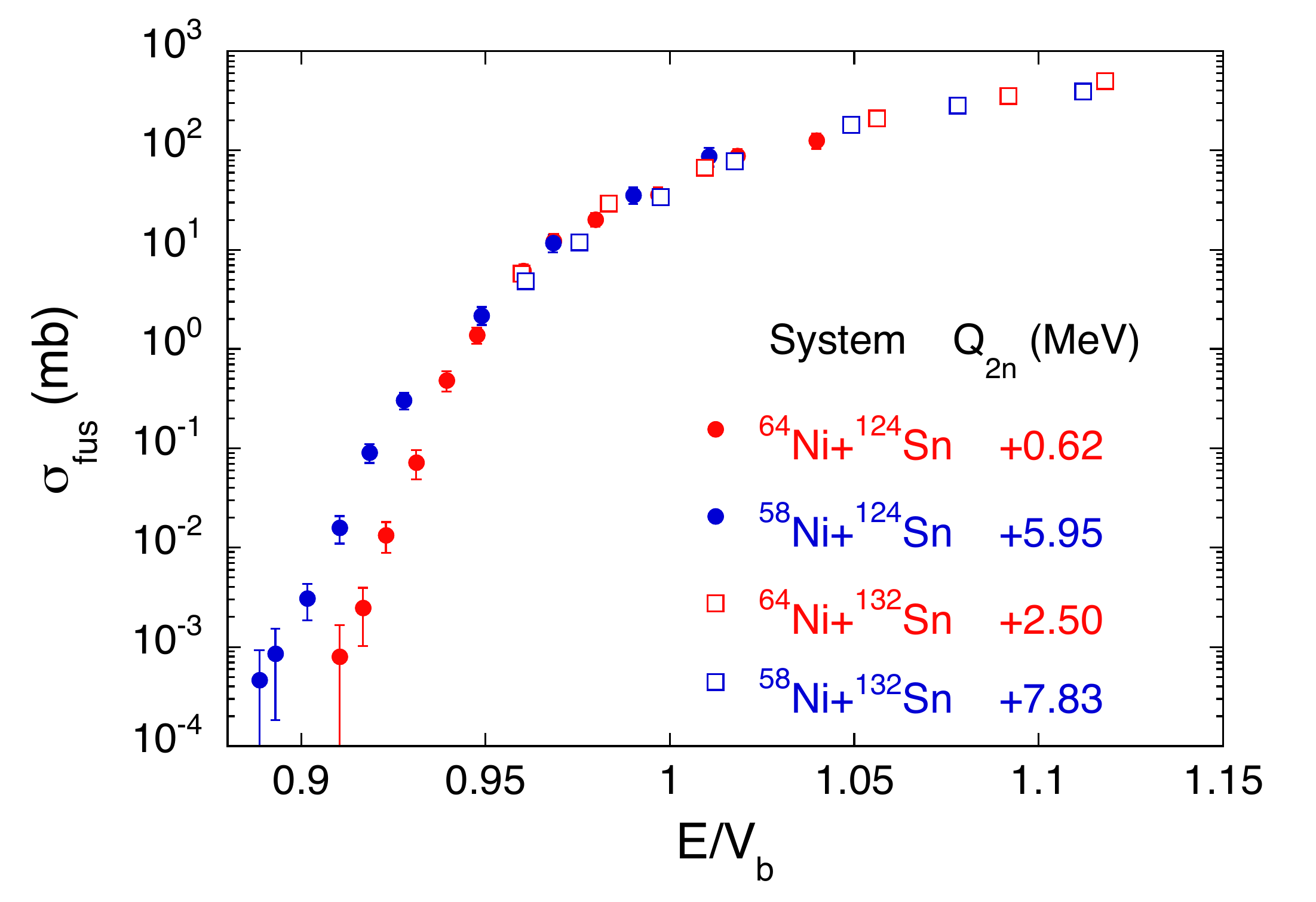}}
\caption{Fusion excitation functions for $^{58,64}$Ni + $^{124}$Sn~\cite{Jiang_15} and for $^{132}$Sn + $^{58,64}$Ni~\cite{Liang03,Liang07}. The abscissa is the distance of the energy from the Aky\"uz-Winther Coulomb barrier.
}
\label{Liang0307}
\end{figure}

First experiments concerned  $^{132}$Sn + $^{64}$Ni~\cite{Liang03,Liang07} and, later on, $^{132}$Sn + $^{58}$Ni and  $^{130}$Te + $^{58,64}$Ni~\cite{Koh11}. The results did not display any particular enhancement that could be attributed to transfer couplings (see the comments in Sect.\ref{Heavier}).  A small systematics is reported in  Fig.~\ref{Liang0307}, showing the excitation functions measured with stable beams~\cite{Jiang_15} and the results obtained using the exotic $^{132}$Sn, for various Ni+Sn systems. The Q-values for two-neutron transfer are also reported for each case, as representative values of the overall Q-value situation. One has clear indication that in these rather heavy systems, the extra-enhancement possibly due to transfer couplings becomes appreciable only at rather low energies (where the cross section is smaller than $\approx 1$ mb).

\begin{figure}[h] 
\centering
\resizebox{0.45\textwidth}{!}{\includegraphics{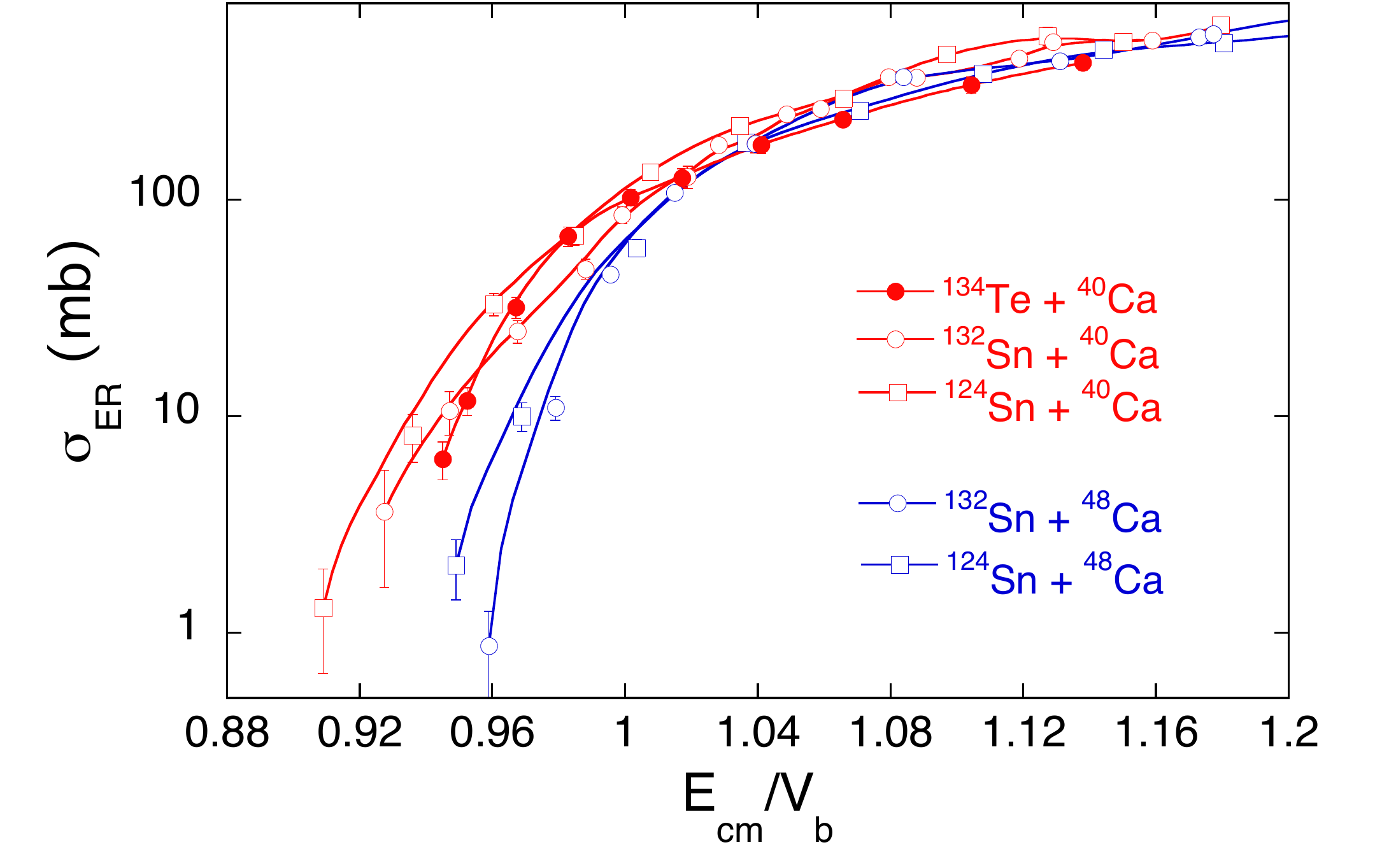}}
\caption{Evaporation residue (ER) cross sections for $^{134}$Te + $^{40}$Ca~\cite{Kohley13} and for $^{132,124}$Sn + $^{40,48}$Ca~\cite{Kolata12}. The energy scale is normalised to the Aky\"uz-Winther Coulomb barrier. Modified and redrawn from Ref.~\cite{Kohley13}.
}
\label{Kohsig}
\end{figure}

More recent measurements were performed using Ca targets. 
In Fig.~\ref{Kohsig} we report the ER cross sections of the $^{134}$Te + $^{40}$Ca~\cite{Kohley13}  and of several Sn + Ca systems involving stable and radioactive Sn isotopes. In that figure the excitation functions  are in an  absolute scale and not normalised to the barrier radius as in the original work~\cite{Kohley13}, because the effect is very small and the conclusions are anyway not modified, but one has the advantage of  directly reading the measured cross sections.

The  excitation functions  are clearly divided in two groups where the systems $^{132,124}$Sn + $^{48}$Ca, with only negative Q-values for neutron transfer channels, have smaller cross section at the same distance from the barrier, with respect to the other cases plotted with red symbols where many 
positive Q-value transfer channels exist. 

\begin{figure}[h] 
\centering
\resizebox{0.45\textwidth}{!}{\includegraphics{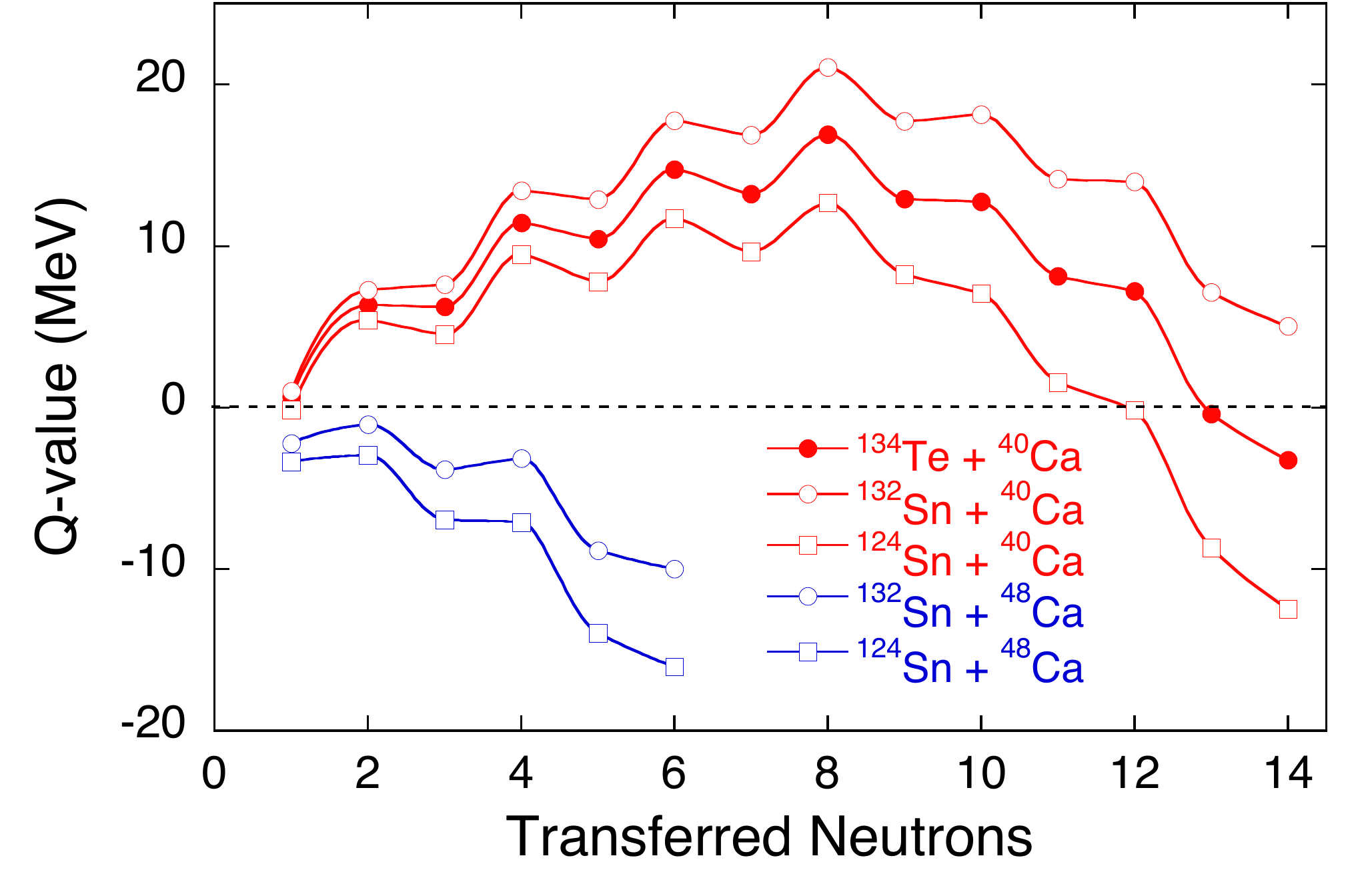}}
\caption{Ground state Q-values for increasing number of stripped neutrons for several systems involving stable ($^{124}$Sn) and exotic ($^{132}$Sn, $^{134}$Te) projectiles on $^{40,48}$Ca~\cite{Kohley13}. Modified and redrawn from Ref.~\cite{Kohley13}.} 
\label{KohQ}
\end{figure} 

The systematics of ground state Q-values is shown in the following Fig.~\ref{KohQ}. One notices that, for example, the Q-value for neutron transfer  in  $^{132}$Sn + $^{40}$Ca  is larger than + 20 MeV for eight neutrons,  and it is still positive at the level of 14 transferred neutrons.

Let us discuss  Fig.~\ref{Kohsig} from a purely qualitative point of view. The evidence is that
coupling to transfer channels strongly influences sub-barrier cross sections in the Ca + Sn systems but only when the Q-value is positive. This was already investigated and clarified in studies with stable beams (we have already  seen in  Sect.~\ref{InfTransf} that for example the systems $^{40}$Ca + $^{124}$Sn,  $^{96}$Zr have very large enhancements below the barrier). The  availability of very neutron rich exotic  beams like $^{132}$Sn and $^{134}$Te has allowed to confirm this evidence,  but it also appears that larger and larger positive Q values do not necessarily imply additional enhancement. Indeed, among the group of systems with positive Q-values, $^{40}$Ca + $^{132}$Sn, $^{134}$Te do not display larger sub-barrier cross sections than $^{40}$Ca + $^{124}$Sn.

In any case the nuclear structure continues to be very important because we have discussed above that in the experiments where $^{132}$Sn was sent on $^{58,64}$Ni targets~\cite{Liang03,Liang07} no particular effect due to transfer could be noticed because, in such heavy systems, it becomes observable only at the level of 1 mb or less (see the recent data of Ref.~\cite{Jiang_15}, Fig.~\ref{Liang0307}). At higher energies couplings to strong collective modes are dominant. Obviously measurements of very small cross sections  would require  beam intensities higher than what is presently available with radioactive ions, at the operating facilities.

 
\subsection{Facilities and experimental set-ups}
\label{future}

Several radioactive beam facilities are presently under construction all around the world, that will be able to produce heavy exotic beams with rather high intensities, within a few years. One can find updated situations of the various facilities  in Ref.~\cite{RIB}.
In particular we would like to mention, because of its  peculiarity, the project CARIBU (CAlifornium Rare Isotope Breeder Upgrade)
 that presently is under development at ANL~\cite{CARIBU}. Indeed,  its originality relies in the fact that no primary accelerator is employed, rather, the short-lived  neutron-rich  isotopes  will be  produced by a 1 Ci  
$^{252}$Cf  fission  source  located  in  a  large  gas  catcher.
 The  fission fragments  will be  transferred  to  an  ECR  ion  source  for  charge  breeding,  before  acceleration  in  the  
ATLAS  superconducting  linac.  Beam intensities in excess of 10$^{5}$  ions per second on  target are expected, at  energies  $\simeq$7-17 MeV/A, for a wide range of neutron rich nuclei.

We present here below a selection of new or recent experimental set ups, that in our opinion are relevant for the topics covered by this review. 
 We are aware that this selection is far from being exhaustive, however we have chosen to limit ourselves to a few laboratories where significant research activity in the field has been carried out in recent years.
 Some of these set ups are in the phase of commissioning or have recently come into operation producing interesting results.


AGFA is a gas-filled separator very recently developed and installed at Argonne, based on an innovative quadrupole-dipole design, with an overall length of $\simeq$ 4 m, with the following main features: 1) high efficiency (up to $\approx$70$\%$) for evaporation residues detection,  2) small image size ($\sim$64$\times$64 $mm^2$) at the focal plane, where a large double-sided Si strip detector is mounted, 3) a maximum B$\rho$ of 2.5 Tm (bending angle 38$^{\circ}$)
 and 4) the ability to work in a combined set-up with Gammasphere and/or with a gas catcher for the production of exotic beams of radioactive ions. The solid angle of AGFA may exceed 40 msr in stand-alone mode. First experiments are expected to run within 2017.

\begin{figure}[h]
\centering
\resizebox{0.40\textwidth}{!}{\includegraphics{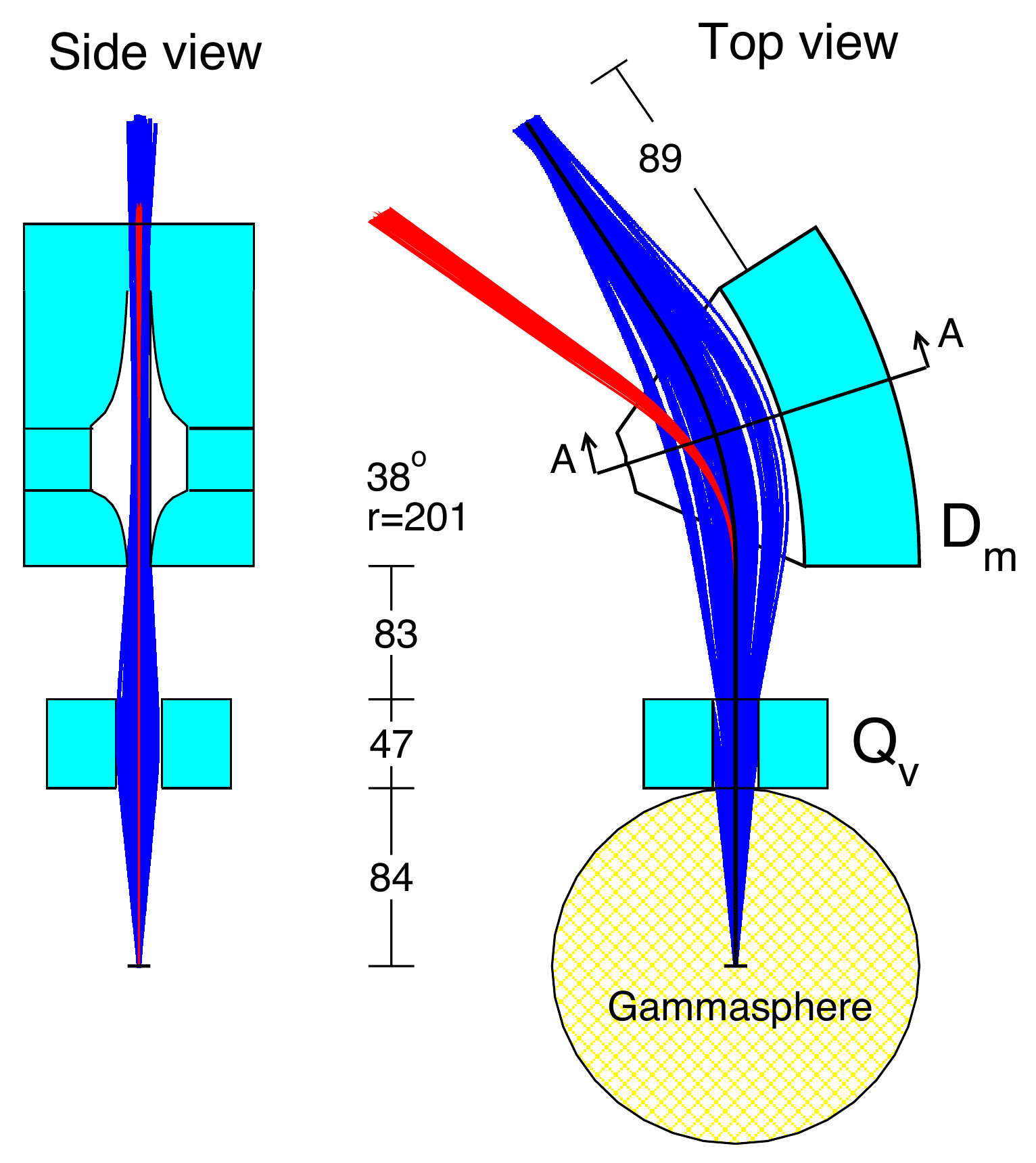}}
\caption{Schematic view of the Argonne Gas-Filled Fragment Analyzer (AGFA)~\cite{AGFA}. This  spectrometer is a new instrument installed at ANL, presently available to operate in conjunction with Gammasphere or in stand-alone mode.}
\label{AGFA}     
\end{figure}

\begin{figure}[h]
\centering
\resizebox{0.40\textwidth}{!}{\includegraphics{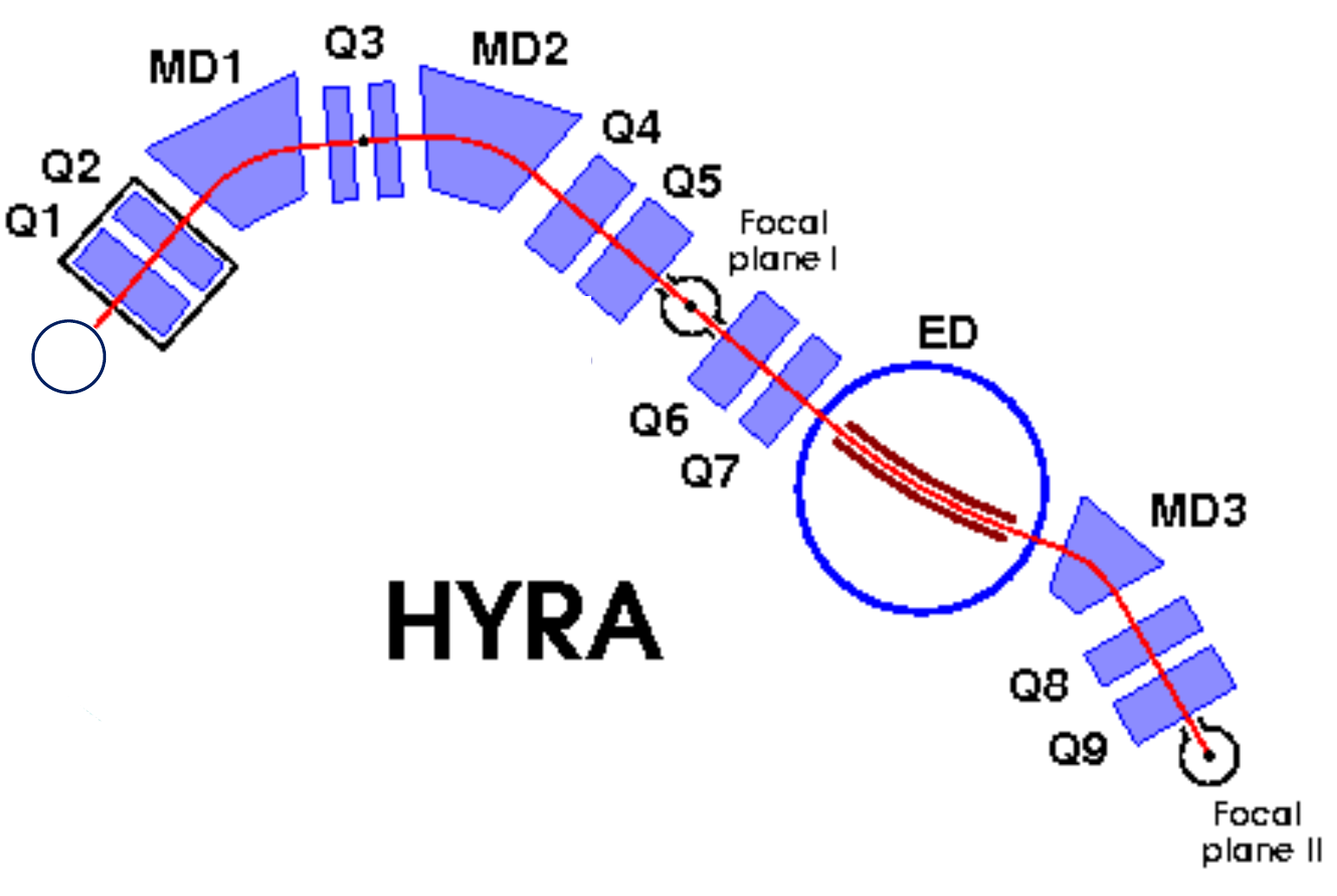}}
\caption{The hybrid recoil mass analyzer (HYRA)~\cite{HYRA} at the IUAC in New Dehli.}
\label{Hyra}     
\end{figure}

We show in Fig.~\ref{Hyra} HYRA which is a  HYbrid Recoil mass Analyzer  downstream of the superconducting linear accelerator, at the Inter University Accelerator Centre in New Dehli~\cite{HYRA}. It is a dual mode and dual stage spectrometer/separator where the first stage can operate in gas-filled mode in normal kinematics (to identify heavy nuclei $\geq$200 amu) and both stages can work in vacuum mode using inverse kinematics (to identify nuclei with N $\sim$ Z up to 100 amu  and to provide light, secondary beams produced in direct reactions). 

The first stage of HYRA has the configuration Q1Q2-MD1-Q3-MD2-Q4Q5 (we indicate with Q the
magnetic quadrupoles and with MD the magnetic dipoles,
respectively). It operates with momentum dispersion in gas-filled mode or as a momentum achromat in vacuum mode. The second stage consists of Q6Q7-ED-MD3-Q8Q9, producing a  mass dispersion by the cancellation of energy dispersion at a fixed focal point.

 The first stage of HYRA in gas-filled mode, in particular,  was used for several  measurements  in the last few years.
We would like to mention the measurements
of spin distributions and cross sections of evaporation residues in  $^{28}$Si +$^{176}$Yb by Sudarshan et al.~\cite{Suda},
and the experiments on fusion-evaporation concerning the $^{16}$O + $^{194}$Pt~\cite{Pras},
$^{19}$F + $^{194,196,198}$Pt~\cite{Vari} and $^{31}$P + $^{170}$Er~\cite{Moha} systems.

\begin{figure}[h]
\centering
\resizebox{0.45\textwidth}{!}{\includegraphics{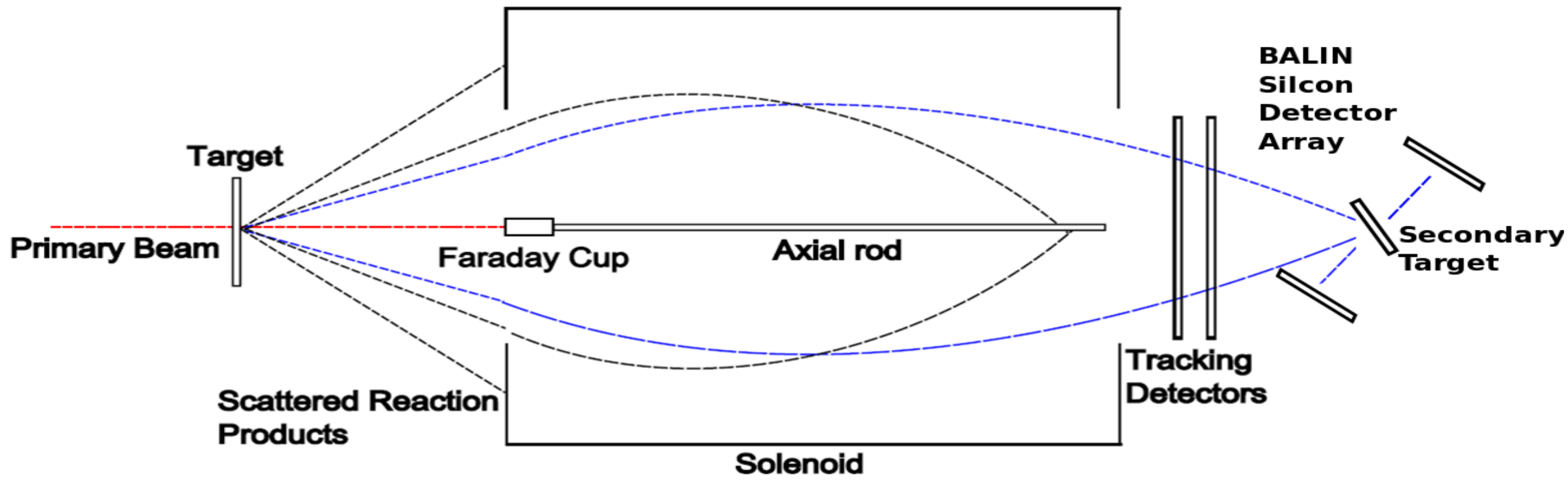}}
\caption{Scheme of the SOLEROO separator (SOLenoidal Exotic Rare isOtOpe separator). It is used to separate RIB species from other reaction products, and elastically scattered particles are stopped on the axial rod~\cite{Carter}.
}
\label{SOLEROO}     
\end{figure}

In Sect.~\ref{Heavier} we have mentioned the new apparatus SOLITAIRE dedicated to the investigation of fusion reactions at ANU, based on a superconducting solenoid.  The recent development of SOLITAIRE for the production of light radioactive ion beams is called  SOLEROO which is described in detail  in  Ref.~\cite{Rafiei} (see Fig.~\ref{SOLEROO} for a schematic picture). It uses a 6.5 T superconducting solenoid as the separator element to reject the large background of primary-beam particles from the radioactive species of interest.
 First measurements  were carried out using elastic scattering of a $^8$Li radioactive ion beam from a $^{197}$Au target~\cite{Carter}.

\begin{figure}[h]
\centering
\resizebox{0.45\textwidth}{!}{\includegraphics{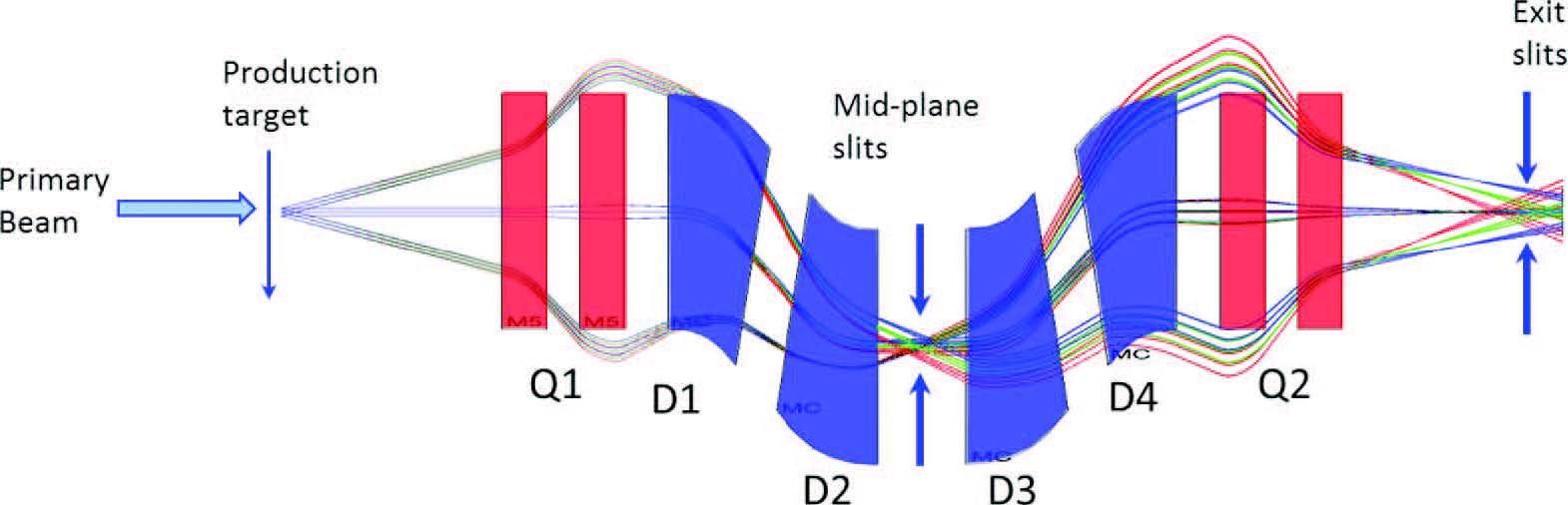}}
\caption{Scheme of the planned In-flight Radioactive Ion Separator AIRIS at Argonne National Laboratory. Adapted from Ref.~\cite{CARIBU}.}
\label{AIRIS}     
\end{figure}

 AIRIS is a dedicated mass separator including a magnetic chicane, an RF sweeper, and a buncher/rebuncher resonator, under development at ANL~\cite{CARIBU}. It will allow for improved transmission of radioactive beams produced in-flight providing access to new regions of secondary beams both higher in mass ($\leq$60) and further from stability than previously possible at ATLAS. A schematic picture of AIRIS is shown in Fig.~\ref{AIRIS}.

\section{Concluding remarks}
\label{concluding}

This review has been focused on recent results obtained in the field of heavy-ion fusion near and below the Coulomb barrier.  
The dynamics of heavy-ion fusion is a matter of continuing experimental and theoretical studies, because it allows a deep insight into the fundamental problem of quantum tunnelling of many-body systems in the presence of coupled degrees of freedom. 
A variety of phenomena have been observed, originating from the close link existing between nuclear structure and reaction dynamics.
Due to the intrinsic complexity of the phenomenon and to the several experimental challenges that it implies, many issues are still awaiting a clear solution within theoretical models.

It is well established that fusion cross sections in that energy range are primarily determined by couplings of the relative motion of the two colliding nuclei to their internal degrees of freedom like stable deformations, low-energy collective surface vibrations and, in some cases, to nucleon transfer channels~\cite{BBB,AnnuRev}.
Such couplings effectively split the original
Coulomb barrier into a distribution of barriers~\cite{rowBD}, where some barriers are lower than the original one. This causes orders-of-magnitude enhancements of the near- and sub-barrier  cross sections. 

Fusion barrier distributions have been extracted from excitation functions for many systems, providing a fingerprint of the couplings involved in several different conditions. The complementary method of measuring backward-angle scattering yields for the determination of barrier distributions, has been successfully used for medium mass or medium-light systems, where the method is justified to a great degree.

Detailed information on the various coupling effects can be extracted, closely correlating 
with the variations of nuclear structure when the magic numbers of the shell model are encountered, or when collective modes dominate, or when going from stable to exotic neutron-rich or proton-rich colliding nuclei. Indeed, the availability of radioactive beams is beginning to open unprecedented possibilities for the study of fusion reactions, despite the rather low intensity of the beams available so far.
\par
We have described the results of the measurements carried out in recent years at the HRIBF of ORNL using heavy exotic beams in the vicinity of $^{132}$Sn.  Those results have been of great help for clarifying the role of nucleon transfer in sub-barrier cross sections. Nevertheless, we have seen, on the basis of several examples, that a quantitative description of the influence of transfer on fusion is still elusive.
Further important help in this sense will come from the use of exotic beams with higher intensities than available now, and experiments using  lighter neutron-rich {\it and} proton-rich ions will certainly bring valuable contributions.
\par
From the theoretical point of view, as an alternative to the CC model, the approach of performing microscopic calculations using the TDHF method is increasingly used in recent years. This method is a promising tool for the description of heavy-ion fusion, even if difficulties for its application to energies below the barrier, have still to be overcome.
\par
Additional phenomena show up in heavy-ion fusion because the ion-ion potential is not known ``a priori", and is modified by nuclear structure and dynamically by the interaction. This is the case of the fusion hindrance effect far below the barrier, that has been the object of several investigations in recent years. In this case, the early claim that low-energy hindrance is a general phenomenon in heavy-ion fusion is probably correct, however, its strength and features can greatly vary from one system to another, and it appears that further detailed studies are needed. 
\par
For medium-light system, in particular, where $Q_{fus}$  are positive, the existence of a maximum of the $S$ factor as a function of the energy (not mathematically necessary) is still awaiting a clear-cut experimental proof, as well as the shape of the $S$ factor close to the maximum. This holds also for the cases where no maximum develops in the measured energy range, but hindrance has anyway been established on the basis of the comparison with the cross sections predicted by standard CC calculations. 
\par
The significance of the possible existence of hindrance for the late evolution of certain classes of stars has been emphasised  and studies of light systems are in progress,  paving the way to reliable extrapolations toward systems of astrophysical interest like $^{12}$C + $^{12}$C and  $^{16}$O + $^{16}$O.
\par
It is not obvious that the hindrance phenomenon is better described within a so-called sudden model using a CC approach and a two-body potential producing a shallow potential pocket, or if an adiabatic approach is more appropriate, 
taking into account the value of the potential at the touching point, and a damping of the coupling form factors for lower energies. Discriminating between these two models will require challenging measurements of very small cross sections and/or equally difficult experiments aiming at the determination of the angular momentum of the compound nucleus, in the sub-barrier energy range. 
\par
An interesting connection is being investigated between fusion hindrance, the influence of transfer reactions on fusion and the Pauli exclusion principle. We have seen that fusion hindrance is not observed down to very low energies in systems with large and positive Q-values for transfer (e.g. $^{40}$Ca + $^{96}$Zr~\cite{esbcazr}),  i.~e., in systems where the nucleons  can flow from one nucleus to the other without been hindered by the Pauli blocking. 

This is a hint that fusion hindrance and Pauli blocking are someway connected.  
In Ref.~\cite{esbcazr} it was pointed out that a theoretical procedure to explicitly calculate the repulsive part of the ion-ion potential by using the Pauli exclusion principle, would be very useful. A very recent paper of Simenel et al.~\cite{Sime17} goes in this direction because  the bare nucleus-nucleus potential  they have calculated by means of the density-constrained frozen Hartree-Fock  method, has a repulsive core inside the Coulomb barrier due to the Pauli blocking. This would lead to the conclusion that this effect  is responsible, at least partially, for the fusion hindrance phenomenon. 

As an overall final observation, and an outlook for the future, we remark that, in spite of the large amount of experimental and theoretical work that has been dedicated to heavy-ion fusion near and below the Coulomb barrier in the last three decades, there are several open questions and problems at issue in this field. Heavy-ion fusion dynamics is intimately linked to nuclear structure, and its complexity is evidenced by the variety of
features that are still missing a satisfactory theoretical explanation. It appears that in the next few years further experiments will have to be performed using advanced set-ups and high-quality exotic beams, and, extrapolating from what happened in the past, we may also anticipate that unexpected phenomena and new effects will show up.

\section{Acknowledgements}

We are grateful to all colleagues  of the Laboratori Nazionali di Legnaro of INFN, without whose professional and invaluable help all our research activity and this review paper would have been impossible to perform and report. Special thanks are addressed to  L. Corradi, E. Fioretto, F. Scarlassara, S. Szilner, whose collaboration has been essential for performing the experiments at LNL. 
We also wish to acknowlegde several other experimental and theoretical colleagues for their valuable contributions to several aspects of the work presented in this article.

\end{document}